\documentclass[10pt,a4paper]{article}

\usepackage[table]{xcolor}
\usepackage{enumitem}
\usepackage{tabularx}
\usepackage{ragged2e}

\usepackage{arxiv}
\usepackage{tikz}
\usetikzlibrary{shapes.geometric}
\usetikzlibrary{arrows.meta}
\usepackage{tikz-3dplot}
\usepackage{geometry}
\usepackage[utf8]{inputenc}
\usepackage{placeins}
\usepackage{amsmath}
\usepackage{physics}
\usepackage{amssymb}
\usepackage[makeroom]{cancel}
\usepackage{graphicx}
\usepackage{float}
\usepackage{multicol}
\usepackage{tabularx}
\usepackage{multirow}
\usepackage{caption}
\usepackage{subcaption}
\usepackage{epstopdf}
\usepackage{comment}
\usepackage{bm}
\DeclareMathOperator*{\argmin}{arg\,min}

\usepackage{tikz, amssymb,bm,color}
\usetikzlibrary{circuits.logic.US,circuits.logic.IEC,fit}
\newcommand\addvmargin[1]{
  \node[fit=(current bounding box),inner ysep=#1,inner xsep=0]{};
}

\usepackage{makecell}

\usepackage{arydshln}

\title{Extreme sparsification of physics-augmented neural networks for interpretable model discovery in mechanics}
\author{  Jan Niklas Fuhg$^{\star}$ \\
  Sibley School of Mechanical and Aerospace Engineering \\
  Cornell University, 
   NY 14850, USA \\
  \texttt{jf853@cornell.edu} \\
   \And
   Reese E. Jones \\ 
 Sandia National Laboratories\\ Livermore, CA 94551, USA \\
   \And
 Nikolaos Bouklas \\
  Sibley School of Mechanical and Aerospace Engineering\\
  Center for Applied Mathematics\\
  Cornell University,
   NY 14850, USA 
 }
\date{}
\usepackage[numbers]{natbib}
\newcommand{\cref}[1]{Ref.\,\cite{#1}}

\begin{document}

\maketitle

\begin{abstract}
Data-driven constitutive modeling with neural networks has received increased interest in recent years due to its ability to easily incorporate physical and mechanistic constraints and to overcome the challenging and time-consuming task of formulating phenomenological constitutive laws that can accurately capture the observed material response.
However, even though neural network-based constitutive laws have been shown to generalize proficiently, the generated representations are not easily interpretable due to their high number of trainable parameters. Sparse regression approaches exist that allow to obtaining interpretable expressions, but the user is tasked with creating a library of model forms which by construction limits their expressiveness to the functional forms provided in the libraries.
In this work, we propose to train regularized physics-augmented neural network-based constitutive models utilizing a smoothed version of $L^{0}$-regularization. This aims to maintain the trustworthiness inherited by the physical constraints, but also enables interpretability which has not been possible thus far on any type of machine learning-based constitutive model where model forms were not assumed a-priory but were actually discovered.
During the training process, the network simultaneously fits the training data and penalizes the number of active parameters, while also ensuring constitutive constraints such as thermodynamic consistency.
We show that the method can reliably obtain interpretable and trustworthy constitutive models for compressible and incompressible hyperelasticity,  yield functions, and hardening models for elastoplasticity, for synthetic and experimental data. 
\end{abstract}


\section{Introduction}
In continuum mechanics, material-specific constitutive models are a necessary closure relationship to describe the motion of solid bodies. In contrast to balance laws and kinematic equations, these mathematical descriptions of material behavior do not directly follow from a physical law \cite{fish2013practical}. Nevertheless, mechanistic assumptions, mathematical well-posedness, and physical understanding still constrain the formulations to adhere to objectivity, material symmetry, and thermodynamic consistency considerations, among others \cite{holzapfel2002nonlinear}.

With the recent advances in manufacturing technologies including additive manufacturing, we are experiencing a rise of highly specific materials with advanced requirements \cite{li2019progress}. The systematic satisfaction of these design requirements has led to a pressing need for advanced predictive capabilities, and automation directly linking experimental characterization and modeling. More specifically, specialized
constitutive models are required that focus on constructing mathematical models capable of describing relevant physical phenomena \cite{francois2017modeling} that can eventually enable computations at the structural level.
In the last decades, phenomenological models have been the main driver of progress in constitutive modeling. Phenomenological models are derived and developed based on knowledge of the material response, are based on a limited number of user-chosen functional forms,  and are characterized by a limited number of unknown material parameters often directly tied to specific experiments  \cite{ottosen2005mechanics}. They aim to describe complex behaviors with as few parameters as
possible \cite{fuhg2023enhancing} and are generally interpretable and the user has knowledge about their extrapolation behavior \cite{flaschel2021unsupervised}. Extrapolation capabilities are important as traditional mechanical experiments provide stress-strain data (or rates of these quantities for rate-dependent responses) only for limited stress- or strain-states, e.g. uniaxial, biaxial, simple shear and  hydrostatic, leaving the majority of the strain- or stress- space virtually unexplored. Crucially, this has been accomplished by strictly enforcing thermodynamic constraints and mechanistic assumptions in phenomenological constitutive models.
However, due to the specific user-chosen model form, phenomenological models in general experience a type of model-form error \cite{frankel2022machine}, i.e. the model often is not descriptive enough to fully fit the data which results from incomplete knowledge about the material response \cite{ozturk2021uncertainty}.

\begin{figure}
    \centering
    \includegraphics[scale=0.45]{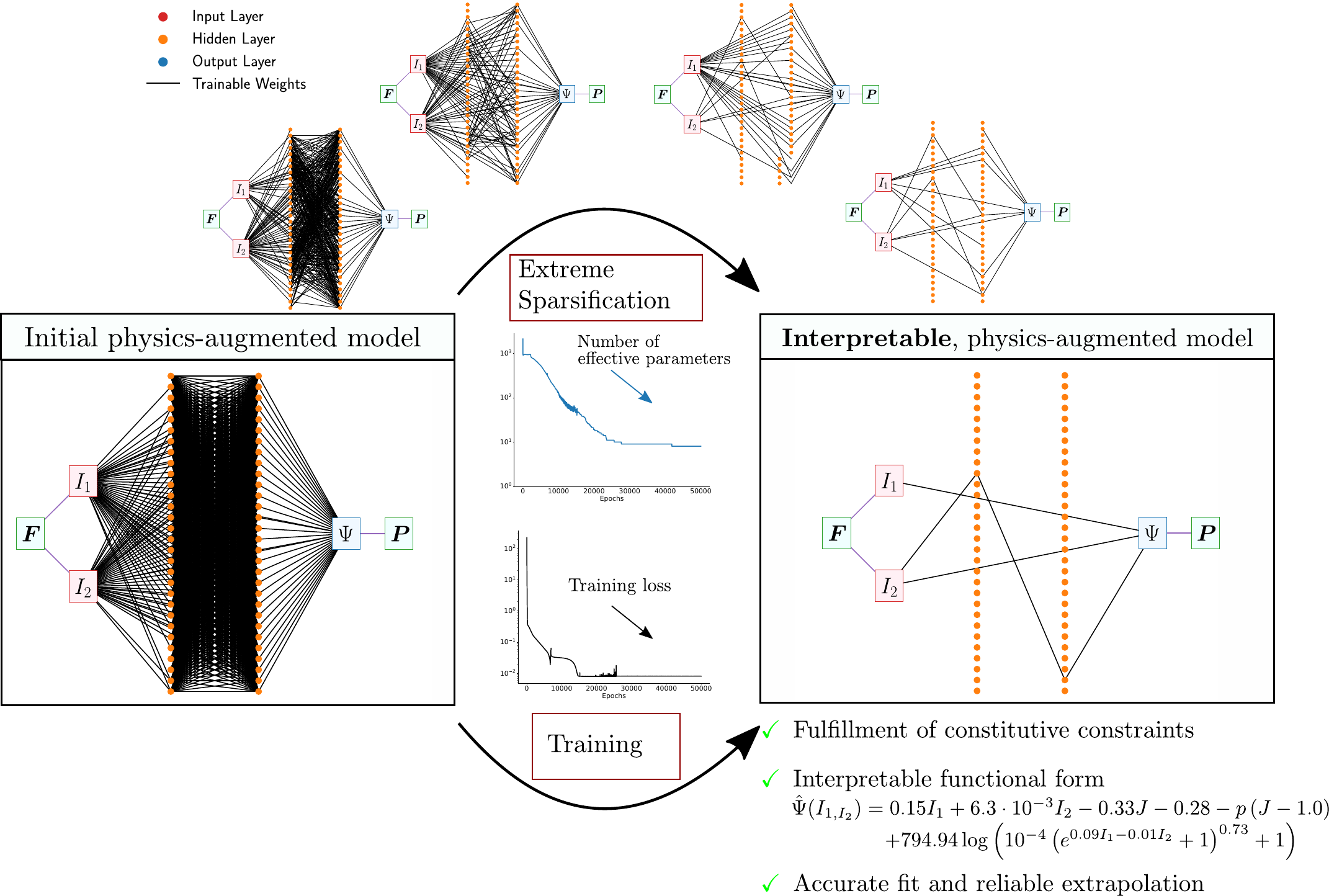}
    \caption{Extreme sparsification of a physics-augmented neural network constitutive model to improve their interpretability. Note the final input convex network utilizes the pass-through of $I_1$ and $I_2$ terms.}
    \label{fig:enter-label}
\end{figure}

In order to improve the restrictions of functional form selections of classical constitutive models and to automate their formulation, machine learning (ML)-based constitutive models have been popularized in recent years \cite{wei2019machine}.
They have been used to model hyperelasticity
\cite{frankel2020tensor,fuhg2021local, vlassis2021sobolev}
viscoelasticity \cite{devries2017enabling, hosseini2021optimized,jones2022neural} and plasticity \cite{jones2018machine,huang2020machine,vlassis2022component}. 
Recently, these representations have been extended to enforce
objectivity- and material symmetry-constraints \cite{frankel2020tensor,fuhg2021physics}, polyconvexity
\cite{klein2022polyconvex, FUHG2022105022, kalina2022automated,linden2023neural} and thermodynamical consistency \cite{masi2021thermodynamics,tacc2023data, fuhg2023modular,upadhyay2023physics} in a physics-augmented manner.
Many of these approaches are built around the flexibility of neural networks that allow them to fulfill the constitutive constraints by construction. For example, hyperelastic models as suggested by \cref{klein2022polyconvex} are designed to be polyconvex and rely on input convex neural networks \cite{amos2017input}; an architecture which can strictly enforce the development of a strain energy density that is polyconvex with respect to its inputs.
However, even though these proposed ML-based formulations are designed to comply with mechanistic and thermodynamic constraints, they  are either not easily interpretable due to the high dimensionality of parameters required in these representations (e.g. often in the order of thousands or more), or might be too restrictive if a model form (or a model-form library) is already pre-selected \cite{flaschel2023automated,flaschel2023automatedBrain,marino2023automated}. The former is also problematic in the low and limited data domain because, although the physics augmentation can act as a regularizer in some cases, the models are still overfitting as will be shown in this work.
Lastly, neural network models take longer to implement in existing commercial and open-source computational infrastructure (e.g. finite element modeling platforms) that is not directly coupled to machine-learning frameworks \cite{suh2023publicly}.
To combat all of these points and to efficiently discover interpretable constitutive models without restrictive assumptions about specific model forms, we introduce extreme sparsification to physics-augmented neural network-based constitutive models. Our approach is based on network pruning, which is illustrated in Figure \ref{fig:enter-label}. 

In recent years, neural network pruning -- the reduction of the network size by removing parameters -- has received increased interest in the machine-learning community \cite{blalock2020state,hoefler2021sparsity}.
In this context, the pruning is mostly utilized to enable the networks to be deployed in real-time on mobile devices.
In general, there are two ways to prune a neural network
\begin{itemize}
    \item In a first step train a full neural network without any regularization. Then in the second step remove all trainable parameters that are below a threshold and then, in the third step, retrain the model. These greedy phases of pruning and retraining may then be repeated until the required balance between performance and network size is reached \cite{han2015learning}.
    \item Regularize the neural network in an eager manner directly during training without the need for any postprocessing or iterative steps \cite{zhang2019eager}. This approach of course reduces the training time to obtain a pruned network but typically has a hyperparameter controlling the influence of the secondary sparsity objective.
\end{itemize}
In this work, we follow the latter since we aim to be as close as possible to the standard training procedure of physics-augmented neural networks for constitutive modeling in mechanics.
In general, most techniques are built around penalizing the $L^{p}$-norm ($p\geq 0$) of the parameters using an additional term in the loss function \cite{han2015learning,anwar2017structured}. 
Typical norms are the $L^{1}$- and $L^{2}$-norm which are known as Lasso- and Ridge-regularization respectively \cite{owen2007robust}. Other methods prune parameters in groups to remove whole neurons or channels \cite{he2017channel}.
Recently, \cref{liu2023seeing} introduced a pruning technique that aims to make neural networks more modular and interpretable by pruning the network during training by encouraging \textit{locality}, i.e. the more neurons communicate the closer they should be in Euclidean space. This is achieved by a local $L^{1}$-regularization measure and by placing neurons with high communication closer together by employing a swapping algorithm.
In this work, we rely on 
\cref{louizos2017learning} which prune the network through a smoothed version of the expected $L^{0}$-regularization. The benefits of this approach include that it enforces sparsity without placing a penalty on the magnitude of the weights and  it allows for parameters to be exactly zero. Hence, no thresholding is necessary.
We remark that our work is related to neural symbolic regression \cite{kim2020integration,udrescu2020ai}, we, however, aim to use standard neural network models that have been used in the data-driven constitutive modeling community  and combine these formulations with sparsification techniques instead of selecting a model from a library of formulations. We choose this approach to enhance the expressiveness of the models and avoid selecting a model-informed basis for our representation.

The paper is structured as follows. 
Section \ref{sec::2} introduces physics-augmented neural networks and explains their sparsification through smoothed $L^{0}$ regularization.
We test the framework's ability to produce interpretable constitutive models for hyperelasticity and elastoplasticity in Section \ref{sec::3}.
The paper is concluded in Section \ref{sec::4}.

\section{Sparsifying physics-augmented neural networks for constitutive modeling}\label{sec::2}
Due to their remarkable flexibility, neural networks have emerged as the most used machine-learning approach for data-driven constitutive modeling. 
In particular, compared to other regression techniques, such as Gaussian process regression or Support Vector Regression that have been applied for constitutive modeling \cite{hartmaier2020data,wang2021metamodeling,fuhg2023enhancing}, neural networks have received more attention with regards to finding a proficient balance between enforcing constitutive-constraints and retaining model expressiveness \cite{swiler2020survey}. In the following, we will show two examples of incorporating constraints into neural networks and discuss a potential way of sparsifying them. We remark that in this context we mean the sparsification of all trainable parameters of a neural network, instead of, e.g., the sparse identification of terms from a user-defined functional library as in SINDy \cite{brunton2016discovering,kaheman2020sindy} or the EUCLID framework \cite{flaschel2021unsupervised,flaschel2022discovering}. 

\subsection{Physics-augmented neural network formulations}
Two classes of physics augmentation have been found to be crucial for constitutive modeling
\begin{itemize}
    \item Input convex/concave functions. Convex functions are needed to model polyconvex strain energy density functions for hyperelasticity \cite{klein2022polyconvex,linden2023neural} and guarantee dissipation requirements when utilizing potentials to model yield functions \cite{fuhg2022machine,vlassis2022component,nascimento2023machine} or hardening behavior \cite{meyer2023thermodynamically,fuhg2023modular}.
    \item Positive, monotonically increasing functions or their counterparts negative, monotonically decreasing functions. These functions are needed to model the derivatives of convex functions \cite{tacc2023data} or where mechanistic assumptions require monotonic effects such as isotropic hardening \cite{fuhg2023modular}.
\end{itemize}
Even though these examples of constraints might sound limited, they form the basis of the framework for constitutive modeling for several classes of materials, whose responses are described via hyperelasticity, viscoelasticity, elastoplasticity, viscoplasticity and damage mechanics \cite{lemaitre1994mechanics}. We will briefly highlight how both of these constraints can be intrinsically incorporated into feed-forward neural networks.

\subsubsection{Input convex neural network}\label{sec:InputConvexNN}
Following \cref{amos2017input}, consider an output $\hat{\bm{y}}\in \mathbb{R}^{n^{L}}$ connected to an input $\bm{x}_{0}\in \mathbb{R}^{n^{0}}$ by the neural network $\mathcal{N}$ given by
\begin{equation}\label{eq:ICNN}
    \begin{aligned}
            \bm{x}_{1} = \sigma_{1} \left( \bm{x}_{0} \bm{W}_{1}^{T} + \bm{b}_{1} \right) &\in \mathbb{R}^{n^{1}} \\
            \bm{x}_{l} = \sigma_{l} \left( \bm{x}_{l-1} \bm{W}_{l}^{T} + \bm{x}_{0} \bm{\mathcal{W}}_{l}^{T} + \bm{b}_{l} \right) &\in \mathbb{R}^{n^{l}}, \qquad l=2, \ldots, L-1 \\
            \hat{\bm{y}} = \bm{x}_{L-1} \bm{W}_{L}^{T} +  \bm{x}_{0} \bm{\mathcal{W}}_{L}^{T} + \bm{b}_{L}, &\in \mathbb{R}^{n^{L}}
    \end{aligned}
\end{equation}
with the weights $\bm{W}$ and $\bm{\mathcal{W}}$, the biases $\bm{b}$ and the activation functions $\sigma$. The weights and biases form the set of trainable parameters $\bm{\theta} = \lbrace \lbrace \bm{W}_{i} \rbrace_{i=1}^{L},  \lbrace \bm{\mathcal{W}}_{i} \rbrace_{i=2}^{L}, \lbrace \bm{b}_{i} \rbrace_{i=1}^{L}  \rbrace$.
The output is then convex with regards to the input if the weights $\lbrace \bm{W}_{i} \rbrace_{i=2}^{L}$ are non-negative and the activation functions $\lbrace \sigma_{i} \rbrace_{i=1}^{L}$ are convex and non-decreasing. For a proof see \cref{amos2017input}.
\subsubsection{Positive, monotonically increasing neural network }\label{sec::PosiMonInc}
Let a positive, monotonically increasing neural network $\mathcal{N}$ be defined by 
\begin{equation}\label{eq:MNN}
    \begin{aligned}
            \bm{x}_{0}  &\in \mathbb{R}^{n^{0}} \\
            \bm{x}_{1} = \sigma_{1} \left( \bm{x}_{0} \bm{W}_{1}^{T} + \bm{b}_{1} \right) &\in \mathbb{R}^{n^{1}} \\
            \bm{x}_{l} = \sigma_{l} \left( \bm{x}_{l-1} \bm{W}_{l}^{T} + \bm{b}_{l} \right) &\in \mathbb{R}^{n^{l}}, \qquad l=2, \ldots, L-1 \\
            \hat{\bm{y}} = \bm{x}_{L-1} \bm{W}_{L}^{T} + \bm{b}_{L}, &\in \mathbb{R}^{n^{L}}
    \end{aligned}
\end{equation}
where $\bm{W}$, $\bm{b}$ and $\sigma$ denote the weights, biases and activation functions respectively. The set of trainable parameters reads $\theta = \lbrace \lbrace \bm{W}_{i=1}^{L} \rbrace , \lbrace \bm{b}_{i} \rbrace_{i=1}^{L} \rbrace$. Then each output value of $\hat{\bm{y}}$ is positive, and monotonically increasing with regard to all input values of $\bm{x}_{0}$ when the trainable parameters are nonnegative and the activation functions are positive and non-decreasing. For a proof, we refer to \cref{fuhg2023modular}.

\subsection{Extreme sparsification with smoothed $L^{0}$ regularization}
Given a data set of input-output pairs $\lbrace \bm{x}^{i}, \bm{y}^{i} \rbrace_{i=1}^{N}$ the trainable parameters of eqs. \eqref{eq:ICNN} and \eqref{eq:MNN} are classically found by minimizing a loss function $\mathcal{R}(\bm{\theta})$
\begin{equation*}
    \bm{\theta}^{\star} = \argmin_{\bm{\theta}} \mathcal{R}(\bm{\theta}) = \argmin_{\bm{\theta}} \frac{1}{N} \sum_{i=1}^{N}  \left[ \mathcal{L}( \mathcal{N}(\bm{x}^{i}; \bm{\theta}) , \bm{y}^{i} )\right] 
\end{equation*}
with some loss function $\mathcal{L}(\bullet)$.
In the following, we discuss how to sparsify the parameters by adding regularization. The approach follows the general idea of a gating system where each trainable parameter is multiplied by a gate value $z\in[0,1]$ of either zero or one depending on whether the parameter should be active or not. Zero reflects an inactive parameter while one defines an active parameter. The number of active gates could then be penalized in the loss function. However, due to the binary nature of the gates, the loss function would not be differentiable. Hence,
following \cref{louizos2017learning} we consider a reparametrization of the trainable parameters using a smoothed "gating" system, i.e. let 
\begin{equation}
    \bm{\theta} = \bm{\overline{\theta}} \odot   \bm{z}, \quad \text{with} \quad \bm{z} = \min (\bm{1}, \max (\bm{0}, \overline{\bm{s}}))
\end{equation}
where $\odot$ denotes the Hadamard product and where 
\begin{equation}
\begin{aligned}
       \overline{\bm{s}} = \bm{s} (\zeta- \gamma) + \gamma,  \quad \bm{s} = \text{Sigmoid}((\log \bm{u} &- \log (1-\bm{u})+ \log \bm{\alpha})/\beta), \quad \bm{u} \sim U(\bm{0},\bm{1}).
\end{aligned}
\end{equation}
Here, $\gamma$, $\beta$, $\zeta$ and $\log \bm{\alpha}$ are user-chosen parameters that define the smoothing of the gate $\bm{z}$. Following the suggestions of the authors \cite{louizos2017learning}, we choose $\gamma = -0.1$, $\zeta = 1.1$, $\beta=2/3$ and $\log \bm{\alpha}$ is obtained by sampling from a normal distribution with zero mean and a standard
deviation of $0.01$.



Since the gate values are dependent on a random variable, we can define a Monte Carlo approximated loss function as  
\begin{equation}
    \mathcal{R}(\overline{\bm{\theta}}) = \frac{1}{L} \sum_{l=1}^{L} \left( \frac{1}{N} \left( \sum_{i=1}^{N} \mathcal{L} (\mathcal{N}(\bm{x}_{i}, \overline{\bm{\theta}} \odot \bm{z}^{l}), \bm{y}_{i}) \right) \right] + \lambda  \sum_{j=1}^{\abs{\theta}} \text{Sigmoid}(\log \alpha_{j} - \beta \log \frac{-\gamma}{\zeta})
\end{equation}
with $L$ being the number of samples of the Monte Carlo approximation and where $\lambda$ is a weighting factor for the regularization.
At test time we can then set the values of the trainable parameters $\bm{\theta}^{\star} = \overline{\bm{\theta}}^{\star} \odot  \hat{\bm{z}}$ where the gates can be obtained as
\begin{equation}
    \hat{\bm{z}} = \min ( \bm{1}, \max (\bm{0}, \text{Sigmoid}(\log \bm{\alpha})(\zeta-\gamma)+ \gamma)).
\end{equation}
In the following section, we show how this idea can be applied to constitutive modeling in computational mechanics.

\section{Results}\label{sec::3}
We consider three different application areas where fitting physics-augmented neural network constitutive laws have been applied in the literature:
\begin{itemize}
    \item Incompressible and compressible hyperelasticity
    \item Yield functions for elastoplasticity
    \item Isotropic hardening for elastoplasticity.
\end{itemize}
For each case, we look at multiple different experimental and synthetic datasets to test the capabilities of the sparsification approach. We have implemented all the models in Pytorch \cite{paszke2019pytorch} and use the Adam optimizer \cite{kingma2014adam} with a learning rate of $10^{-3}$ to solve the optimization problem. We also note that for elastoplasticity, we follow the paradigm suggested in \cite{fuhg2023modular} for modular learning approaches. 

\subsection{Constitutive modeling of hyperelastic potential}
Hyperelasticity enables modeling finite strain, nonlinear elastic material behavior by assuming that a strain-dependent potential function $\Psi(\bm{F})$ exists from which the stress can be derived \cite{holzapfel2002nonlinear}. Here $\bm{F} = \frac{\partial \phi}{\partial \bm{X}}$ is the deformation gradient that depends on $\phi(\bm{X},t)$ which describes the motion of a body from a reference position $\bm{X}$ to its current position at time $t$.  One important strain measure for finite strain modeling is the right Cauchy-Green tensor $\bm{C} = \bm{F}^{T} \bm{F}$. In the following, we restrict ourselves to isotropic material behavior.
The hyperelastic potential is generally subject to constitutive constraints that are derived from thermodynamic considerations and mechanistic assumptions \cite{klein2022polyconvex}. These include 
\begin{itemize}
    \item Objectivity and material symmetry. Let $\bm{R}$ be an orthogonal matrix, the strain energy function then needs to adhere to the assumed isotropic nature of the material behavior, i.e.
    \begin{equation}
        \Psi(\bm{F} \bm{R}^{T} ) = \Psi(\bm{F}).
    \end{equation}
    Furthermore, the energy value needs to be independent of the choice of an observer, which is known as objectivity, and which is defined by 
    \begin{equation}
        \Psi(\bm{R} \bm{F} ) = \Psi(\bm{F}).
    \end{equation}
    By formulating the strain energy function in terms of the invariants of the right Cauchy-Green tensor $\psi(I_{1},I_{2},J)$ where
    \begin{equation}
    \begin{aligned}
        I_{1} = \text{tr} (\bm{C}), \quad 
        I_{2} = \text{tr} (\text{cof} \bm{C}), \quad 
        J = \sqrt{\det (\bm{C})},
    \end{aligned}
\end{equation}
instead of the deformation gradient, the objectivity and the material symmetry conditions are fulfilled.
    \item Normalization condition. It is physically sensible to define the form of the strain energy function such that both its value and the derived stress value are zero at the undeformed configuration $\bm{C}=\bm{I}$ \cite{holzapfel2002nonlinear}.
    \item Polyconvexity. Following Ball \cite{ball1976convexity}, polyconvex hyperelasticity guarantees the existence of minimizers of the underlying potential in finite elasticity. Polyconvexity in this context requires that 
    $\Psi(\bm{F}, \text{cof} \bm{F}, \det \bm{F})$
    is convex in all its arguments     \cite{schroder2003invariant}. This is equivalent to ensuring convexity of the strain energy function with regard to a set of polyconvex invariants.
\end{itemize}
In the following, we discuss neural network formulations that enforce all these constitutive conditions and use regularization to obtain interpretable model forms for different data sets. We differentiate between problems in compressible and incompressible hyperelasticity.

\subsubsection{Compressible hyperelastic materials from numerical data}
In order to generate a compressible, polyconvex, objective, and isotropic material description the strain energy function can be assumed to be a convex function $\psi(I_{1}, I_{2}, J)$.
We then build an input convex neural network, c.f. eq. \eqref{eq:ICNN} that takes these invariants as input and outputs a scalar $\hat{\Psi}^{NN}(I_{1}, I_{2}, J)$. In order to also fulfill the normalization condition, the approach suggested by \cref{linden2023neural} can be used, we thus obtain the strain energy prediction as 
\begin{equation}
    \hat{\Psi}(I_{1}, I_{2}, J) = \hat{\Psi}^{NN}(I_{1}, I_{2}, J) - \hat{\Psi}^{NN}(3, 3, 1) - \psi^{S}(J)
\end{equation}
where $\psi^{S}$ defined as
\begin{equation}
    \psi^{S}= n (J-1) =  \underbrace{\left. \left( 2 \frac{\partial \hat{\Psi}^{NN}}{\partial I_{1}}  + 4 \frac{\partial \hat{\Psi}^{NN}}{\partial I_{2}} + \frac{\partial \hat{\Psi}^{NN}}{\partial J}  \right) \right\rvert_{\bm{C}=\bm{I}}}_{=n}  (J-1)
\end{equation}
enforces the normalization of the stress. For more information see Appendix \ref{app:Comhyperelasticity}.
A prediction of the second Piola-Kirchhoff stress can then be obtained with
\begin{equation}
\begin{aligned}
        \hat{\bm{S}} = 2 \frac{\partial \hat{\Psi}}{\partial \bm{C}} = 2 \left( \sum_{i} \frac{\partial \hat{\Psi}}{\partial I_{i}} \frac{\partial I_{i}}{\partial \bm{C}} \right) 
    = 2 \left(\frac{\partial \hat{\Psi}}{\partial I_{1}} + I_{1} \frac{\partial \hat{\Psi}}{\partial I_{2}} \right) \bm{I} - 2 \frac{\partial \hat{\Psi}}{\partial I_{2}} \bm{C} + J \frac{\partial \hat{\Psi}}{\partial J}  \bm{C}^{-1}.
\end{aligned}
\end{equation}
Using this formulation we can train a sparsified, compressible, physics-augmented neural network model on stress-strain data.
For this application, consider a region in the deformation gradient space of the form 
\begin{equation}
    F_{ij} \in [F_{ij}^{L}, F_{ij}^{U}] = \begin{cases}
        [1-\delta, 1+\delta] & \text{if} \, i=j \\
        [-\delta, +\delta] & \text{else}.
    \end{cases}
\end{equation}
Following \cref{fuhg2023stress} we define a training region with $\delta=0.2$ and a test region with $\delta=0.3$. We then utilize the space-filling sampling algorithm proposed in \cref{fuhg2021physics} to sample $50$ training and $10,000$ testing data points in their respective regions. In terms of the invariants, the 50 data points are summarized in Table \ref{tab:Compressible50}.
We furthermore divide the training dataset in an $80/20$ split into training and validation points.

\begin{figure}
\begin{subfigure}[b]{0.33\linewidth}
        \centering
    \includegraphics[scale=0.22]{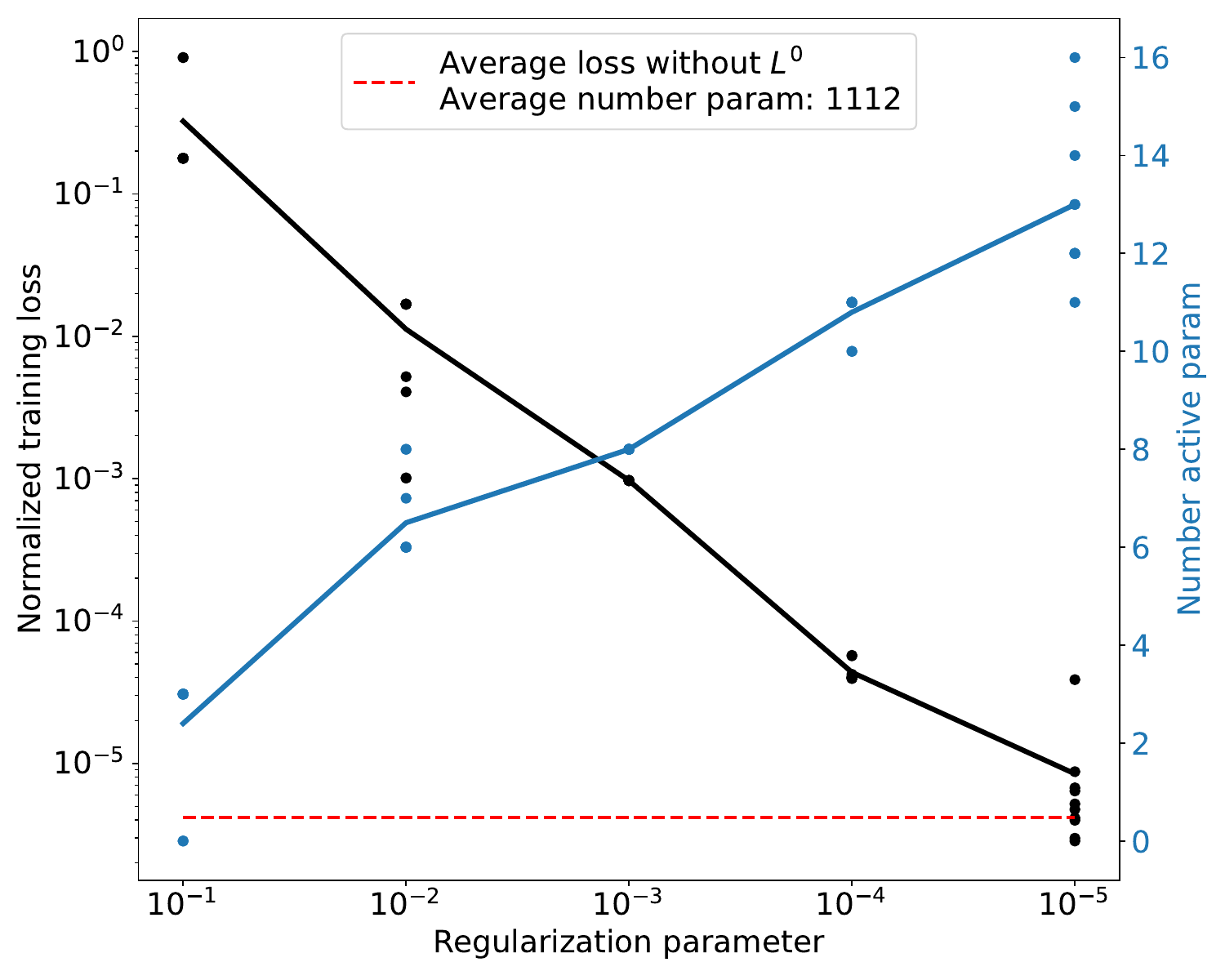}
    \caption{1  hidden  layer - 30 neurons}\label{fig:onlyKina}
\end{subfigure}
\begin{subfigure}[b]{0.33\linewidth}
        \centering
        \includegraphics[scale=0.22]{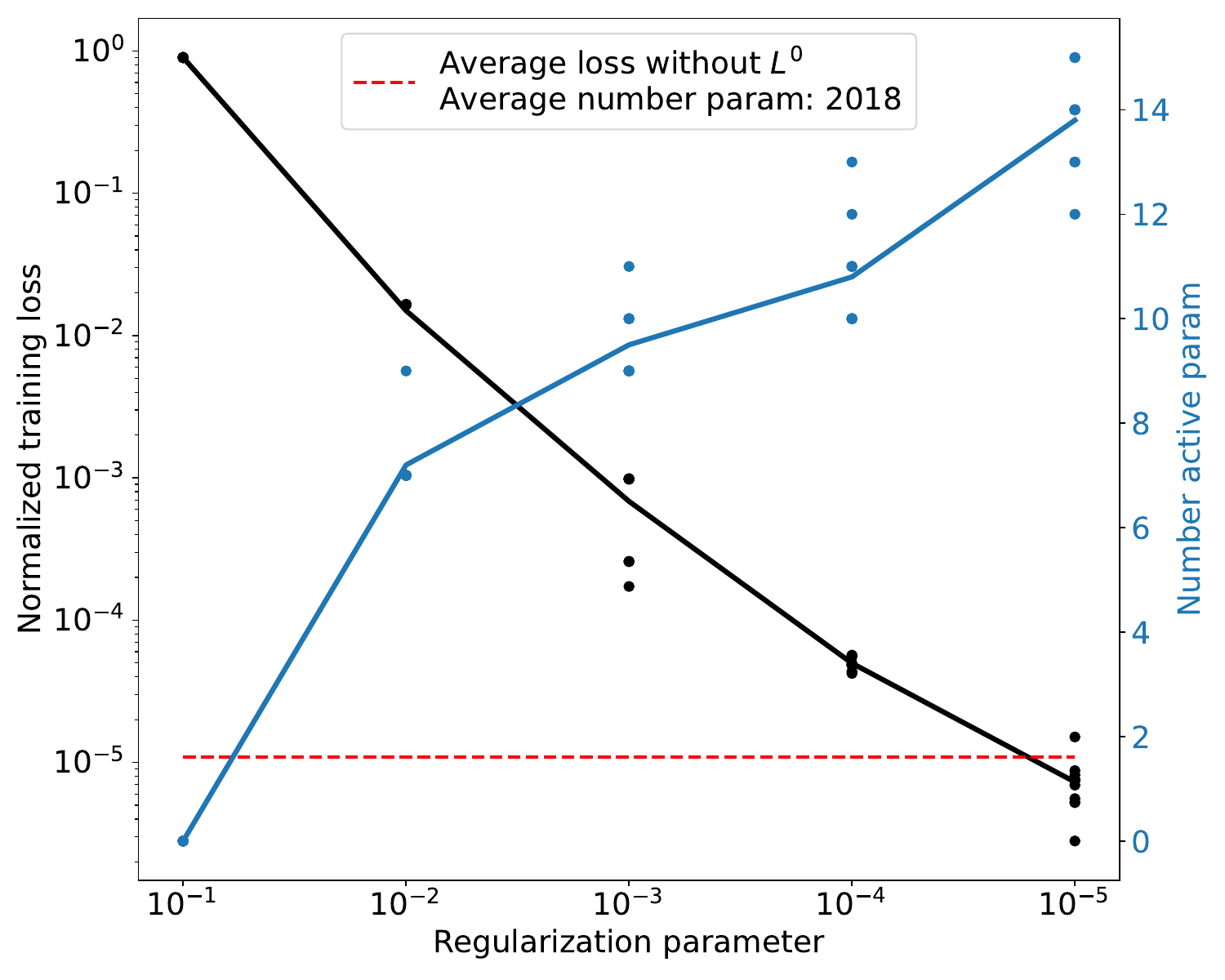}
    \caption{2  hidden  layers - 30 neurons}\label{fig:onlyKinb}
\end{subfigure}
\begin{subfigure}[b]{0.33\linewidth}
        \centering
        \includegraphics[scale=0.22]{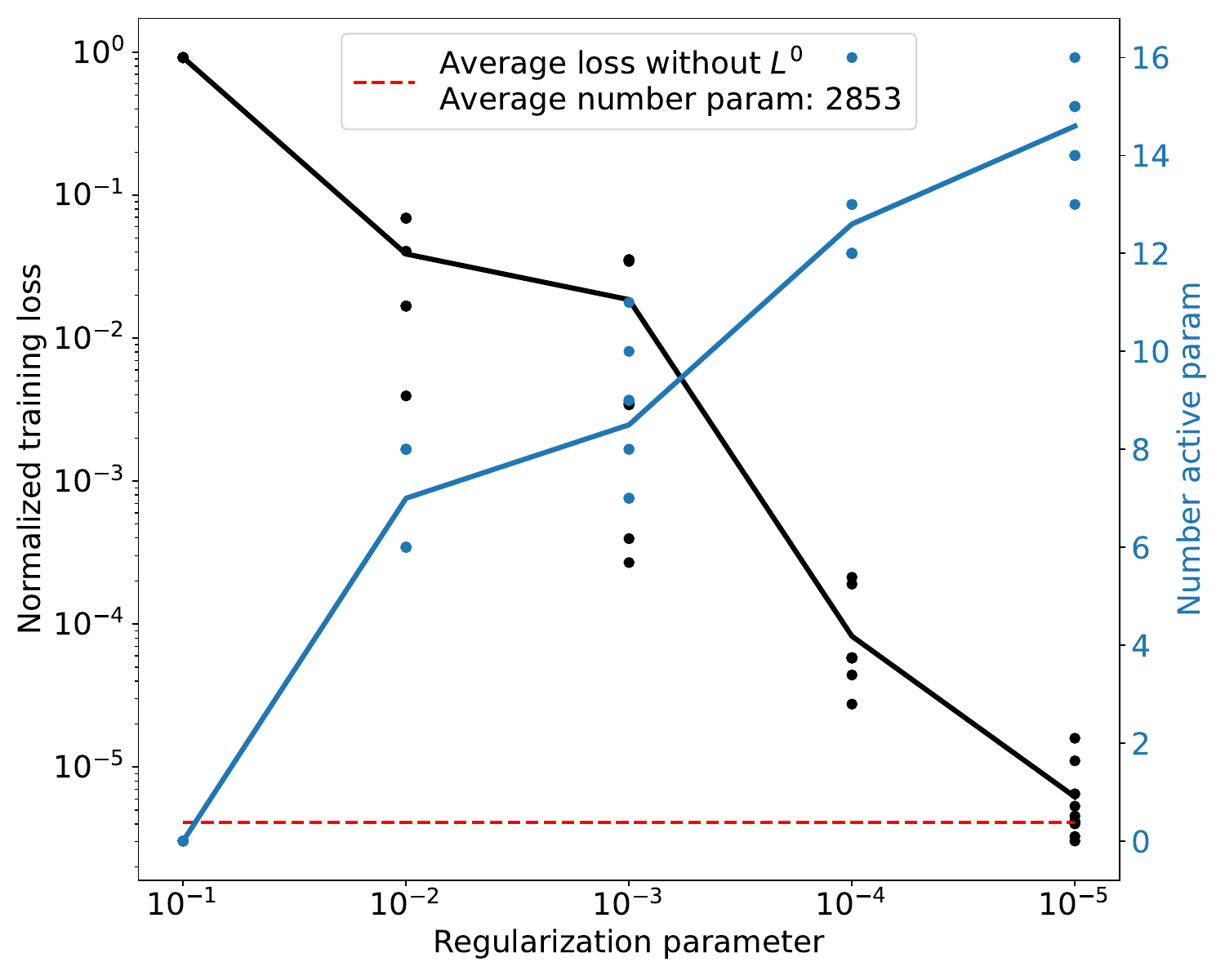}
    \caption{3  hidden  layers - 30 neurons}\label{fig:onlyKinb}
\end{subfigure}
\caption{Gent-Gent law training loss for different architectures. The red dotted line indicates the training loss of the network when no $\mathcal{L}^{0}$-sparsification is employed.}\label{fig:GentTrainingLoss}
\end{figure}

\paragraph{Examples}
Firstly, we study the performance of the proposed approach and how it is influenced by the network size and the weighting factor $\lambda$. Consider a Gent-Gent hyperelastic model \cite{ogden2004fitting} of the form
\begin{equation}
\Psi_{gent} = -\frac{\theta_{1}}{2} J_{m} \log \left( 1 - \frac{I_{1}-3}{J_{m}} \right) - \theta_{2} \log \left( \frac{I_{2}}{J} \right) + \theta_{3} \left( \frac{1}{2} (J^{2}-1) - \log J \right)
\end{equation}
where we employ $\theta_{1} =2.4195$, $J_{m}= 77.931$,  $\theta_{2} =-0.75$ and $\theta_{3} =1.20975 $, see \cref{fuhg2023stress}.
We consider three different architectures: 1 hidden layer with 30 neurons, 2 hidden layers with 30 neurons, and 3 hidden layers with 30 neurons), 5 different regularization parameter values, and repeat the training process $10$ times with different random seeds. Figure \ref{fig:GentTrainingLoss} depicts the average (in solid lines) and each individual data point of the final training loss and the number of parameters with an absolute value greater than $0$ at the end of $100,000$ training epochs for each regularization parameter value and architectural setting. We additionally highlight the average training loss (red dotted line) and the number of active parameters (in the legend) when no regularization was employed. We can see that the final training loss value as well as the number of active training parameters is relatively similar in terms of absolute values for all architectures. As was expected, higher regularization parameters penalize the number of active parameters more, and hence we see can see that the number of active parameters increases with lower regularization parameters. Since the loss is then also dominated by the regularization, the training loss decreases with a decrease in $\lambda$. 
Due to the higher number of tunable parameters in the unregularized models (e.g. $1112$ parameters whose absolute value is greater than zero for the architecture with 1 hidden layer and 30 neurons), these networks achieve a comparatively low training loss. 
Even though, the training loss for the sparsified networks with $\lambda=10^{-5}$ might reach a similar value with around $15$ parameters.

\begin{figure}
\begin{subfigure}[b]{0.33\linewidth}
        \centering
    \includegraphics[scale=0.22]{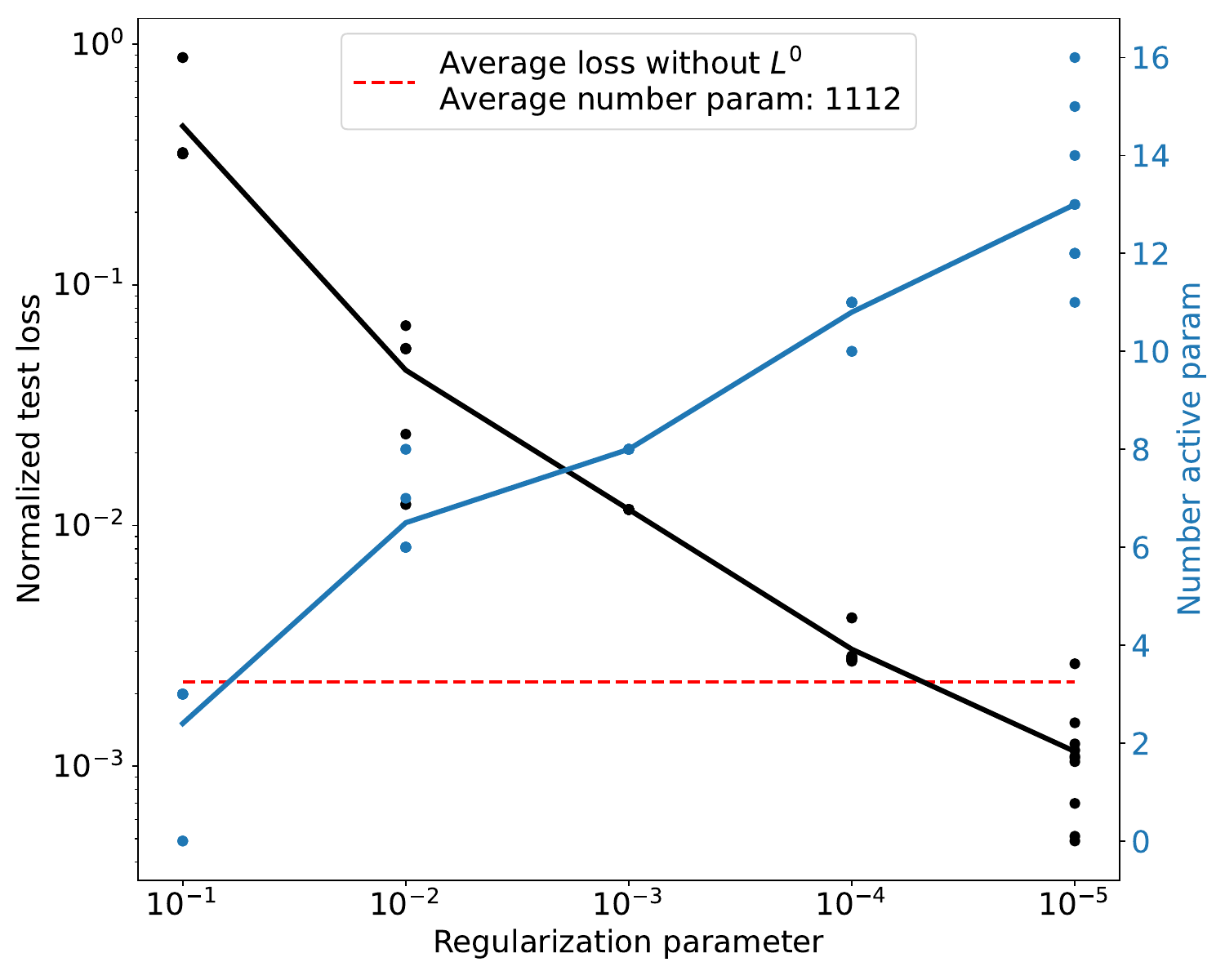}
    \caption{1  hidden  layer - 30 neurons}\label{fig:onlyKina}
\end{subfigure}
\begin{subfigure}[b]{0.33\linewidth}
        \centering
        \includegraphics[scale=0.22]{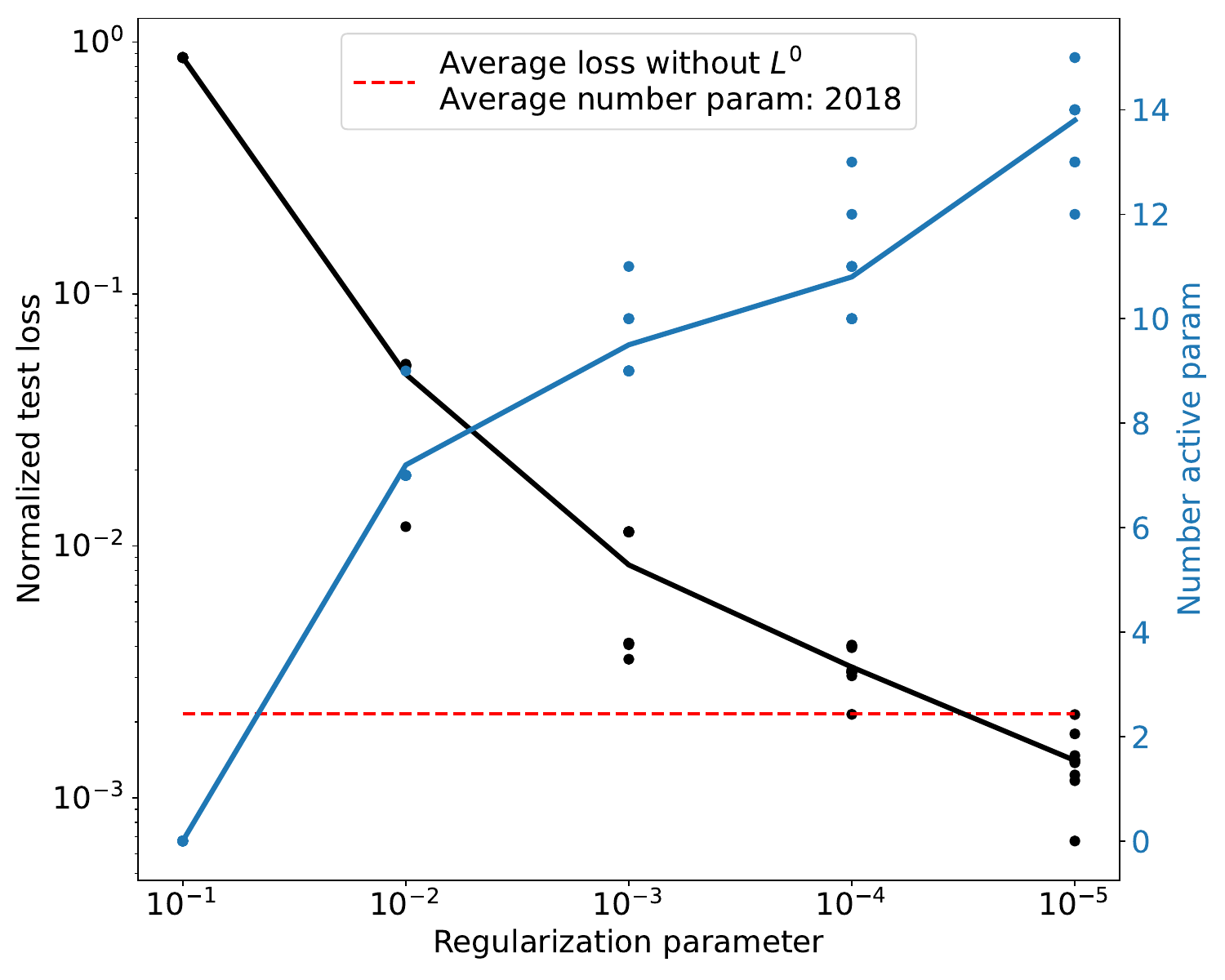}
    \caption{2  hidden  layers - 30 neurons}\label{fig:onlyKinb}
\end{subfigure}
\begin{subfigure}[b]{0.33\linewidth}
        \centering
        \includegraphics[scale=0.22]{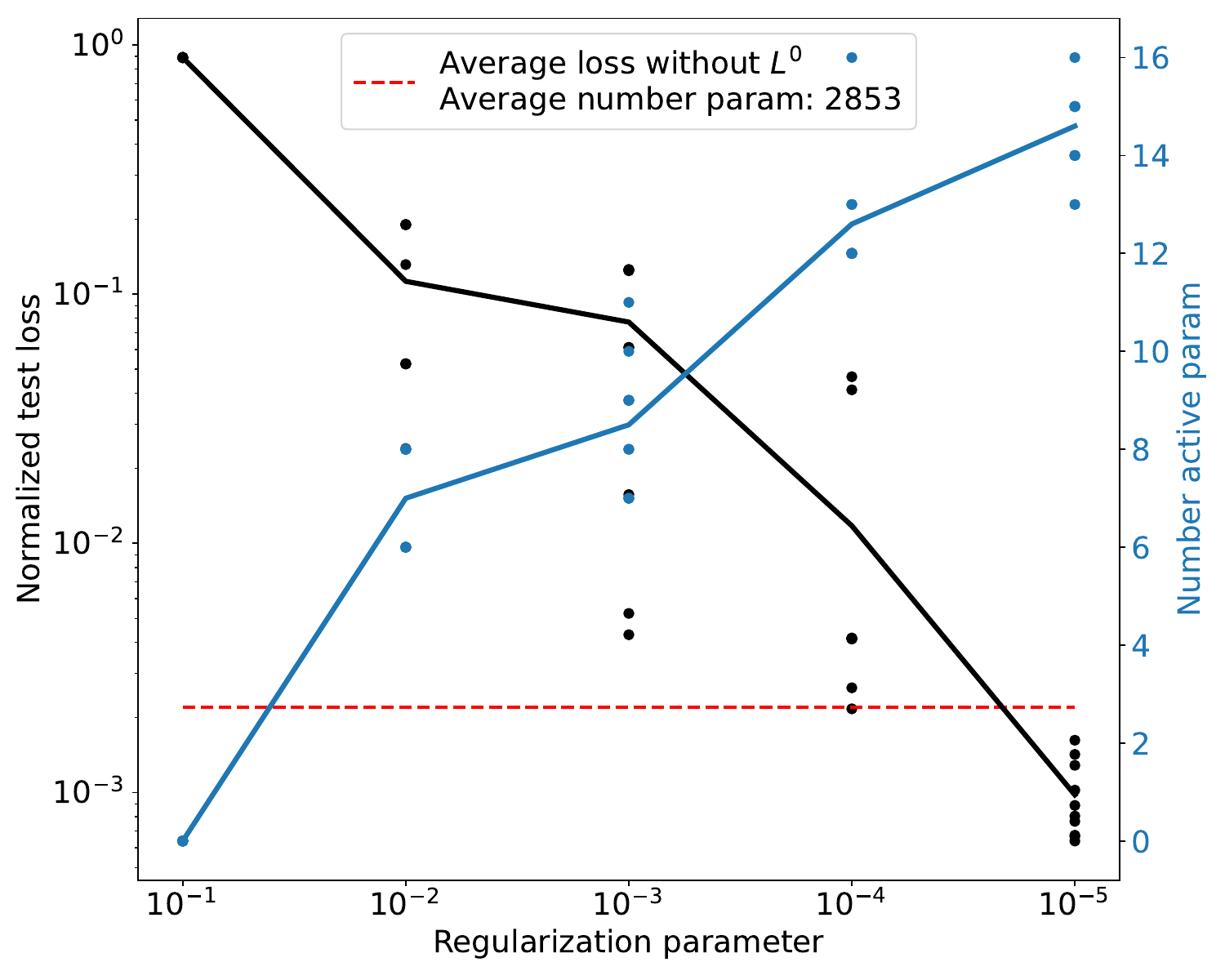}
    \caption{3  hidden  layer - 30 neurons}\label{fig:onlyKinb}
\end{subfigure}
\caption{Gent-Gent law test loss for different architectures. The red dotted line indicates the test loss of the network when no $\mathcal{L}^{0}$-sparsification is employed.}\label{fig:GentTestLoss}
\end{figure}

For the same trained networks as above, Figure \ref{fig:GentTestLoss} summarizes the generalization performance of the networks based on the test loss over the $10,000$ test points. Here, again, the red dotted lines indicate the average test loss when no regularization was employed. Interestingly, the average test loss of the networks with a regularization value of $\lambda=10^{-5}$ is better than that of the full models where $\lambda=0$. This suggests that due to the additional regularization, these models are not only more interpretable but they also generalize more proficiently.
The functional form of the network (1 hidden layer and $\lambda=10^{-4}$) with the median training loss is given by
\begin{equation}
     \hat{\Psi}=  0.398 J + 3.095 \log{\left(\left(1 + e^{- 1.356 I_{2}}\right)^{1.314} \left(e^{0.755 I_{1}} + 1\right)^{0.515} \left(e^{0.135 I_{1} - 0.319 I_{2} - 0.329 J} + 1\right)^{1.874} + 1 \right)} - 6.686.
\end{equation}
The representation consists of 3 terms that can be written out in one line of text, making it easy to implement and allowing for easier interpretation than the equivalent $1112$ parameters of the full model. For example, it is easy to see that the obtained hyperelastic law does not fulfill the coercivity condition \cite{klein2022polyconvex}, i.e., as $J\rightarrow 0^{+}$, we do not get $\hat{\Psi} \rightarrow \infty$. This observation would not have been possible with over $1000$ parameters. \\
We remark that we could have included a functional term in the strain energy function, such that the coercivity would be fulfilled by design, c.f. \cref{klein2022polyconvex}.

\begin{figure}
        \centering
    \includegraphics[scale=0.3]{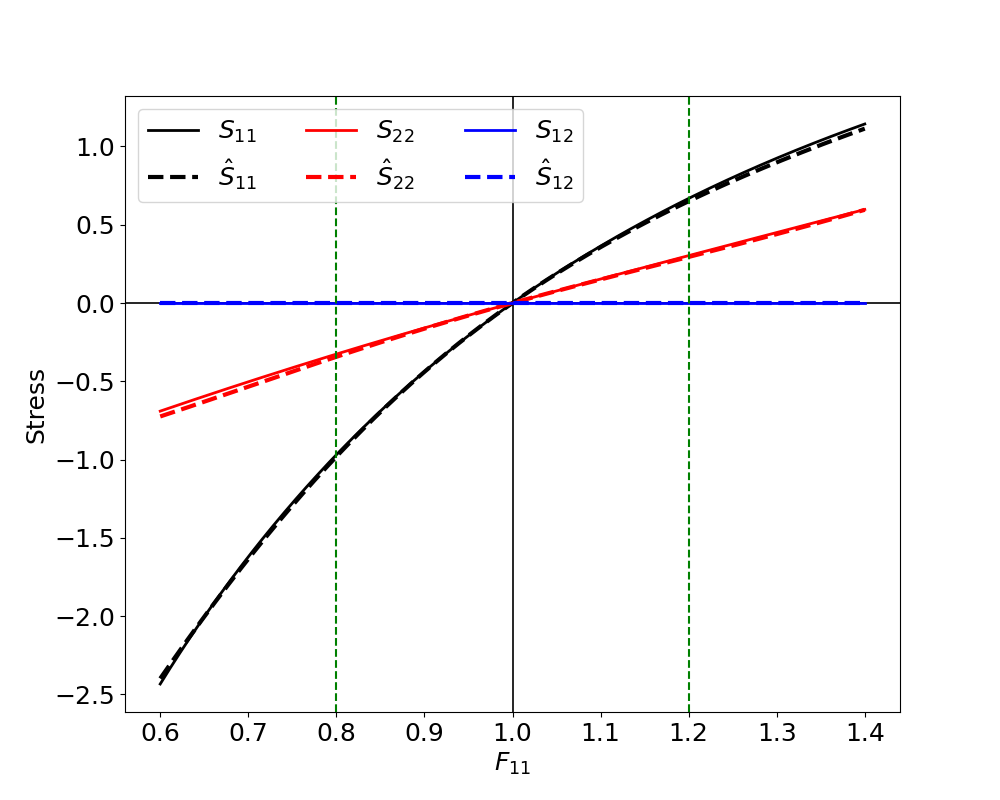}
    \caption{Stress strain curves of fitted Gent-Gent hyperelastic laws. Green dotted lines represent the limits of the training regime.}\label{fig:hyperStressStrainA}
\end{figure}

To highlight how accurate the obtained formulation nevertheless is we can look at the stress-strain behavior inside and outside of the training regime $0.6 \leq F_{11} \leq 1.4$, see Figure \ref{fig:hyperStressStrainA}, where the green lines indicate the training domain. We can see that the network finds a sparse representation that appears to interpolate well and is also proficient in extrapolation.

\begin{figure}
\begin{subfigure}[b]{0.33\linewidth}
        \centering
    \includegraphics[scale=0.22]{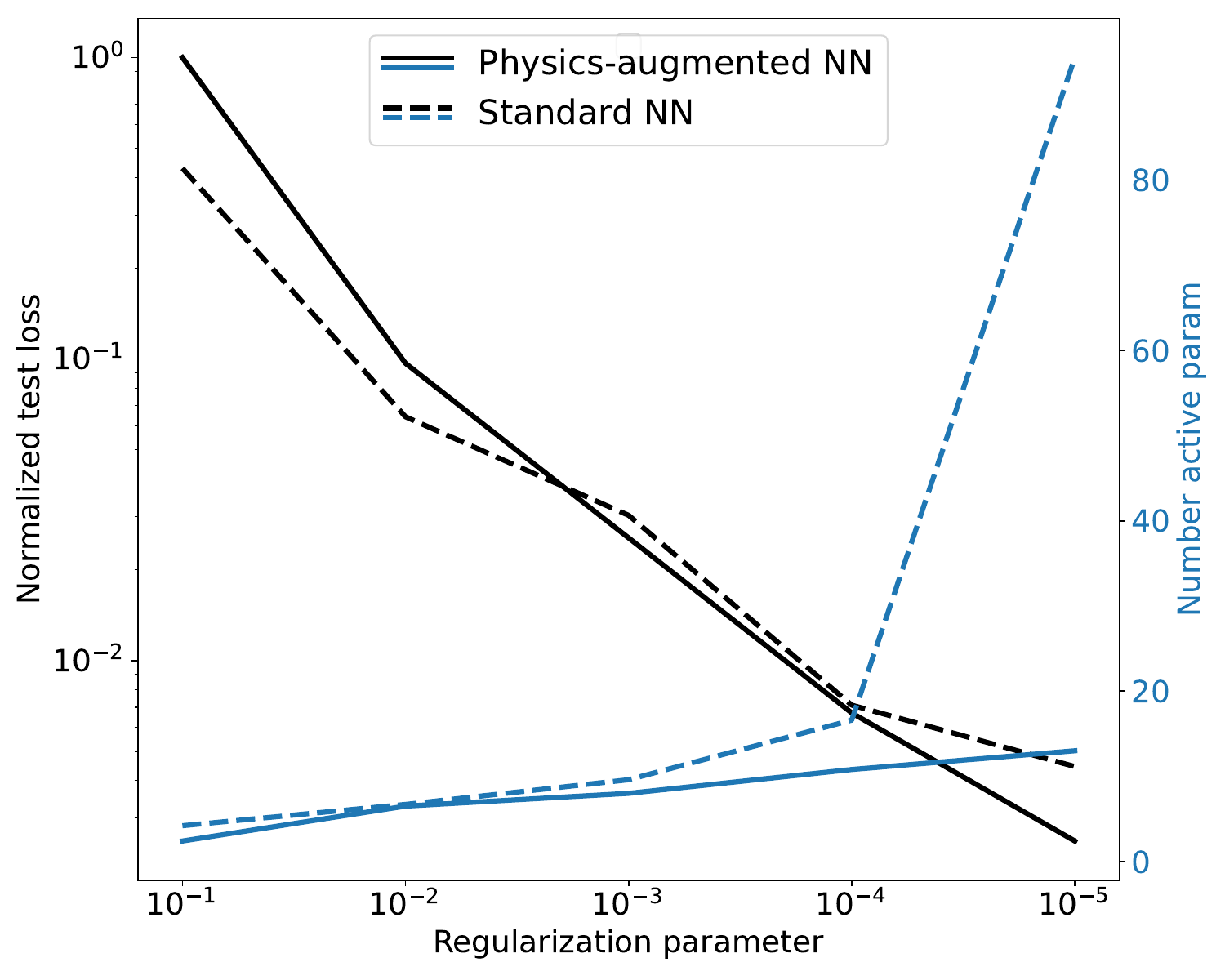}
    \caption{1 hidden layer - 30 neurons}\label{fig:onlyKina}
\end{subfigure}
\begin{subfigure}[b]{0.33\linewidth}
        \centering
        \includegraphics[scale=0.22]{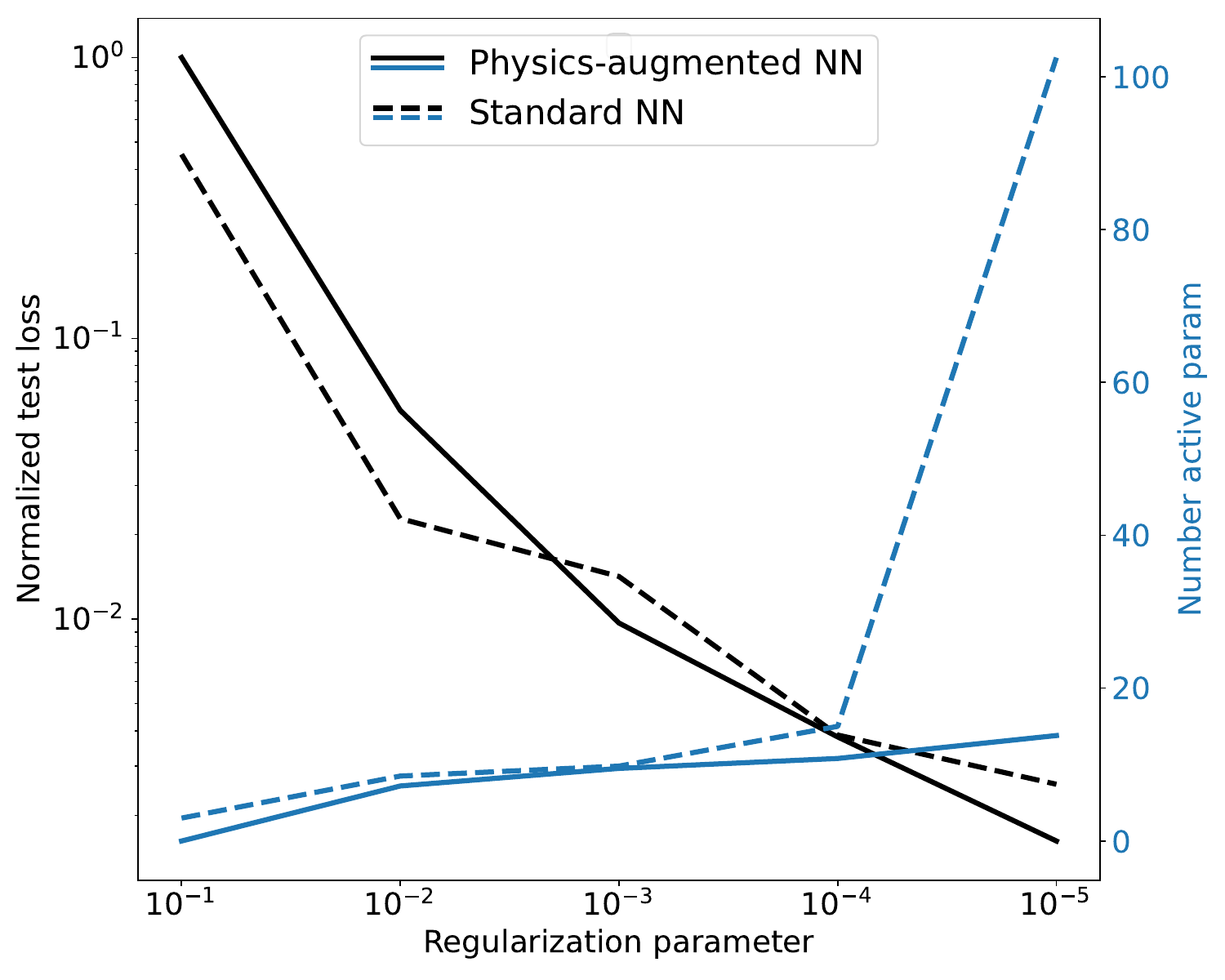}
    \caption{2  hidden layers - 30 neurons}\label{fig:onlyKinb}
\end{subfigure}
\begin{subfigure}[b]{0.33\linewidth}
        \centering
        \includegraphics[scale=0.22]{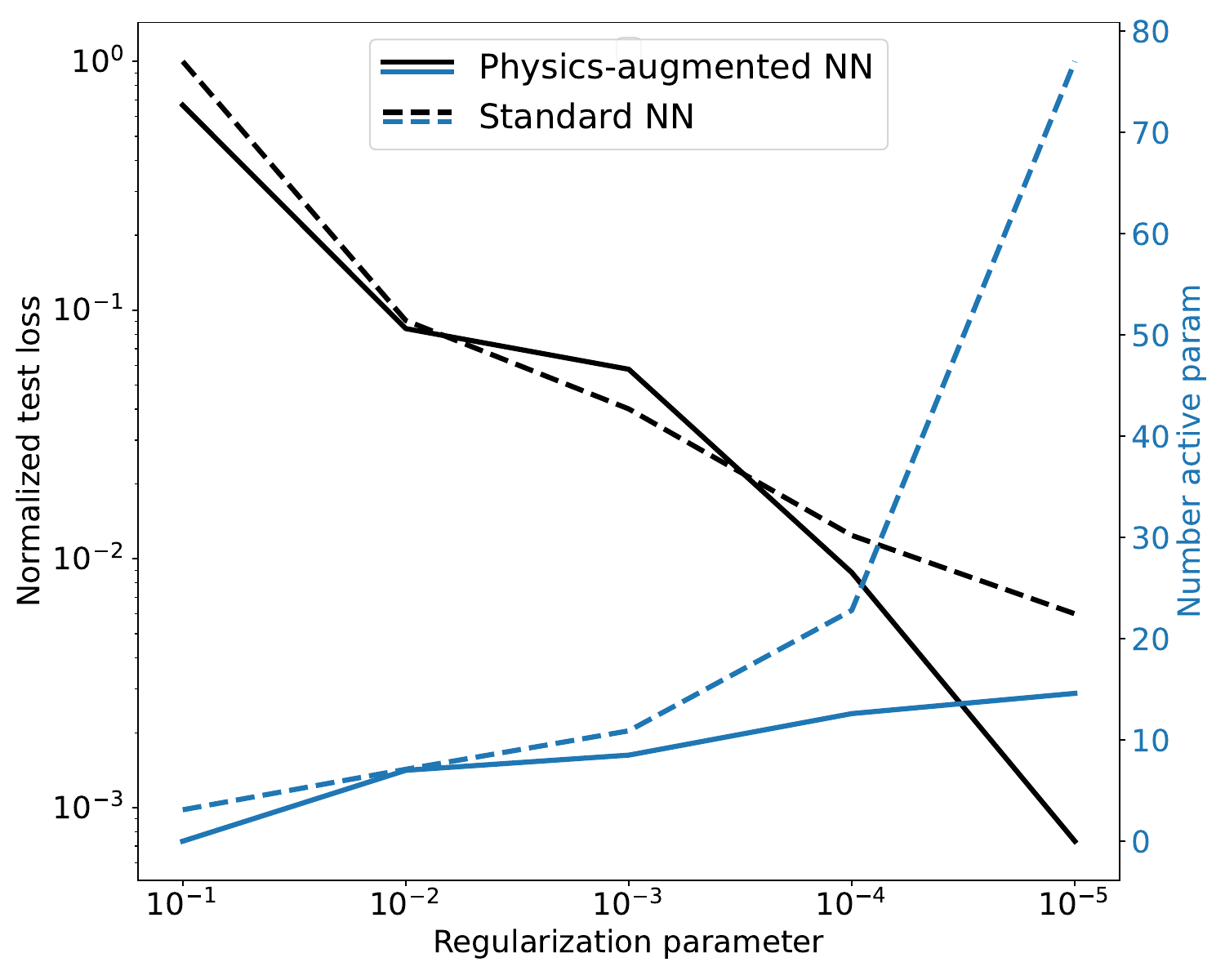}
    \caption{3  hidden layers - 30 neurons}\label{fig:onlyKinb}
\end{subfigure}
\caption{Effect of the network architecture and the $\mathcal{L}^{0}$-regularization parameter on the Gent-Gent law test loss and the number of active parameters for physics-augmented neural networks and standard neural networks. Averaged over 10 runs.}\label{fig:GentTestLossNoICNN}
\end{figure}

Next, we look at the effect of the physics augmentation compared to a standard neural network model. Figure \ref{fig:GentTestLossNoICNN}
depicts the influence of the $\mathcal{L}^{0}$ regularization parameter and the network architecture on the averaged (over 10 runs) test loss and number of active parameters of the Gent-Gent model. Two things are noteworthy: (i) the test loss is generally similar between the two network formulations, and (ii) the average number of active parameters of the standard neural networks seems to be significantly more sensitive to the regularization parameter than its physics-augmented counterparts. Apart from the guaranteed adherence to physical principles and mechanistic assumptions, this seems to be a major positive side effect of the physics augmentation. The robustness of the physics-augmented approach to maintain interpretability irrespective of the choice of the regularization parameter is a key feature of the proposed framework here, and it will positively impact the overall trustworthiness of the approach, as will be discussed further in the manuscript. 

Lastly, to prove the flexibility of the proposed approach, we also consider two more hyperelastic laws, in particular, a
compressible Mooney-Rivlin model \cite{mooney1940theory}
and a polynomial model \cite{flaschel2021unsupervised}. We again ran 10 random realizations of the training process and report the result with the median final training loss after $100,000$ epochs.
The true functional forms as well as their sparsified representations of the neural network trained on 40 training data points are shown in Table \ref{tab:HyperLaws}. We can see that these fitted functional forms are similarly interpretable to the one discussed for the Gent-Gent model. For example, neither of the other two obtained representations fulfill the coercivity condition either.

\begin{table}
\begin{center}
\begin{tabular}{|c|}
                \hline 
                    \\
\textbf{Gent-Gent:}\\
$\Psi_{gent} = -\frac{\theta_{1}}{2} J_{m} \log \left( 1 - \frac{I_{1}-3}{J_{m}} \right) - \theta_{2} \log \left( \frac{I_{2}}{J} \right) + \theta_{3} \left( \frac{1}{2} (J^{2}-1) - \log J \right), $ \\
   $\theta_{1} =2.4195$, $J_{m}= 77.931$,
   $\theta_{2} =-0.75$,
   $\theta_{3} =1.20975 $
\\ \\ \hdashline \\  
\textbf{Gent-Gent fit:}\\
$  \hat{\Psi}=  0.398 J + 3.095 \log{\left(\left(1 + e^{- 1.356 I_{2}}\right)^{1.314} \left(e^{0.755 I_{1}} + 1\right)^{0.515} \left(e^{0.135 I_{1} - 0.319 I_{2} - 0.329 J} + 1\right)^{1.874} + 1 \right)} - 6.686$
\\ \\ \hline  \hline \\
\textbf{Mooney-Rivlin:}\\
$
\Psi_{MR} = \theta_{1} \left( \frac{I_{1}}{J^{2/3}}- 3 \right) + \theta_{2} \left( \frac{I_{2}}{J^{4/3}} -3 \right) + \theta_{3} (J-1)^{2}\ ,
$\\
   $\theta_{1} =9.2\cdot 10^{-4}$, 
   $\theta_{2} =2.37 \cdot 10^{-3}$,
   $\theta_{3} =10.0010 $
\\ \\ \hdashline \\  
\textbf{Mooney-Rivlin fit:}\\
$
      \hat{\Psi}_{MR}=  90.338 J + 17.263 \log{\left(\left(1 + e^{- 0.368 J}\right)^{34.796} + 1 \right)} + 0.276 \log{\left(\left(e^{- 1.671 I_{1} + 1.305 I_{2}} + 1\right)^{1.298} + 1 \right)} - 406.511
$
\\ \\ \hline  \hline \\
                    
                    \textbf{Polynomial:}\\
   $ \Psi_{poly} = \theta_{1} \left( I_{1} -3 \right)^{2} + \theta_{2} \left( I_{1} -3 \right)^{4} + \theta_{3} \left( I_{2} -3 \right)^{2} + \theta_{4} \left( I_{2} -3 \right)^{4} + \theta_{5} \left( I_{3} -1 \right)^{2}$ ,\\
   $\theta_{1} = 0.1$, 
   $\theta_{2} = 0.15$,
   $\theta_{3} = 2\cdot10^{-4}$,
    $\theta_{4} = 1\cdot10^{-4}$,
    $\theta_{5} = 0.125$
   \\ \\ \hdashline \\   \textbf{Polynomial fit:}\\ 
   $\hat{\Psi}_{poly}=- 0.116 J + 9.046 \log{\left(\left(1 + e^{- 0.179 I_{2}}\right)^{3.59} \left(e^{0.193 I_{2}} + 1\right)^{1.927} + 1 \right)} $\\$+ 0.597 \log{\left(\left(e^{- 2.931 I_{1} + 2.272 I_{2}} + 1\right)^{0.696} + 1 \right)} - 33.35
$ \\ \\ \hline 
\end{tabular}
    \end{center}
    \caption{Expressions of the studied hyperelastic laws and examples of their sparse representations.}
    \label{tab:HyperLaws}
\end{table}

\begin{figure}
\begin{subfigure}[b]{0.45\linewidth}
        \centering
        \includegraphics[scale=0.3]{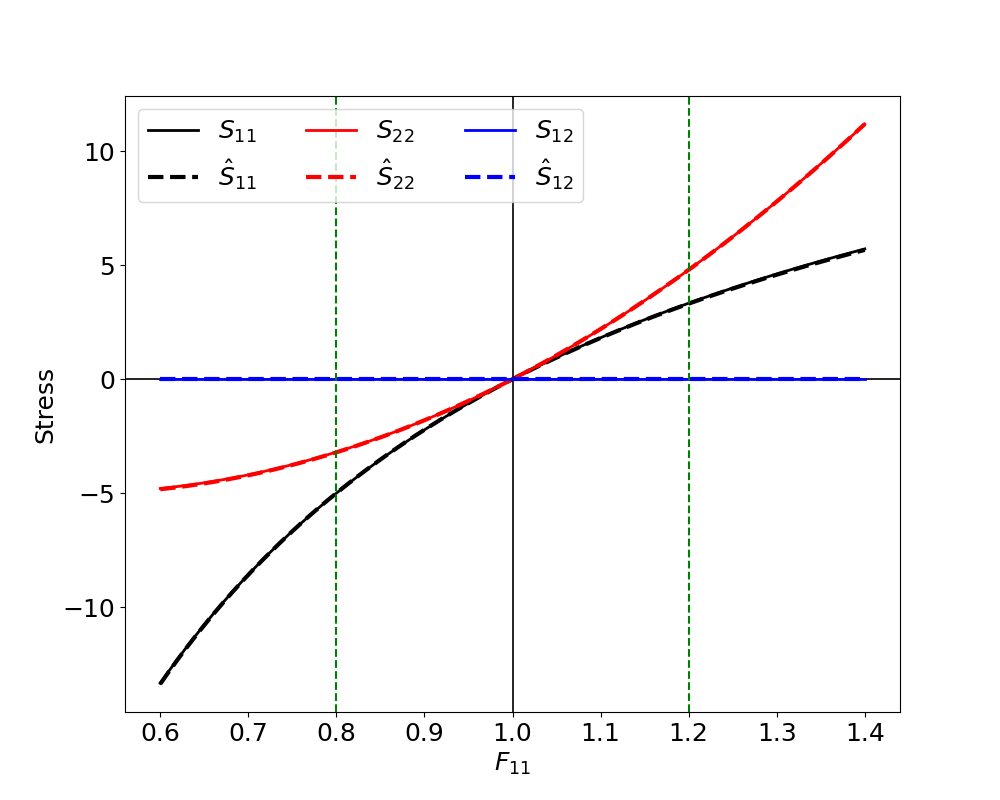}
    \caption{MooneyRivlin}\label{fig:fig:hyperStressStrainB}
\end{subfigure}
\begin{subfigure}[b]{0.45\linewidth}
        \centering
        \includegraphics[scale=0.3]{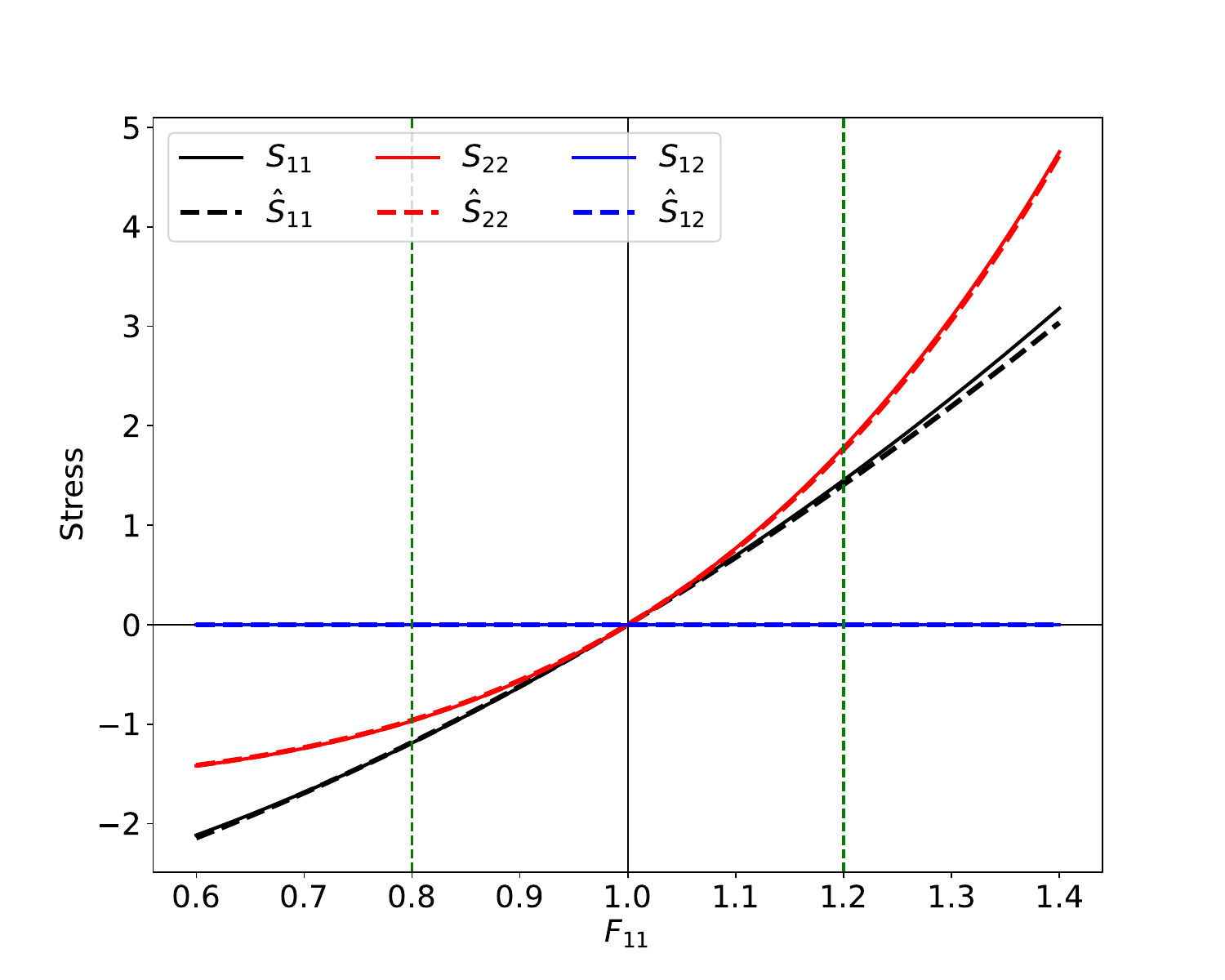}
    \caption{Polynomial}\label{fig:fig:hyperStressStrainC}
\end{subfigure}
\caption{Stress strain curves of fitted hyperelastic laws. Green dotted lines represent the limits of the training regime.}\label{fig:hyperStressStrain}
\end{figure}

Furthermore, similarly, to the Gent-Gent representation, the formulations obtained by these networks also appear accurate inside and even far outside the training regime, see
Figure \ref{fig:hyperStressStrain}, where the training domain is highlighted by the green dotted lines.

\subsection{Incompressible hyperelastic materials from experimental data}
If the material is incompressible, it can accommodate no volumetric deformations, captured by the constraint $J=1$. We can then assume a predictor of the strain energy function given by
\begin{equation}\label{eq:strainIncom}
\begin{aligned}
    \hat{\Psi}(I_{1}, I_{2}) &=\hat{\Psi}^{NN}(I_{1}, I_{2}) - (p+n) (J-1) - \hat{\Psi}^{NN}(3,3)
\end{aligned}
\end{equation}
where the second term on the right-hand side of this expression has no contribution on the strain energy, $p$ is the hydrostatic pressure (remaining to be determined by the boundary value problem at hand) and 
\begin{equation}
    n = 2   \left. \left(\frac{\partial \hat{\Psi}^{NN} }{\partial I_{1}} + 2 \frac{\partial \hat{\Psi}^{NN} }{\partial I_{2}} \right)\right\rvert_{\bm{C}=\bm{I}}
\end{equation}
is determined by enforcing the normalization constraint.
Furthermore $\hat{\Psi}^{NN}(\bullet)$ is the output of an input convex neural network. Formulation \eqref{eq:strainIncom} allows us to obtain a model that fulfills all constitutive constraints described above.
The derivation can be found in Appendix \ref{app:InComhyperelasticity}.

We can then obtain a formulation for the first Piola-Kirchoff stress with
\begin{equation}\label{eq:PForInc}
\begin{aligned}
        \bm{P} = 2 \left( \left[ \frac{\partial \hat{\Psi} }{\partial I_{1}} + I_{1} \frac{\partial \hat{\Psi} }{\partial I_{2}} \right]  \bm{F} - \frac{\partial \hat{\Psi}}{\partial I_{2}}   \bm{F} \bm{C} \right) - (p+n) J \bm{F}^{-T}
\end{aligned}
\end{equation}
which in terms of the principal strains $\lambda_{1} \leq \lambda_{2} \leq \lambda_{3}$ reads
\begin{equation}
    \begin{aligned}
        P_{i} \lambda_{i} =   2 \left( \lambda_{i}^{2} \frac{\partial \hat{\Psi} }{\partial I_{1}} +  \frac{\partial \hat{\Psi} }{\partial I_{2}}  \left[ (\lambda_{1}^{2}+\lambda_{2}^{2}+\lambda_{3}^{2})\lambda_{i}^{2}-  \lambda_{i}^{4}\right] \right) - p- n.
    \end{aligned}
\end{equation}
Analytical forms for the hydrostatic pressure can be found for some simple boundary value problems. 
\begin{table}
\begin{center}
\begin{tabular}{|c|| c||c |}
    \hline
Type & Deformation & Relevant experimental output\\
        \hline
                \hline 
                &   & \\
Uniaxial tension (UT)
\scalebox{0.2}{
\begin{tikzpicture}
[cube/.style={very thick,black},
			grid/.style={very thin,gray},
			axis/.style={->,blue,thick},cubeSmall/.style={thin,black}]
	\draw[cube] (0,0,1) -- (0,2,1) -- (2,2,1) -- (2,0,1) -- cycle;
	\draw[cube] (0,2,0) -- (0,2,1);
	\draw[cube] (2,0,0) -- (2,0,1);
	\draw[cube] (2,2,0) -- (2,2,1);
 	\draw[cube] (0,2,0) -- (2,2,0);
	\draw[cube] (2,2,0) -- (2,0,0);
	\draw[cubeSmall] (0.2,0.2,1) -- (0.2,1.8,1);
 	\draw[cubeSmall] (0.4,0.2,1) -- (0.4,1.8,1);
 	\draw[cubeSmall] (0.6,0.2,1) -- (0.6,1.8,1);
  	\draw[cubeSmall] (0.8,0.2,1) -- (0.8,1.8,1);
  	\draw[cubeSmall] (1.0,0.2,1) -- (1.0,1.8,1);
  	\draw[cubeSmall] (1.2,0.2,1) -- (1.2,1.8,1);
  	\draw[cubeSmall] (1.4,0.2,1) -- (1.4,1.8,1);
    \draw[cubeSmall] (1.6,0.2,1) -- (1.6,1.8,1);
    \draw[cubeSmall] (1.8,0.2,1) -- (1.8,1.8,1);
    \draw [-{Stealth[scale=2]},line width=1.5pt] (1.0,-0.2,0.5) -- (1.0,-1.5,0.5); 
    \draw [-{Stealth[scale=2]},line width=1.5pt] (1.0,2.0,0.5) -- (1.0,3.5,0.5); 
    \addvmargin{1mm};
  \end{tikzpicture}
  }
& $\bm{F} = \begin{bmatrix}
        \lambda_1 & 0  & 0 \\
        0 & \frac{1}{\sqrt{\lambda_{1}}} & 0 \\
        0 & 0 & \frac{1}{\sqrt{\lambda_{1}}}    \end{bmatrix}$ &   $P_{1} 
= 2 \left( \frac{\partial \hat{\Psi} }{\partial I_{1}} + \frac{1}{\lambda_{1}} \frac{\partial \hat{\Psi} }{\partial I_{2}} \right) \left[ \lambda_{1} - \frac{1}{\lambda_{1}^{2}}\right]$ \\ & &\\
   \hline 
  & &  \\
Equibiaxial tension (ET) \scalebox{0.2}{
\begin{tikzpicture}
[cube/.style={very thick,black},
			grid/.style={very thin,gray},
			axis/.style={->,blue,thick},cubeSmall/.style={thin,black}]
	\draw[cube] (0,0,1) -- (0,2,1) -- (2,2,1) -- (2,0,1) -- cycle;
	\draw[cube] (0,2,0) -- (0,2,1);
	\draw[cube] (2,0,0) -- (2,0,1);
	\draw[cube] (2,2,0) -- (2,2,1);
 	\draw[cube] (0,2,0) -- (2,2,0);
	\draw[cube] (2,2,0) -- (2,0,0);
	\draw[cubeSmall] (0.2,0.2,1) -- (0.2,1.8,1);
 	\draw[cubeSmall] (0.4,0.2,1) -- (0.4,1.8,1);
 	\draw[cubeSmall] (0.6,0.2,1) -- (0.6,1.8,1);
  	\draw[cubeSmall] (0.8,0.2,1) -- (0.8,1.8,1);
  	\draw[cubeSmall] (1.0,0.2,1) -- (1.0,1.8,1);
  	\draw[cubeSmall] (1.2,0.2,1) -- (1.2,1.8,1);
  	\draw[cubeSmall] (1.4,0.2,1) -- (1.4,1.8,1);
    \draw[cubeSmall] (1.6,0.2,1) -- (1.6,1.8,1);
    \draw[cubeSmall] (1.8,0.2,1) -- (1.8,1.8,1);
    \draw [-{Stealth[scale=2]},line width=1.5pt] (1.0,-0.2,0.5) -- (1.0,-1.5,0.5); 
    \draw [-{Stealth[scale=2]},line width=1.5pt] (1.0,2.0,0.5) -- (1.0,3.5,0.5); 
    \draw [-{Stealth[scale=2]},line width=1.5pt] (2.0,1.0,0.5) -- (3.5,1.0,0.5); 
    \draw [-{Stealth[scale=2]},line width=1.5pt] (0.0,1.0,0.5) -- (-1.5,1.0,0.5); 
    \addvmargin{1mm};
  \end{tikzpicture}
  }&  
   $\bm{F} = \begin{bmatrix}
        \lambda_1 & 0  & 0 \\
        0 & \lambda_{1} & 0 \\
        0 & 0 & \frac{1}{\lambda_{1}^2}    \end{bmatrix}$ 
  & $P_{1} = P_{2}= 2 \left( \frac{\partial \hat{\Psi} }{\partial I_{1}} + \lambda_{1}^{2}  \frac{\partial \hat{\Psi} }{\partial I_{2}} \right) \left[ \lambda_{1} - \frac{1}{\lambda_{1}^{5}}\right]$ \\& &   \\
   \hline 
  & &  \\
Pure shear stress (PS) \scalebox{0.2}{
\begin{tikzpicture}
[cube/.style={very thick,black},
			grid/.style={very thin,gray},
			axis/.style={->,blue,thick},cubeSmall/.style={thin,black}]
	\draw[cube] (0,0,1) -- (0,2,1) -- (2,2,1) -- (2,0,1) -- cycle;
	\draw[cube] (0,2,0) -- (0,2,1);
	\draw[cube] (2,0,0) -- (2,0,1);
	\draw[cube] (2,2,0) -- (2,2,1);
 	\draw[cube] (0,2,0) -- (2,2,0);
	\draw[cube] (2,2,0) -- (2,0,0);
	\draw[cubeSmall] (0.2,0.2,1) -- (0.2,1.8,1);
 	\draw[cubeSmall] (0.4,0.2,1) -- (0.4,1.8,1);
 	\draw[cubeSmall] (0.6,0.2,1) -- (0.6,1.8,1);
  	\draw[cubeSmall] (0.8,0.2,1) -- (0.8,1.8,1);
  	\draw[cubeSmall] (1.0,0.2,1) -- (1.0,1.8,1);
  	\draw[cubeSmall] (1.2,0.2,1) -- (1.2,1.8,1);
  	\draw[cubeSmall] (1.4,0.2,1) -- (1.4,1.8,1);
    \draw[cubeSmall] (1.6,0.2,1) -- (1.6,1.8,1);
    \draw[cubeSmall] (1.8,0.2,1) -- (1.8,1.8,1);
    \draw [-{Stealth[scale=2]},line width=1.5pt] (0.0,-0.5,0.5) -- (2.0,-0.5,0.5); 
    \draw [-{Stealth[scale=2]},line width=1.5pt] (2.0,2.5,0.5) -- (0.0,2.5,0.5); 
    \draw [-{Stealth[scale=2]},line width=1.5pt] (-.5,2.0,0.5) -- (-.5,0.0,0.5); 
    \draw [-{Stealth[scale=2]},line width=1.5pt] (2.5,0.0,0.5) -- (2.5,2.0,0.5); 
    \addvmargin{1mm};
  \end{tikzpicture}
  }& 
   $\bm{F} = \begin{bmatrix}
        \lambda_1 & 0  & 0 \\
        0 & 1 & 0 \\
        0 & 0 & \frac{1}{\lambda_{1}}    \end{bmatrix}$ 
  & \multicolumn{1}{r|}{\makecell{ $    P_{1} = 2 \left(\frac{\partial \hat{\Psi} }{\partial I_{1}} + \frac{\partial \hat{\Psi} }{\partial I_{2}} \right) \left[ \lambda_{1} - \frac{1}{\lambda_{1}^{3}} \right]$\\ $   P_{2} = 2 \left( \frac{\partial \hat{\Psi} }{\partial I_{1}} + \lambda_{1}^{2} \frac{\partial \hat{\Psi} }{\partial I_{2}} \right) \left[ 1 - \frac{1}{\lambda_{1}^{2}} \right]$}} \\& &   \\
   \hline 
  & &  \\
Simple shear deformation (SS) \scalebox{0.2}{
\begin{tikzpicture}
[cube/.style={very thick,black},
			grid/.style={very thin,gray},
			axis/.style={->,blue,thick},cubeSmall/.style={thin,black}]
	\draw[cube] (0,0,1) -- (0,2,1) -- (2,2,1) -- (2,0,1) -- cycle;
	\draw[cube] (0,2,0) -- (0,2,1);
	\draw[cube] (2,0,0) -- (2,0,1);
	\draw[cube] (2,2,0) -- (2,2,1);
 	\draw[cube] (0,2,0) -- (2,2,0);
	\draw[cube] (2,2,0) -- (2,0,0);
	\draw[cubeSmall] (0.2,0.2,1) -- (0.2,1.8,1);
 	\draw[cubeSmall] (0.4,0.2,1) -- (0.4,1.8,1);
 	\draw[cubeSmall] (0.6,0.2,1) -- (0.6,1.8,1);
  	\draw[cubeSmall] (0.8,0.2,1) -- (0.8,1.8,1);
  	\draw[cubeSmall] (1.0,0.2,1) -- (1.0,1.8,1);
  	\draw[cubeSmall] (1.2,0.2,1) -- (1.2,1.8,1);
  	\draw[cubeSmall] (1.4,0.2,1) -- (1.4,1.8,1);
    \draw[cubeSmall] (1.6,0.2,1) -- (1.6,1.8,1);
    \draw[cubeSmall] (1.8,0.2,1) -- (1.8,1.8,1);
    \draw [-{Stealth[scale=2]},line width=1.5pt] (0.0,2.5,0.5) -- (2.0,2.5,0.5); 
    \draw[cubeSmall] (0.0,0.0,1) -- (-.2,-0.2,1);
    \draw[cubeSmall] (0.2,0.0,1) -- (-.0,-0.2,1);
    \draw[cubeSmall] (0.4,0.0,1) -- (.2,-0.2,1);
    \draw[cubeSmall] (0.6,0.0,1) -- (.4,-0.2,1);
    \draw[cubeSmall] (0.8,0.0,1) -- (.6,-0.2,1);
    \draw[cubeSmall] (1.0,0.0,1) -- (.8,-0.2,1);
    \draw[cubeSmall] (1.2,0.0,1) -- (1.0,-0.2,1);
    \draw[cubeSmall] (1.4,0.0,1) -- (1.2,-0.2,1);
    \draw[cubeSmall] (1.6,0.0,1) -- (1.4,-0.2,1);
    \draw[cubeSmall] (1.8,0.0,1) -- (1.6,-0.2,1);
    \draw[cubeSmall] (2.0,0.0,1) -- (1.8,-0.2,1);
    \addvmargin{1mm};
  \end{tikzpicture}
  }& $\bm{F} =\begin{bmatrix}
        1 & \gamma  & 0 \\
        0 & 1 & 0 \\
        0 & 0 & 1    \end{bmatrix}$ & $      P_{12} = 2 \gamma \left(  \frac{\partial \hat{\Psi} }{\partial I_{1}} +  \frac{\partial \hat{\Psi} }{\partial I_{2}} \right) $ \\& &   \\
   \hline 
  & &  \\
   Simple torsion \scalebox{0.2}{\begin{tikzpicture}[
cube/.style={very thick,black},
			grid/.style={very thin,gray},
			axis/.style={->,blue,thick},cubeSmall/.style={thin,black}]
\node[cylinder, 
draw,
    minimum size = 2cm] (c) at (0.5,1,0.5) {};
    \draw[cubeSmall] (-0.2,0.4,1) -- (-0.4,0.6,1);
    \draw[cubeSmall] (-0.2,0.6,1) -- (-0.4,0.8,1);
    \draw[cubeSmall] (-0.2,0.8,1) -- (-0.4,1.0,1);
    \draw[cubeSmall] (-0.2,1.0,1) -- (-0.4,1.2,1);
    \draw[cubeSmall] (-0.2,1.2,1) -- (-0.4,1.4,1);
    \draw[cubeSmall] (-0.2,1.4,1) -- (-0.4,1.6,1);
    \draw[cubeSmall] (-0.2,1.6,1) -- (-0.4,1.8,1);
    \draw[cubeSmall] (-0.2,1.8,1) -- (-0.4,2.0,1);
    \draw[cubeSmall] (-0.2,2.0,1) -- (-0.4,2.2,1);
    \tdplotsetmaincoords{60}{110}
    \tdplotsetthetaplanecoords{12} 
\tdplotdrawarc[tdplot_rotated_coords,-{Stealth[scale=2]},line width=1.5pt]{(2.0,1.5,2.3)}{1.0}{110}{460}{anchor=south west,color=black}{}
    \addvmargin{1mm};
  \end{tikzpicture}}
  & $ \bm{F} = \begin{bmatrix}
        1& 0 & 0 \\
        0 & 1& \rho \, \phi \\
        0 &0 & 1
    \end{bmatrix}$ & $        \tau = \int_{0}^{1} 4 \pi \rho^{3} \phi \left( \frac{\partial \hat{\Psi}}{\partial I_{1}} + \frac{\partial \hat{\Psi}}{\partial I_{2}} \right) d \rho$ \\& &   \\
   \hline 
\end{tabular}
    \end{center}
    \caption{Studied deformation modes with relevant output}
    \label{tab:deformationModes}
\end{table}

 In the following, we fit curves to data from uniaxial tension (UT), equibiaxial tension (ET), pure shear stress (PS), simple shear deformation (SS), and simple torsion (ST). The relevant deformations and quantities of interest that were used to fit the data are summarised in Table \ref{tab:deformationModes} and described in detail in Appendix \ref{app:InComhyperelasticity}.
 
\subsection{Examples}
 In the following, we analyze and study the performance of the sparsified physics-augmented ML model on different sets of experimental data. We have repeated the training process 10 times with different random seeds and in the following only report the results corresponding to the median training loss after (depending on the dataset) $50,000$ or $80,000$ epochs. We utilize an architecture of 1 hidden layer with 30 neurons and $\lambda=10^{-3}$ for all the following results.
\subsubsection{Model for rubber}
\begin{figure}
\begin{subfigure}[b]{0.45\linewidth}
        \centering
        \includegraphics[scale=0.3]{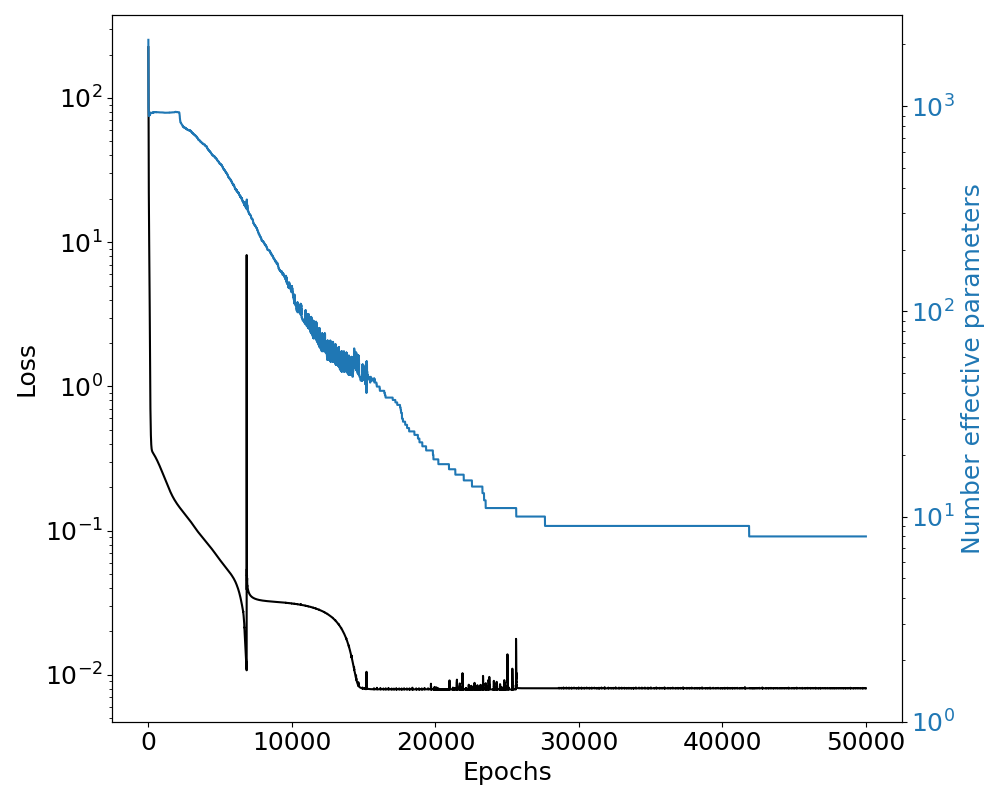}
    \caption{}\label{fig:Treloar20A}
\end{subfigure}
\begin{subfigure}[b]{0.45\linewidth}
        \centering
    \includegraphics[scale=0.3]{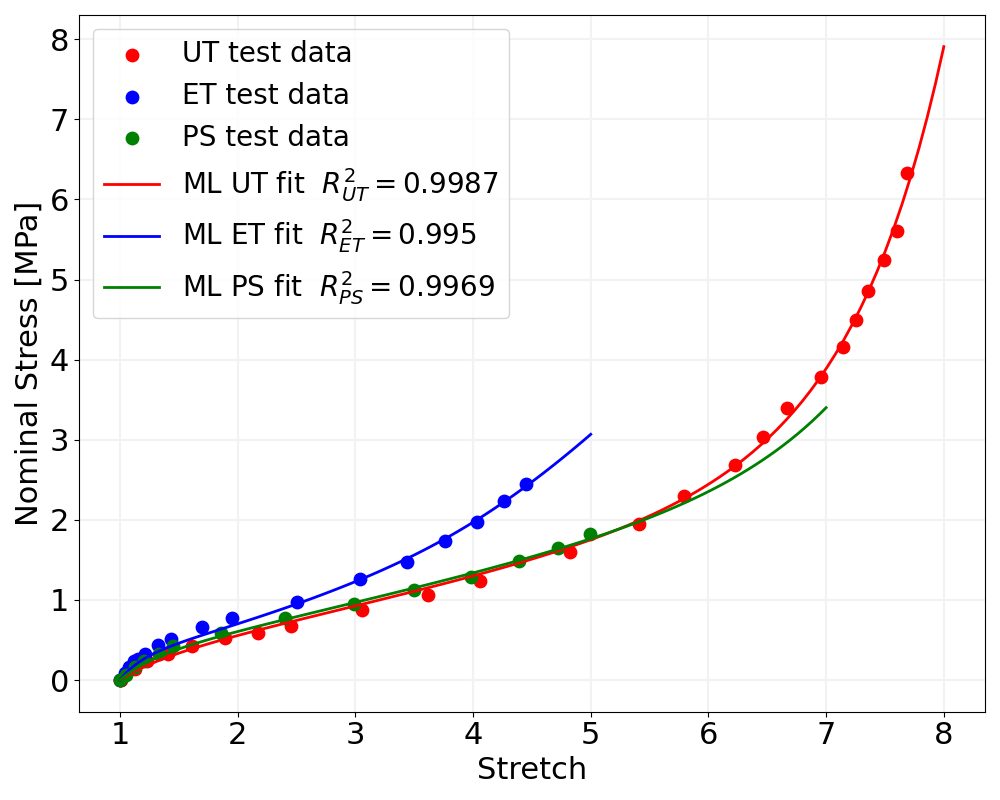}
    \caption{}\label{fig:Treloar20B}
\end{subfigure}
\caption{Rubber response $20^{\circ} C$. (a) Training loss and number of active parameters over the training process, (b) Data fit and $R^{2}$ errors.}\label{fig:Treloar20}
\end{figure}

One of the most used datasets for incompressible hyperelasticity goes back to Treloar \cite{treloar1944stress} who reported the uniaxial, equibiaxial and pure shear responses of vulcanized rubber for $20^{\circ} C$ and $50^{\circ} C$.
Following \cref{linka2023new} this data is summarized in Table \ref{tab:treloarRubber}. In the following, we use the UT and ET data as training data and test the performance on the PS dataset. 
\paragraph{$20^{\circ} C$ dataset}
For the $20^{\circ} C$ dataset, the evolution of the median training loss and the median number of active parameters is shown in  
Figure \ref{fig:Treloar20A}. It can be seen that both values reduce simultaneously and that the number of active parameters reduces from an initial value greater than $1000$ to around $10$ over the training process. The training data, the predicted stress responses, and the $R^{2}$-scores are depicted in Figure \ref{fig:Treloar20B}. The predicted responses on the training data are proficiently accurate. Further, even on the unseen PS data the model reaches an accuracy $R^{2}>0.99$.
The obtained sparsified representation for the compressible hyperelastic law reads 
\begin{equation}
\begin{aligned}
       \hat{\Psi} = 0.15 I_{1} + 6.3\cdot 10^{-3} I_{2} - 0.33 J - 0.28 - p \left(J - 1.0\right) 
       + 794.94 \log{\left(10^{-4} \left(e^{0.09 I_{1} - 0.01 I_{2}} + 1\right)^{0.73} + 1 \right)}
\end{aligned}
\end{equation}
which fulfills all the discussed constraints such as convexity and the normalization conditions. 

\begin{figure}
\begin{subfigure}[b]{0.45\linewidth}
        \centering
        \includegraphics[scale=0.3]{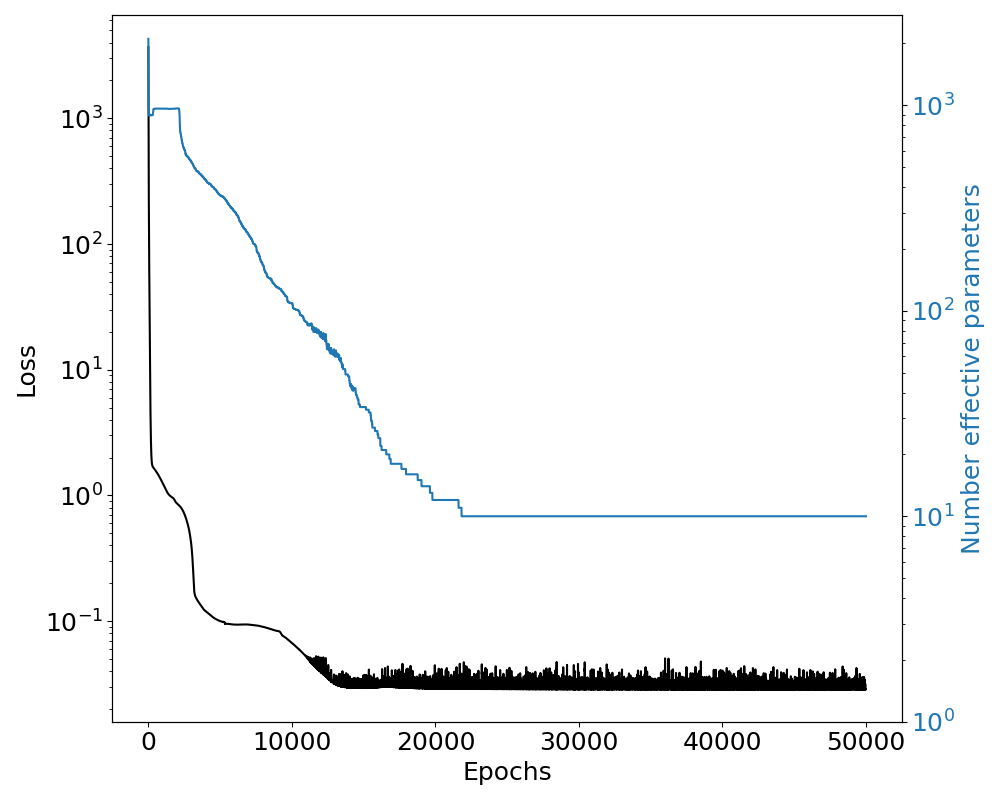}
    \caption{}\label{fig:Treloar50A}
\end{subfigure}
\begin{subfigure}[b]{0.45\linewidth}
        \centering
    \includegraphics[scale=0.3]{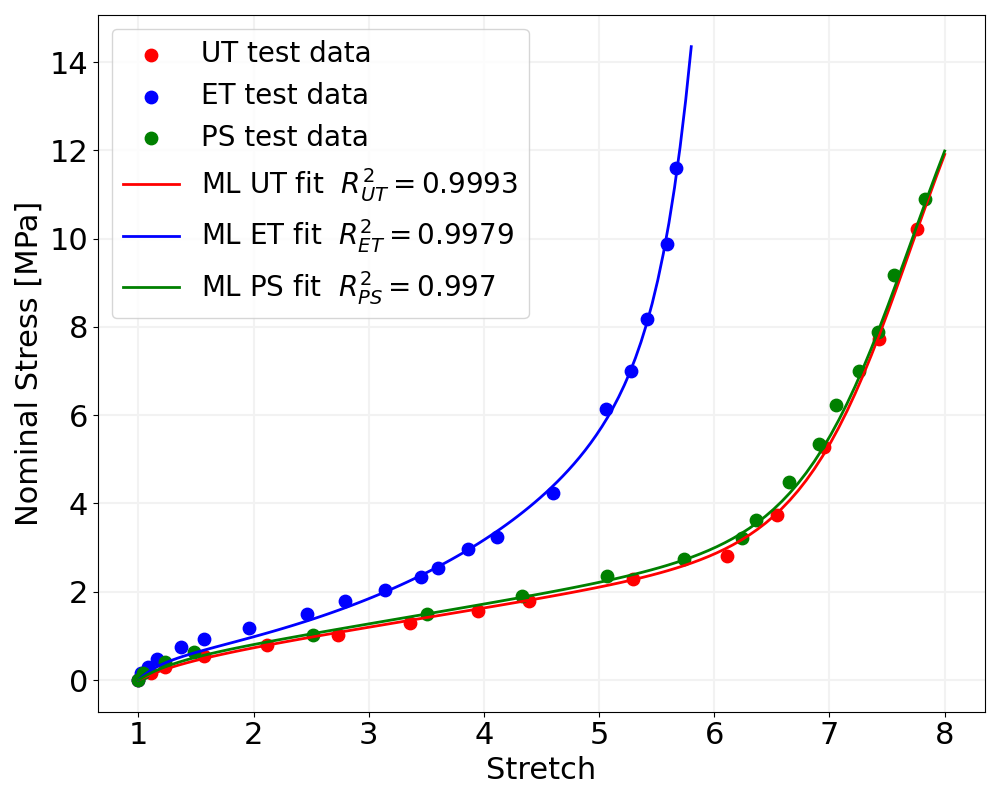}
    \caption{}\label{fig:Treloar50B}
\end{subfigure}
\caption{Rubber response $50^{\circ} C$. (a) Training loss and number of active parameters over the training process, (b) Data fit and $R^{2}$ errors.}\label{fig:Treloar50}
\end{figure}

\begin{figure}
\begin{subfigure}[b]{0.45\linewidth}
        \centering
        \includegraphics[scale=0.3]{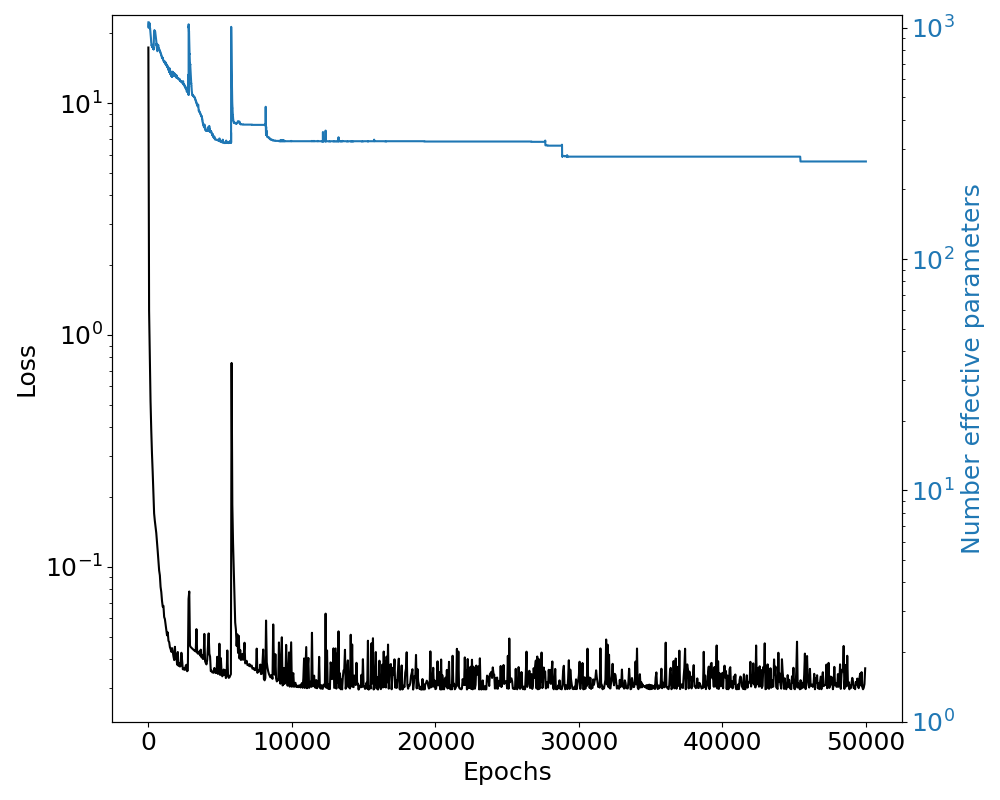}
    \caption{}\label{fig:Treloar50_WithoutL0A}
\end{subfigure}
\begin{subfigure}[b]{0.45\linewidth}
        \centering
    \includegraphics[scale=0.3]{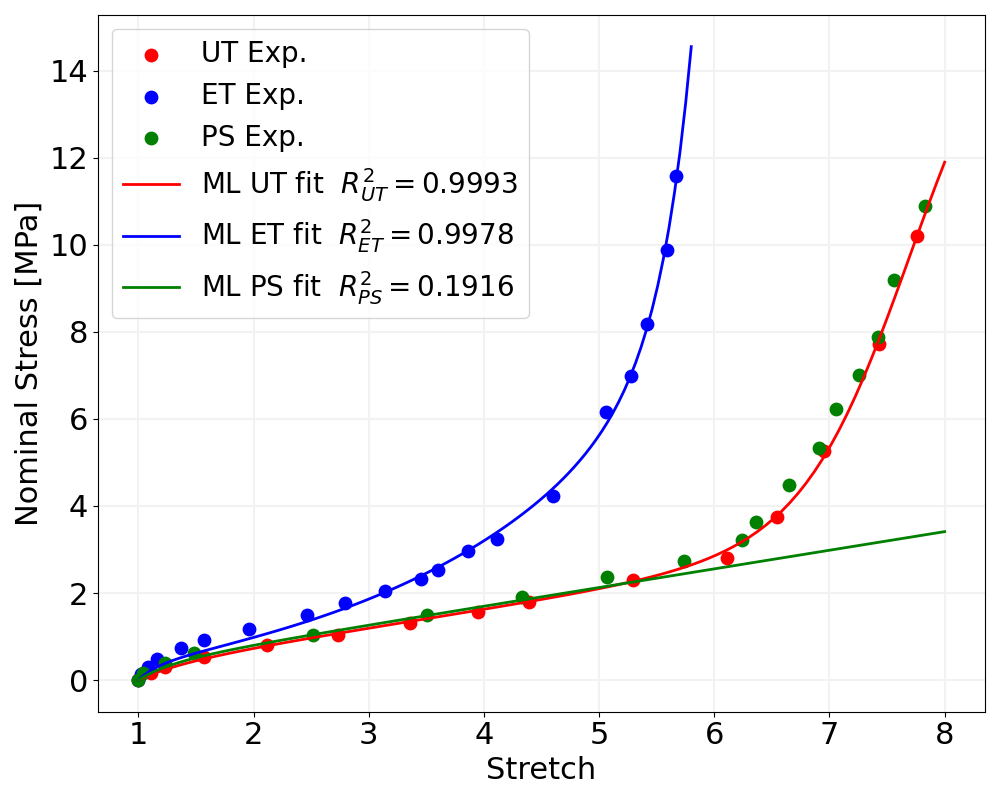}
    \caption{}\label{fig:Treloar50_WithoutL0B}
\end{subfigure}
\caption{Rubber response $50^{\circ} C$ using a physics-augmented neural network without $\mathcal{L}^{0}$-regularization. (a) Training loss and number of active parameters over the training process, (b) Data fit and $R^{2}$ errors.}\label{fig:Treloar50_WithoutL0}
\end{figure}

\paragraph{$50^{\circ} C$  dataset}
For the $50^{\circ} C$  dataset, the evolution of the training loss and the number of active parameters is plotted in 
Figure \ref{fig:Treloar50A}. At the end of the training process, the discovered strain formulation is given by
\begin{equation}
    \begin{aligned}
      \hat{\Psi}&=  0.2025 I_{1} + 0.0114 I_{2} - 0.4512 J - p \left(J - 1.0\right) + 5.1533 \log{\left(0.0009 \left(0.2155 e^{0.1852 I_{1} - 0.0028 I_{2}} + 1\right)^{0.7778} + 1 \right)} \\
      &+ 12.5236 \log{\left(0.0009 e^{0.0052 I_{2}} + 1 \right)} - 0.2076.
    \end{aligned}
\end{equation}
Using this formulation, Figure \ref{fig:Treloar50B} shows the stress responses, the training data points, and the $R^{2}$-scores. Here again, the predicted responses are accurate and the model shows it can generalize well.
For the same dataset Figure \ref{fig:Treloar50_WithoutL0} depicts the effect of the regularization. We can see that if we train a physics-augmented neural network with 1 hidden layer with $30$ neurons without reducing the number of trainable parameters with the $\mathcal{L}^{0}$-regularizer, the number of trainable parameters of the run with the median final training loss (over 10 runs) is still in the hundreds at the end of the training process. Furthermore, while the training data is fitted accurately in a comparable manner to the obtained model with $\mathcal{L}^{0}$-regularization, the model generalizes significantly worse on the unseen PS data, c.f. Figure \ref{fig:Treloar50_WithoutL0B}. We can hence see that in the limited data domain, the regularization not only increases the interpretability of the model but improves its generalization performance as well.

\subsubsection{Model for human brain tissue }
Next, we employ the $\mathcal{L}^{0}$-regularized neural network approach to find incompressible hyperelastic formulations for human brain tissues. In particular, we look at four datasets of mechanical tests on the cortex, the corona radiata and two different sections of the midbrain.

\begin{figure}
\begin{subfigure}[b]{0.45\linewidth}
        \centering
        \includegraphics[scale=0.3]{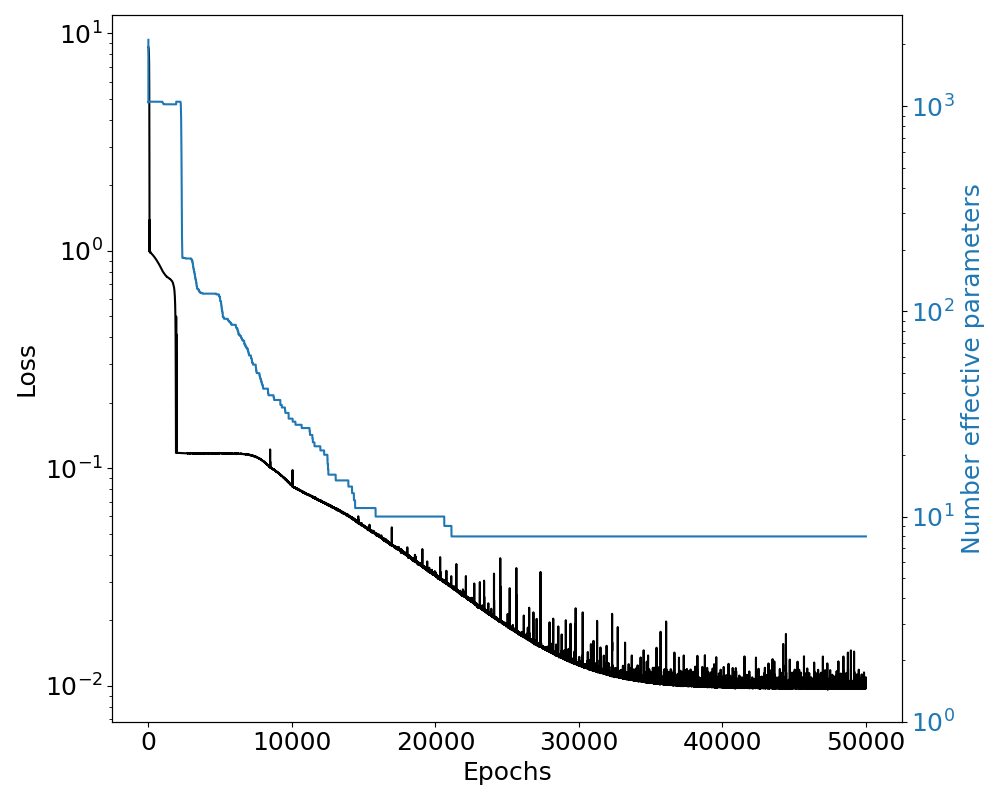}
    \caption{}\label{fig:CortexA}
\end{subfigure}
\begin{subfigure}[b]{0.45\linewidth}
        \centering
    \includegraphics[scale=0.3]{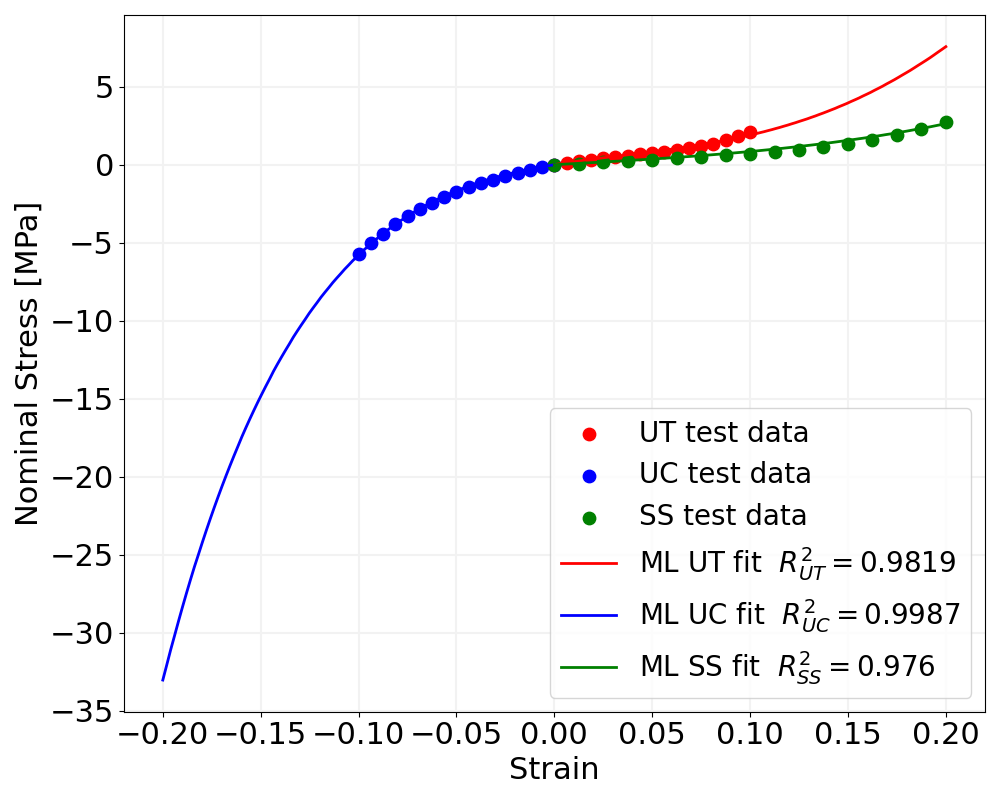}
    \caption{}\label{fig:CortexB}
\end{subfigure}
\caption{Cortex dataset. (a) Training loss and number of active parameters over the training process, (b) Data fit and $R^{2}$ errors.}\label{fig:Cortex}
\end{figure}

\paragraph{Cortex and Corona Radiata}
The first two datasets respectively contain UT, ET and SS data for the cortex and the corona radiata which were adopted from \cref{budday2017mechanical} and \cref{pierre2023principal}. The data is summarized in Table \ref{tab:CortexRadiataData};
For both cases, we use the UT and ET data as training data and test the model performance on the SS dataset.
Figure \ref{fig:CortexA} shows the evolution of the training loss and the number of active parameters over the training process for the cortex. We can see that over the training period of $50,000$ epochs both values significantly decrease and we end up with around $10$ trainable parameters. The final functional form of the strain energy functions reads 
\begin{equation}
    \begin{aligned}
        \hat{\Psi}_{Cortex} &= - 53.409 J - p \left(J - 1.0\right) - 6.954 \\
        &+ 1589.41 \log{\left(0.001 \left(1 + 0.001 e^{- 0.602 I_{1}}\right)^{10722.09}
        \left(1 + 0.001 e^{- 0.595 I_{1}}\right)^{10243.530} \left(0.001 e^{0.34 I_{1} + 0.220 I_{2}} + 1\right)^{569.93} + 1 \right)}
    \end{aligned}
\end{equation}
we argue that its complexity is (roughly) in line with other reported forms used to model human brain tissues, e.g. compare \cref{wang2023modified} which use a 3-term Ogden model \cite{ogden1972large}.
Figure \ref{fig:CortexB} depicts the data points, the predicted responses outside, and the $R^{2}$-error values. We can see that not only the training data is fitted proficiently well, and the model is also highly accurate on the shear stress test data. 

\begin{figure}
\begin{subfigure}[b]{0.45\linewidth}
        \centering
        \includegraphics[scale=0.3]{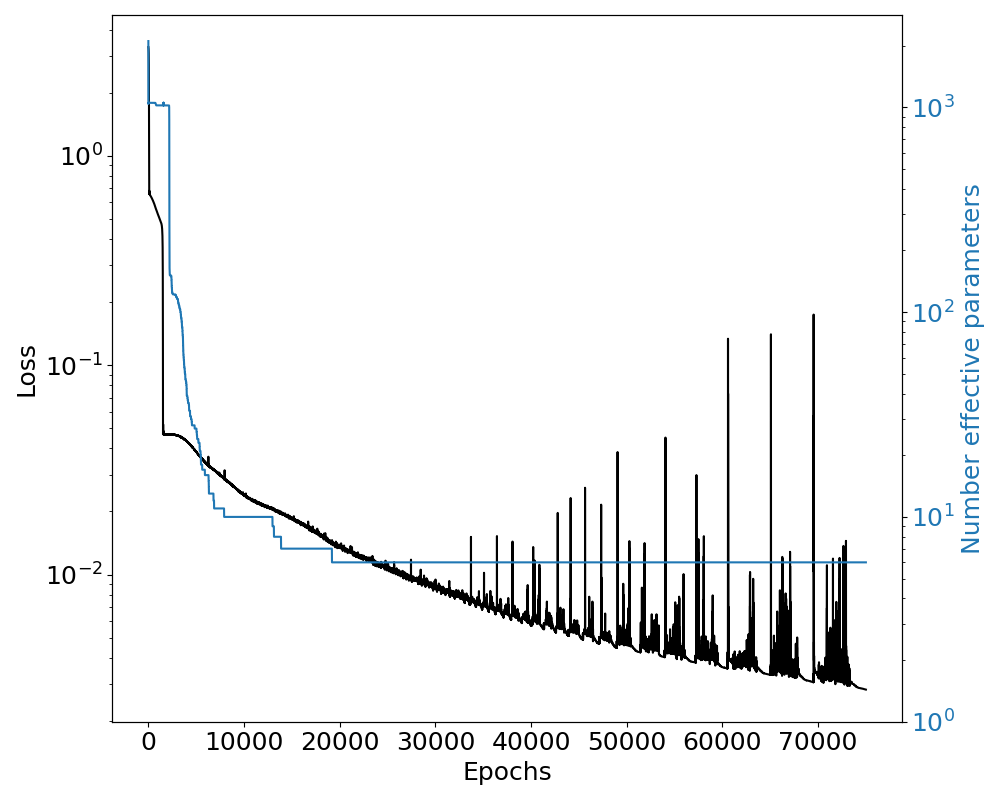}
    \caption{}\label{fig:CoronaRadiataA}
\end{subfigure}
\begin{subfigure}[b]{0.45\linewidth}
        \centering
    \includegraphics[scale=0.3]{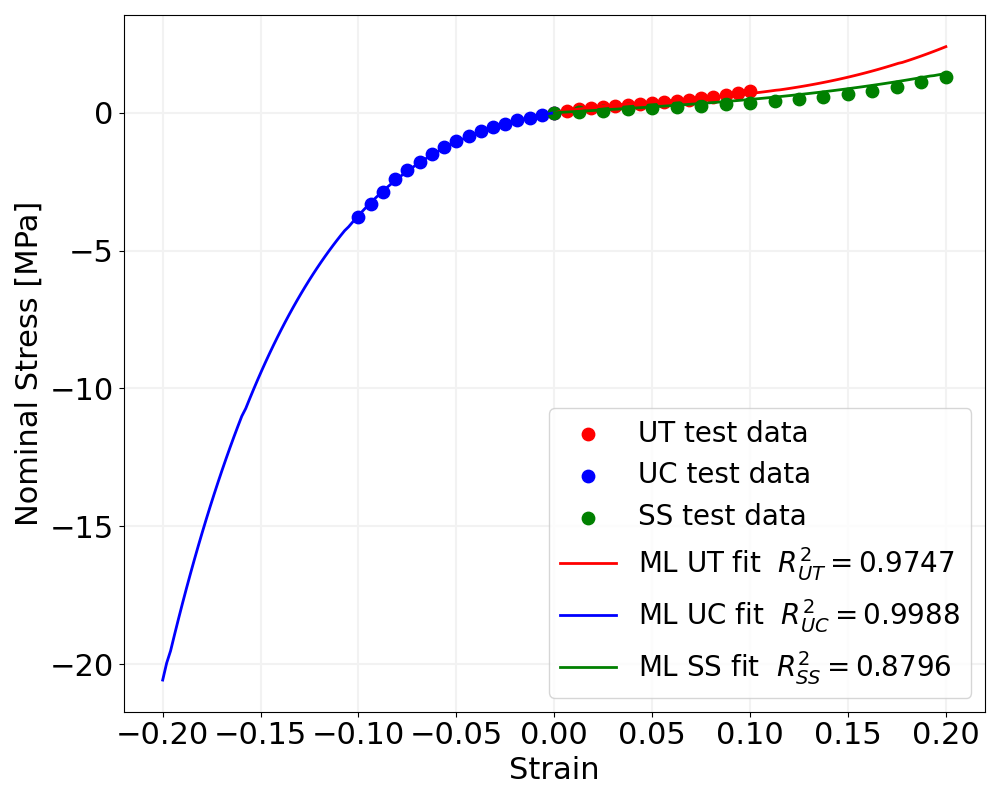}
    \caption{}\label{fig:CoronaRadiataB}
\end{subfigure}
\caption{Corona radiata dataset. (a) Training loss and number of active parameters over the training process, (b) Data fit and $R^{2}$ errors.}\label{fig:CoronaRadiata}
\end{figure}

The loss behavior and the number of active parameters of the network fitted on the Corona Radiata are shown in 
Figure \ref{fig:CoronaRadiataA}. We can see a similar performance to the Cortex example. The final functional form reads
\begin{equation}
\begin{aligned}
       \hat{\Psi}_\text{CRadiata} &= - 43.126 J - p \left(J - 1.0\right) - 56.327 \\
       &+ 4962.449 \log{\left(0.014 \left(1 + 0.0001 e^{- 1.294 I_{1}}\right)^{66688.773} \left(0.0001 e^{0.841 I_{2}} + 1\right)^{189.231} + 1 \right)}.
\end{aligned}
\end{equation}
The fitted responses and the $R^{2}$-score are shown in Figure \ref{fig:CoronaRadiataB} where again proficient generalization of the obtained model can be noted.

\begin{figure}
\begin{subfigure}[b]{1.0\linewidth}
        \centering
        \includegraphics[scale=0.3]{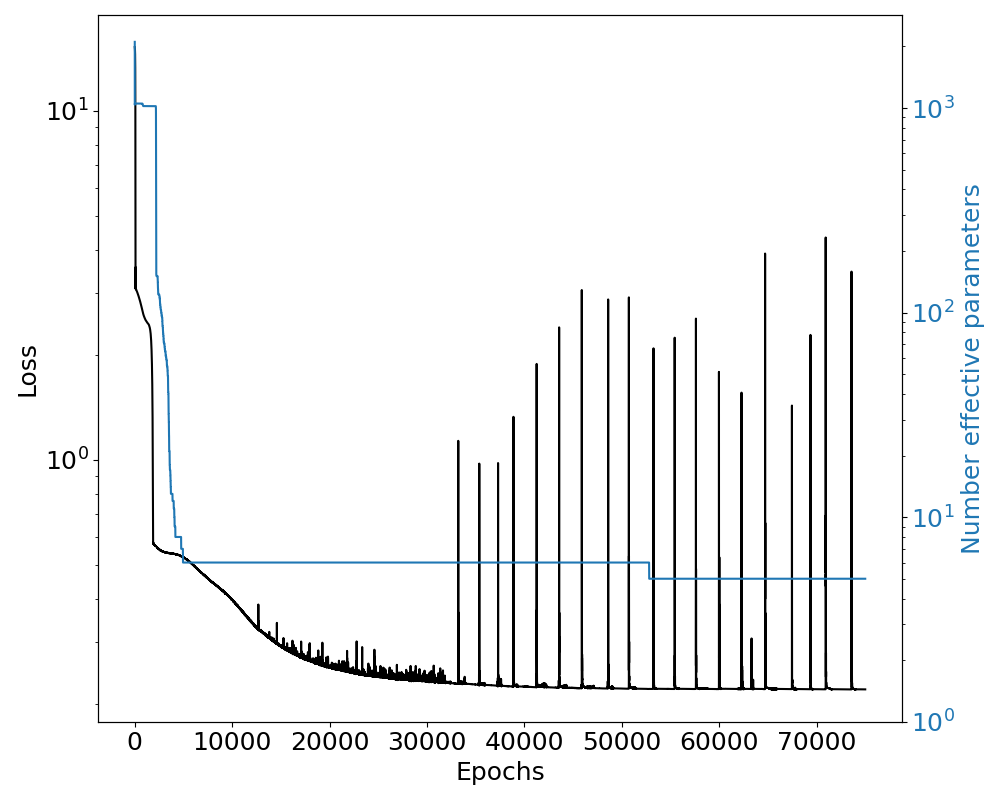}
    \caption{}\label{fig:MBSec1A}
\end{subfigure}

\begin{subfigure}[b]{0.45\linewidth}
        \centering
    \includegraphics[scale=0.3]{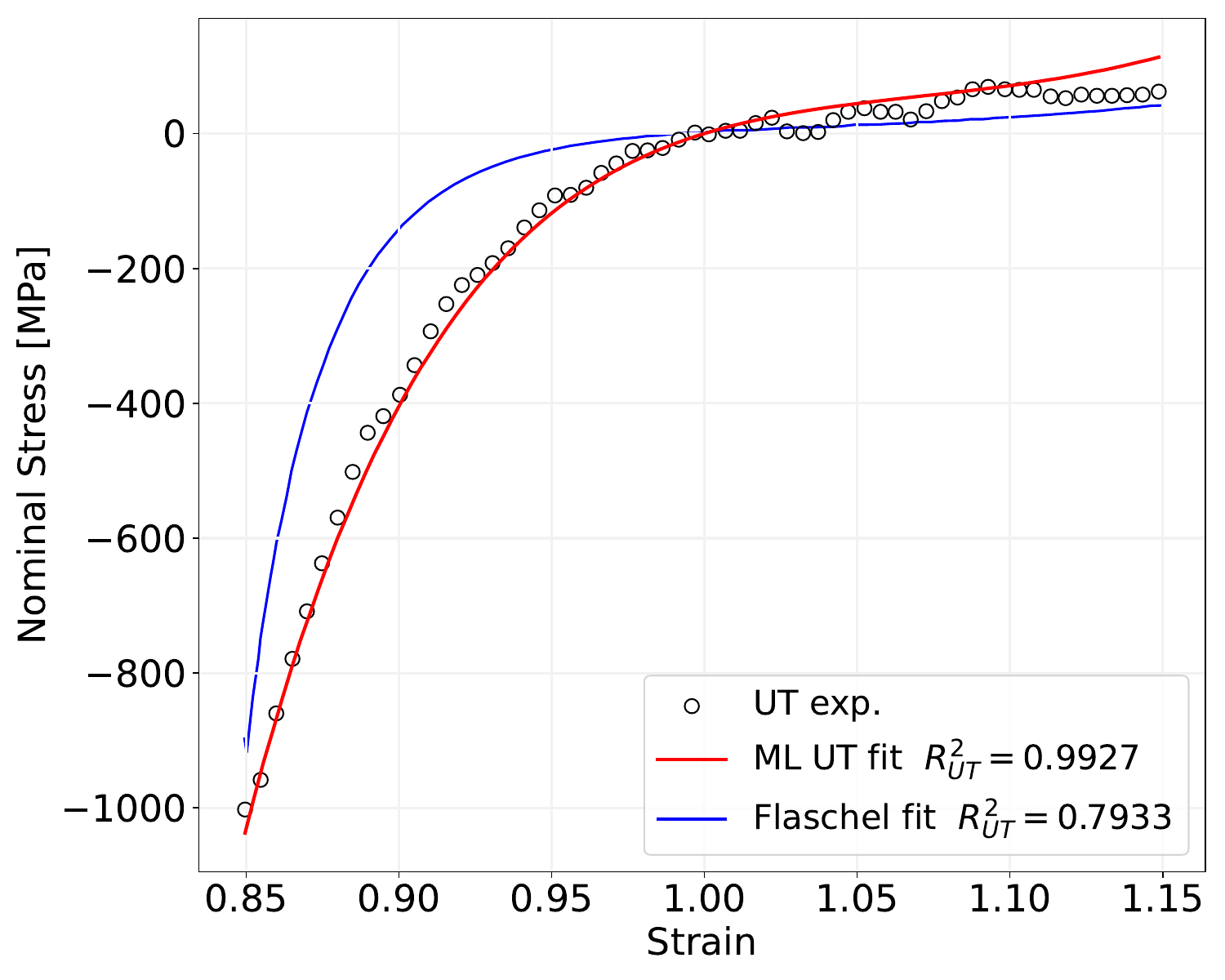}
    \caption{}\label{fig:MBSec1B}
\end{subfigure}
\begin{subfigure}[b]{0.45\linewidth}
        \centering
        \includegraphics[scale=0.3]{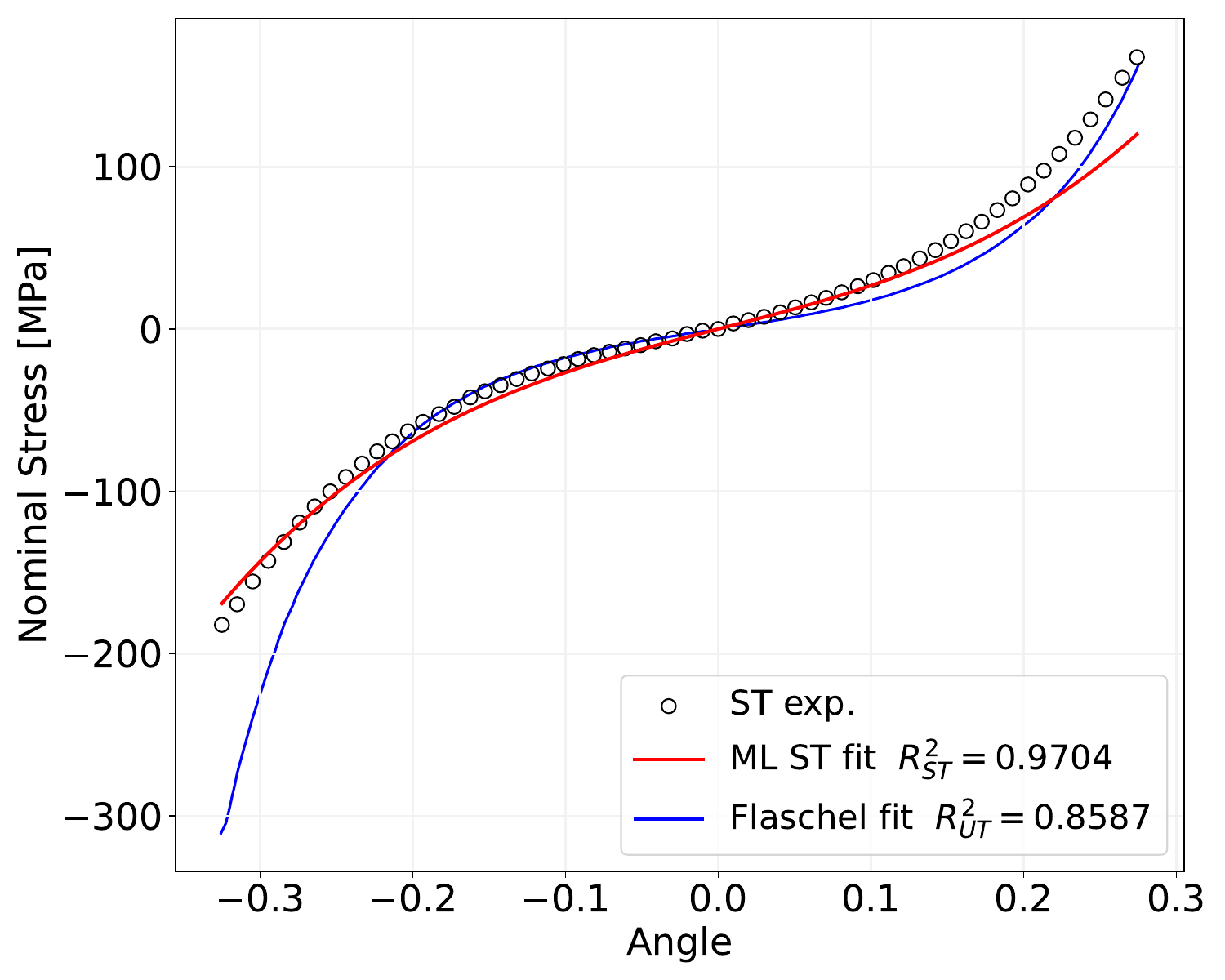}
    \caption{}\label{fig:MBSec1C}
\end{subfigure}
\caption{Midbrain Section 1 response. (a) Training loss and number of active parameters over the training process, (b) Uniaxial tension data, the fit of the proposed approach, and fit of \cref{flaschel2023automatedBrain}, (c) torsion data, the fit of the sparsified neural network approach, compared to the fit of \cref{flaschel2023automatedBrain}.}\label{fig:MBSec1}
\end{figure}

\begin{figure}
\begin{subfigure}[b]{1.0\linewidth}
        \centering
        \includegraphics[scale=0.3]{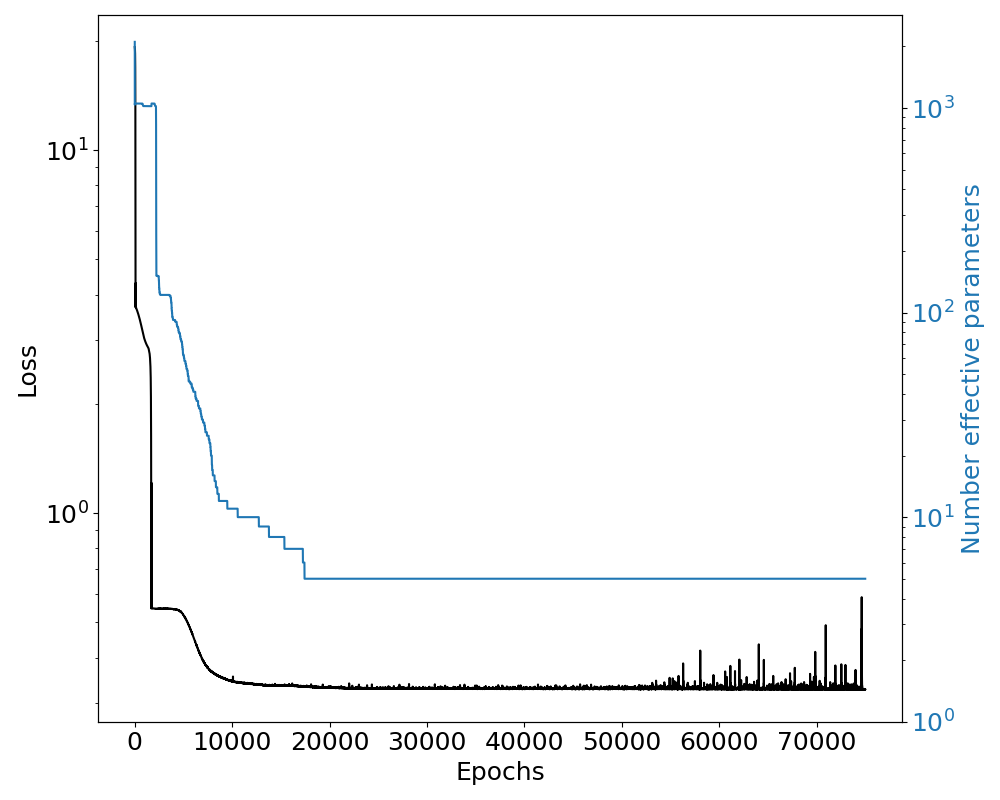}
    \caption{}\label{fig:MBSec3A}
\end{subfigure}
\begin{subfigure}[b]{0.45\linewidth}
        \centering
    \includegraphics[scale=0.3]{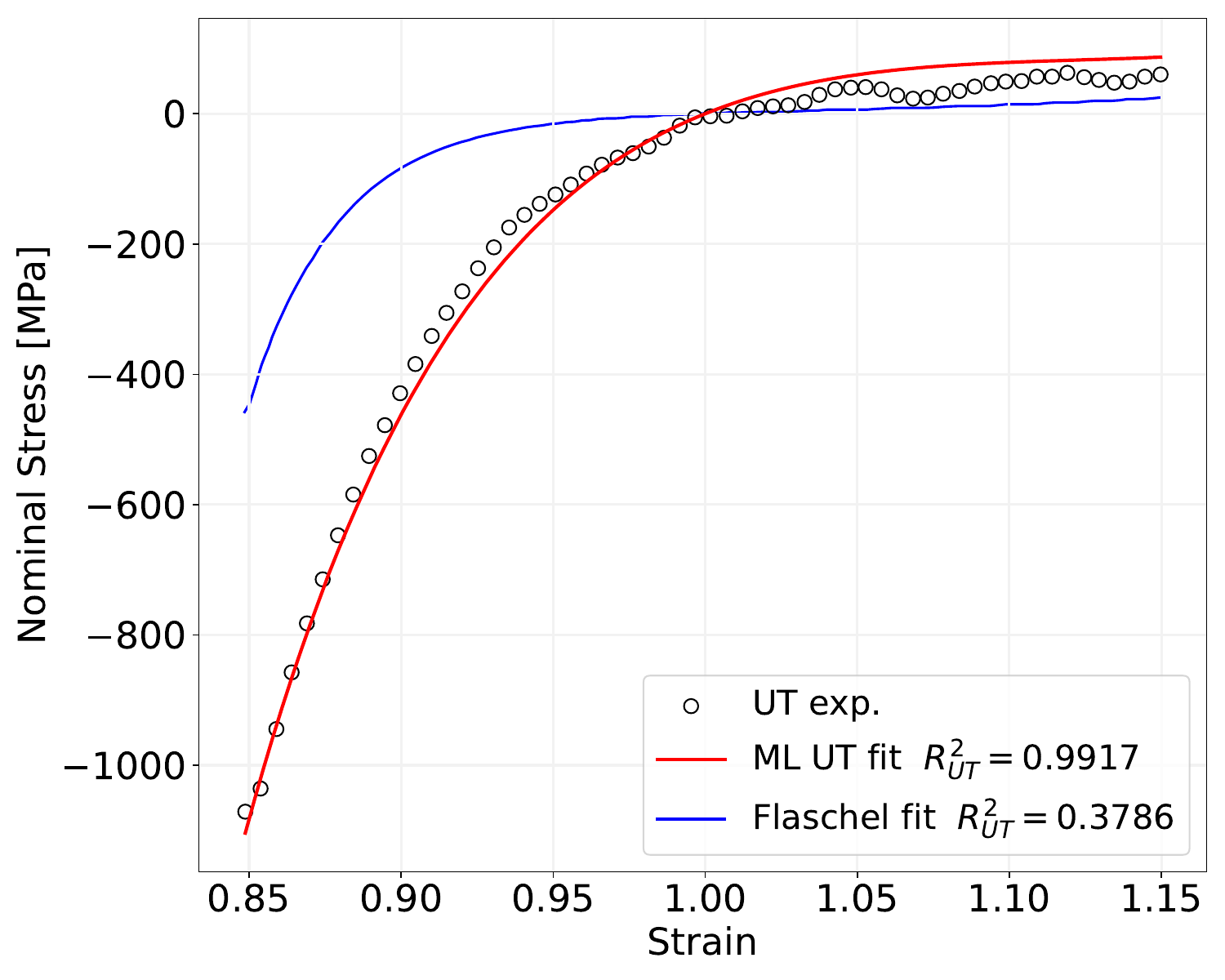}
    \caption{}\label{fig:MBSec3B}
\end{subfigure}
\begin{subfigure}[b]{0.45\linewidth}
        \centering
        \includegraphics[scale=0.3]{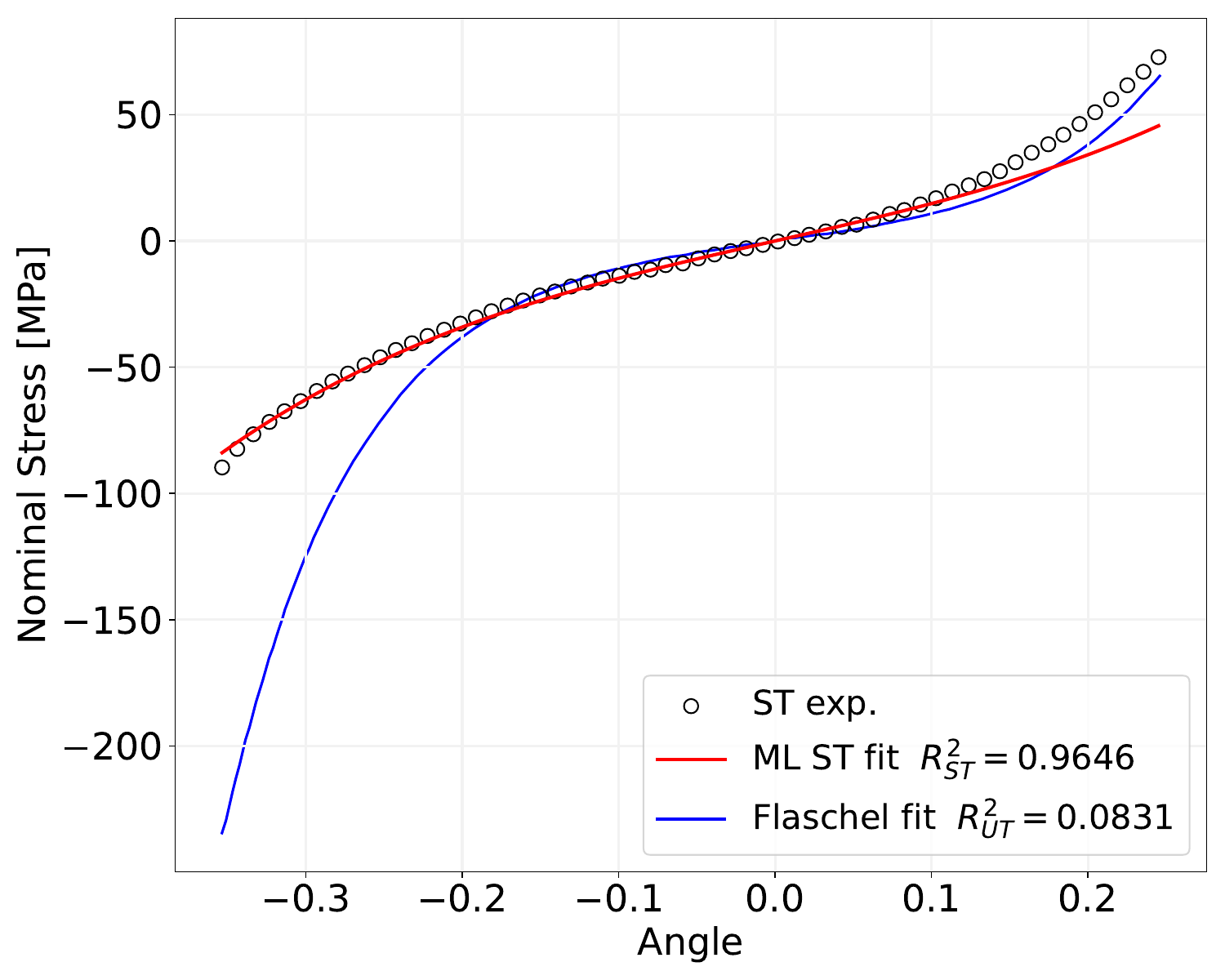}
    \caption{}\label{fig:MBSec3C}
\end{subfigure}
\caption{Midbrain Section 2 response. (a) Training loss and number of active parameters over the training process, (b) Uniaxial tension data, the fit of the proposed approach, and fit of \cref{flaschel2023automatedBrain}, (c) torsion data, the fit of the sparsified neural nework approach, compared to fit of \cref{flaschel2023automatedBrain}.}\label{fig:MBSec3}
\end{figure}

\paragraph{Midbrain}
Next, we fit a model to UT and ST data for two sections of the midbrain. The datsets were adopted from \cref{flaschel2023automatedBrain} and are listed in Table \ref{tab:Midbrain}. For the first section, Figure \ref{fig:MBSec1A} shows the training loss corresponding to the training run with the median final training loss and the respective evolution of the number of active parameters. 
The obtained model form for this section of the midbrain reads
\begin{equation}\label{eq:MBSec1Formula}
    \begin{aligned}
       \Psi_\text{MB1} &= - 47.0604 J - p \left(J - 1.0\right) + 28.8387 \\
       &+ 9.9229 \log{\left(0.0048 \left(1 + 0.0001 e^{- 0.4491 I_{1}}\right)^{194016.0037} \left(0.0001 e^{1.299 I_{2}} + 1\right)^{399.3928} + 1 \right)}.
    \end{aligned}
\end{equation}
We remark that due to the lower number of parameters the loss appears to have single optimization steps where the training loss sharply increases but then immediately reverts. This could potentially be avoided by a lower learning rate, which is however out of the scope of the current paper, and appears to have no effect on the final accuracy of the obtained model. The latter is highlighted in 
Figures \ref{fig:MBSec1B} and \ref{fig:MBSec1C} that show the data and the model response for UT and ST, respectively. We furthermore highlight the $R^{2}$ score which for both cases is $>0.97$ and depict the fit of Flaschel et. al. \cite{flaschel2023automatedBrain} which was obtained using a sparse regression framework called EUCLID \cite{flaschel2021unsupervised}.
\cref{flaschel2023automatedBrain} obtain a strain energy function of the form
\begin{equation}
    \Psi_\text{Fl} = \frac{2 \cdot 0.01}{(-78.58)^{2}} \left( \lambda_{1}^{-78.58} +  \lambda_{2}^{-78.58} +  \lambda_{3}^{-78.58} -3 \right)+ \frac{2 \cdot 90.36}{(-28.71)^{2}} \left( \lambda_{1}^{-28.71} +  \lambda_{2}^{-28.71} +  \lambda_{3}^{-28.71} -3 \right)
\end{equation}
which could potentially be perceived as slightly more interpretable than the strain energy function of eq. \eqref{eq:MBSec1Formula}, which however has a significantly worse accuracy with $R^{2}$-scores below $0.86$.
The loss and parameter evolution of the network trained on the data of the second midbrain section is depicted in 
Figure \ref{fig:MBSec3A}. The final model form reads
\begin{equation}
    \begin{aligned}
       \Psi_\text{MB2} &=  - 57.0237 J - p \left(J - 1.0\right) +  29.8625 \\
       &+ 87.1968 \log{\left(0.0015 \left(1 + 0.0003 e^{- 0.2335 I_{1}}\right)^{24578.9446} \left(0.0003 e^{0.7607 I_{2}} + 1\right)^{437.4412} + 1 \right)}
    \end{aligned}
\end{equation}
which is of similar complexity to the model obtained for the first midbrain section.
Figures \ref{fig:MBSec3B} and \ref{fig:MBSec3C} plot the data points, our fit and the fit of \cref{flaschel2023automatedBrain}. Again the proposed approach shows a proficient accuracy which is better than the one provided by the EUCLID method.

\subsection{Constitutive modeling of elastoplastic material responses}
Elastoplasticity is a framework to model history-dependent nonlinear materials. In the small strain regime, we assume that we can 
split the strain into elastic and plastic strains $\bm{\epsilon}= \bm{\epsilon}^{e} + \bm{\epsilon}^{p}$. We can then postulate the existence of a
free energy function $\Psi(\bm{\epsilon}^{e}, r)= \Psi^{e}(\bm{\epsilon}^{e})+ \Psi^{r}(r)$ that is decomposed into its elastic and part $\Psi^{e}$ and a history-dependent component that models isotropic hardening $\Psi^{r}$ which is dependent on the internal variable $r$. 
To be consistent with thermodynamics, the intrinsic dissipation has to be non-negative \cite{coleman1967thermodynamics}, i.e.
\begin{equation}\label{eq:Dissp}
\begin{aligned}
        \mathcal{D}_{int} &= \bm{\sigma}: \dot{\bm{\epsilon}} - \dot{\Psi} \geq 0  =
        &= \left( \bm{\sigma} - \frac{\partial \Psi^{e}}{\partial \bm{\epsilon}^{e}} \right) :\dot{\bm{\epsilon}}^{e}  + \bm{\sigma}: \dot{\bm{\epsilon}}^{p} - R \dot{r}\geq 0
\end{aligned}
\end{equation}
where $R = \frac{\partial \Psi^{p}}{\partial r}$ is the thermodynamic force conjugate to $r$. In order to guarantee the fulfillment of eq. \eqref{eq:Dissp} we set $\bm{\sigma}=\frac{\partial \Psi^{e}}{\partial \bm{\epsilon}^{e}}$ and introduce the yield function $f(\bm{\sigma}, R)$ which is required to be a non-negative convex function of its arguments and zero-valued at the origin, i.e. $f(\bm{0},0)=0$, see \cref{de2011computational}. This function allows us to derive the evolution equations 
\begin{equation}
    \dot{\bm{\epsilon}}^{p} = \dot{\lambda} \frac{\partial f}{\partial \bm{\sigma}}, \quad \dot{r} =  -\dot{\lambda}\frac{\partial f}{\partial \bm{R}}
\end{equation}
with the consistency parameter $\dot{\lambda}$.
Many materials (such as metals or rocks) are characterized by pressure-independence and by linear elastic behavior in the elastic regime, i.e. where $f(\bm{\sigma},R)<0$. The latter allows us to define the elastic component of the free energy function as 
\begin{equation}
    \Psi^{e}(\bm{\epsilon}^{e}) = \frac{1}{2} \bm{\epsilon}^{e}: \mathbb{C}: \bm{\epsilon}^{e}
\end{equation}
where $\mathbb{C}$ is a fourth-order material tangent specifying the (elastic) anisotropy. 
Under the assumption of isotropic yielding and due to the pressure independence of metals, the yield function can be rewritten as a function of two $\pi$-plane components 
\begin{equation}\label{eq:yieldFinalIso}
f(\frac{1}{R} \pi_{1}, \frac{1}{R} \pi_{2})    
\end{equation}
with 
\begin{equation}
    \begin{bmatrix}
        \pi_{1} \\
        \pi_{2} \\
        \pi_{3}
    \end{bmatrix} = \begin{bmatrix}
        \sqrt{\frac{2}{3}} & - \sqrt{\frac{1}{6}} & - \sqrt{\frac{1}{6}} \\
        0 & \sqrt{\frac{1}{2}} & - \sqrt{\frac{1}{2}} \\
        \sqrt{\frac{1}{3}} & \sqrt{\frac{1}{3}} & \sqrt{\frac{1}{3}}
    \end{bmatrix}
    \begin{bmatrix}
        \sigma_{1} \\
        \sigma_{2} \\
        \sigma_{3}
    \end{bmatrix}
\end{equation}
where $\sigma_{i}$, $i=1,2,3$ are the principal stresses and where $R$ now acts as the ratio of a homothetic transformation of the initial yield function ($R=1$). For $0<R<1$ the yield function is expanding and isotropic hardening can be represented.  In the following, we will use the sparse regression framework to find representations for the initial yield function $f(\pi_{1},\pi_{2})$ and for the hardening function $R(r)$, respectively. 
\subsection{Yield function from numerical data}
We aim to represent an isotropic, pressure-independent initial yield function of the form $f(\pi_{1},\pi_{2})$. Following \cref{lippmann1970matrixungleichungen} we can prove that for thermodynamic consistency the yield function is convex with regards to its inputs $\pi_{1}$ and $\pi_{2}$. We can therefore train a sparse input convex neural network (c.f. Section \ref{sec:InputConvexNN}) and find an interpretable formulation for the yield functions. Since we focus on pressure-independent yield functions we introduce the deviatoric stress $\bm{s}$ as well the two invariants $J_{2}$ and $J_{3}$, i.e.
\begin{equation}
  \bm{s} = \bm{\sigma}- \frac{1}{3} \text{tr}(\bm{\sigma}) \bm{I}, \qquad  J_{2} = \frac{1}{2}\text{tr}(\bm{s}^{2}), \qquad J_{3}  = \frac{1}{3}\text{tr}(\bm{s}^{3}).
\end{equation}
To make them easier to read the following representations are rounded to three decimal places. For all the examples the yield functions are trained on 30 training points, shown in Table \ref{tab:yieldFunData}.

\subsection{Drucker}
\begin{figure}
\begin{subfigure}[b]{0.45\linewidth}
        \centering
        \includegraphics[scale=0.3]{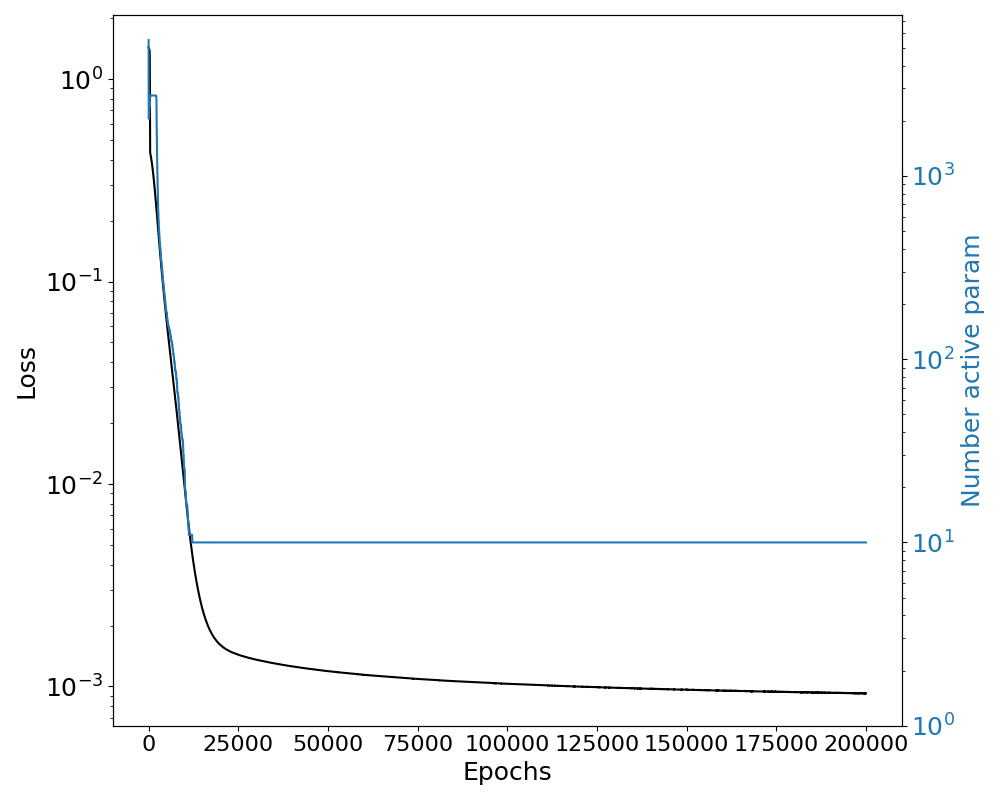}
    \caption{}\label{fig:DruckerFitA}
\end{subfigure}
\begin{subfigure}[b]{0.45\linewidth}
        \centering
    \includegraphics[scale=0.3]{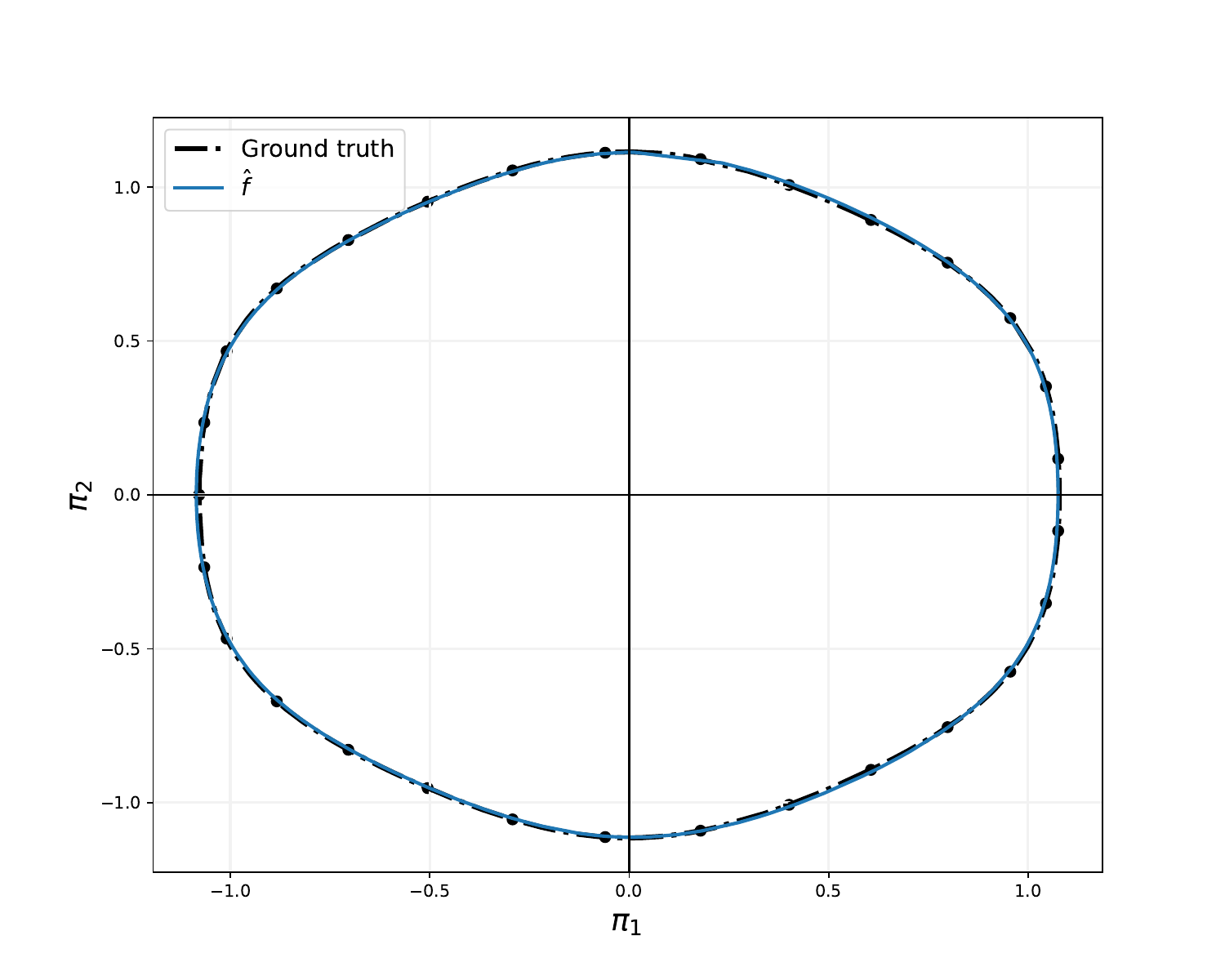}
    \caption{}\label{fig:DruckerFitB}
\end{subfigure}
\caption{Fit of a convex neural network to the Drucker yield function. (a) Training loss behavior and number of active ($\abs{w}>0$) trainable parameters; (b)Data points (black dots), true curve (black dotted), and predicted curve (blue). }\label{fig:DruckerFit}
\end{figure}

The first example focuses on the well-known Drucker yield function \cite{drucker1949relation} which involves both invariants of the Cauchy stress deviator and is here specified as
\begin{equation}
    f_{D} = J_{2}^{3} + 1.5 \, J_{3}^{2} - (0.24)^{6}.
\end{equation}
Figure \ref{fig:DruckerFitA} shows the training loss and the number of active parameters over the training process of the training run with the median final loss. We can see that the final model contains roughly $10$ parameters at the end. The obtained, interpretable functional form reads
\begin{equation}\label{eq:DruckerFormFitted}
    \begin{aligned}
      \hat{f}_{D} &=    0.068 \log{\left(\left(e^{- 2.122 \pi_{1} + 3.648 \pi_{2}} + 1\right)^{5.238} + 1 \right)} + 0.068 \log{\left(\left(e^{- 2.121 \pi_{1} - 3.648 \pi_{2}} + 1\right)^{5.229} + 1 \right)} \\&+ 0.296 \log{\left(e^{4.922 \pi_{1}} + 1 \right)} - 1.699.
    \end{aligned}
\end{equation}
Using this form the approximated yield surface and the raw data are shown in Figure \ref{fig:DruckerFitB}. We can see that the model fits the data proficiently well.
Given the simplicity of the model we can furthermore, for example, use the series expansion around $x=0$ of
\begin{equation}
    \log \left( \frac{(e^{ax}+1)^{y} +1}{(e^{-ax}+1)^{y} +1} \right) \approx -\frac{a 2^{y} y x}{2^{y}+1}
\end{equation}
 to show that
\begin{equation}
\begin{aligned}
       R&= \hat{f}_{D}(0,\pi_{2}) - \hat{f}_{D}(0,-\pi_{2}) \\
       &=0.068 \log{\left( \frac{\left(e^{ 3.648 \pi_{2}} + 1\right)^{5.238} + 1}{\left(e^{ -3.648 \pi_{2}} + 1\right)^{5.238} + 1} \right)} +  0.068 \log{\left( \frac{\left(e^{ -3.648 \pi_{2}} + 1\right)^{5.229} + 1}{\left(e^{ +3.648 \pi_{2}} + 1\right)^{5.229} + 1} \right)} \\
        &\approx -0.068 \frac{2.122 2^{5.238} 5.238 \pi_{2}}{2^{5.238}+1} + 0.068 \frac{2.121 2^{5.229} 5.229 \pi_{2}}{2^{5.229}+1}\\
        &\approx -0.0018 \pi_{2}
\end{aligned}
\end{equation}
the obtained yield function is roughly symmetric along $\hat{f}_{D}(0,\pi_{2})$.

\subsection{Cazacu}
\begin{figure}
\begin{subfigure}[b]{0.45\linewidth}
        \centering
        \includegraphics[scale=0.3]{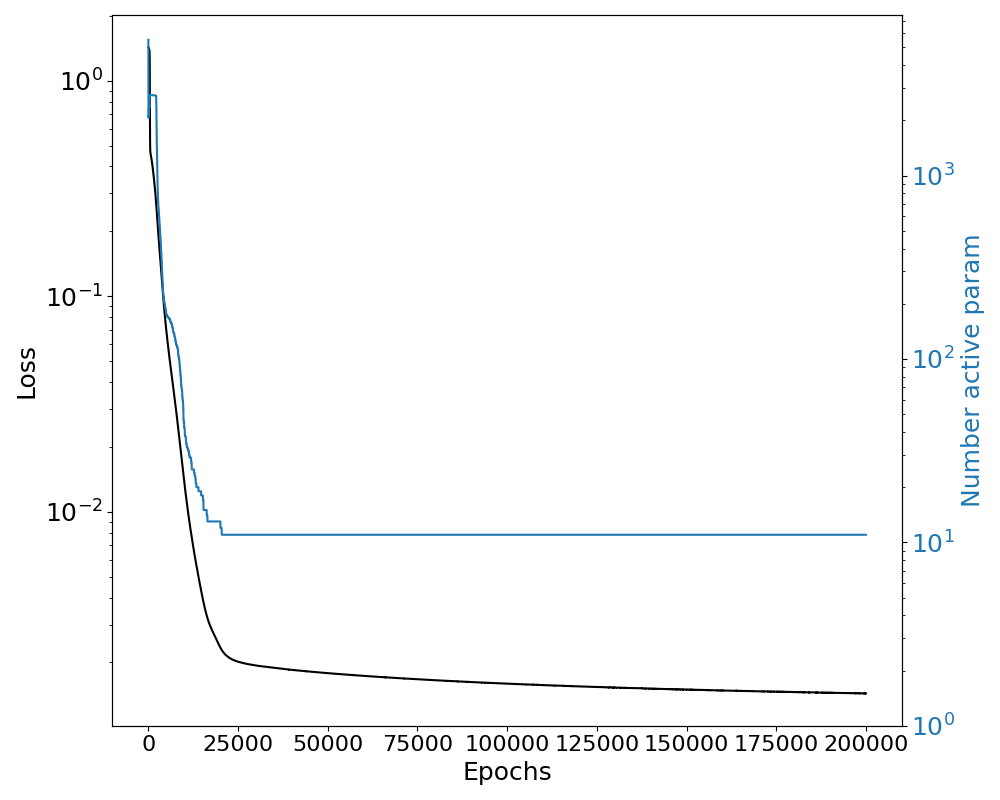}
    \caption{}\label{fig:CacazuFitA}
\end{subfigure}
\begin{subfigure}[b]{0.45\linewidth}
        \centering
    \includegraphics[scale=0.3]{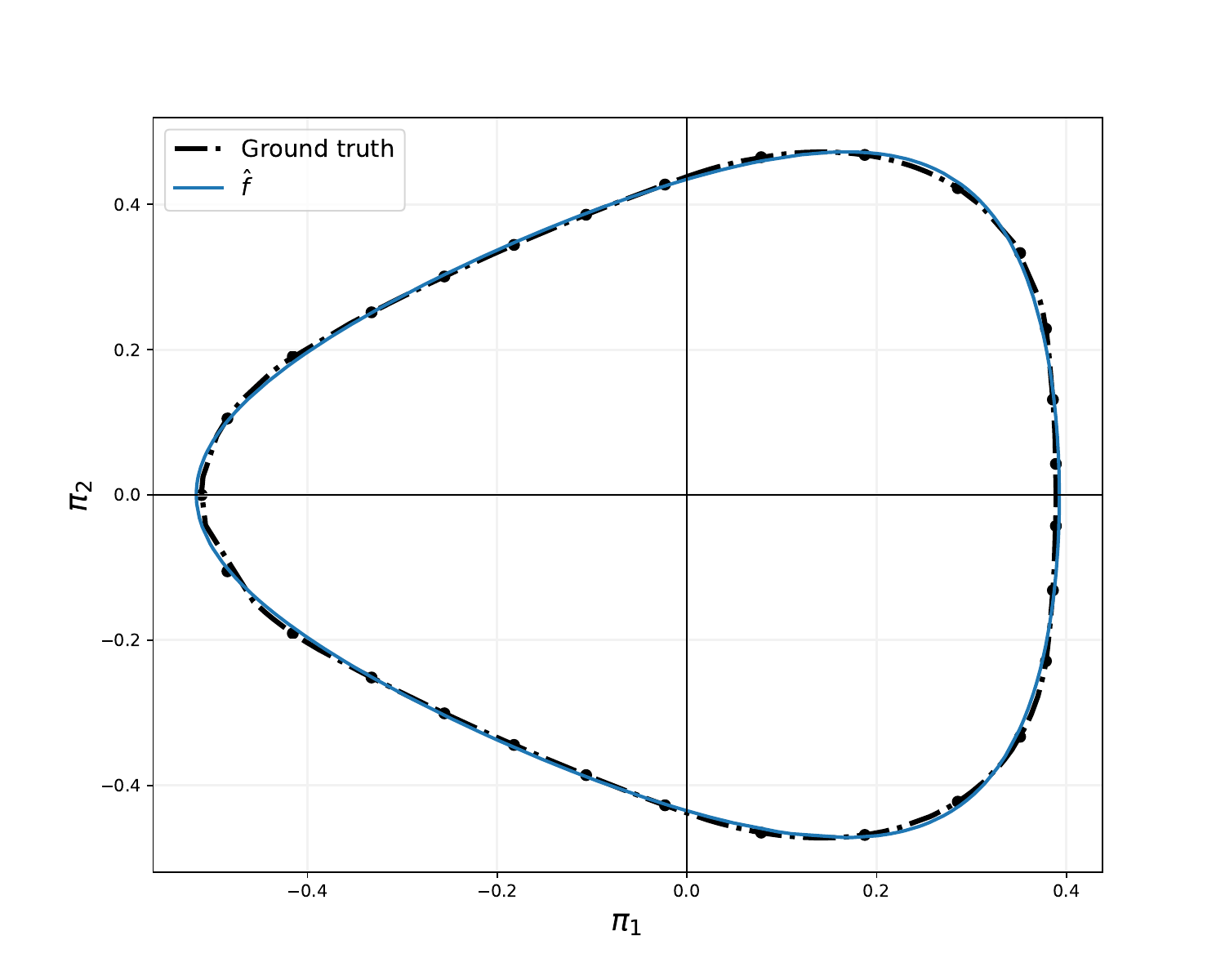}
    \caption{}\label{fig:CacazuFitB}
\end{subfigure}
\caption{Fit of a convex neural network to the Cazacu yield function. (a) Training loss behavior and number of active ($\abs{w}>0$) trainable parameters; (b)Data points (black dots), true curve (black dotted), and predicted curve (blue).}\label{fig:CacazuFit}
\end{figure}

The Drucker yield function is symmetric in the $\pi_{1}$-$\pi_{2}$-plane. To highlight that the model is also able to
obtain an accurate model when tension-compression asymmetries are presented we consider the yield function suggested by Cazacu et. al. \cite{cazacu2006orthotropic} which we specify to be
\begin{equation}
    f_{C} =  \left( \abs{s_{1}} + 0.5 s_{1}\right)^{2}+ \left( \abs{s_{2}} + 0.5 s_{2}\right)^{2} + \left( \abs{s_{3}} + 0.5 s_{3}\right)^{2} -0.24.
\end{equation}
We again obtain 30 data points on the yield limit of this model and fit a $\mathcal{L}^{0}$-regularized input convex neural network.
Figure \ref{fig:CacazuFitA} shows the loss and parameter evolution. The functional form of the final model is given by
\begin{equation}
    \begin{aligned}
        \hat{f}_{C}&=17.327 \log{\left(\left(1 + e^{- 0.364 \pi_{1}}\right)^{0.939} \left(1 + e^{- 0.313 \pi_{2}}\right)^{1.852} + 1 \right)}  \\
        &+1.119 \log{\left(\left(e^{- 7.341 \pi_{1} + 4.18 \pi_{2}} + 1\right)^{0.95} e^{6.953 \pi_{1} + 3.93 \pi_{2}} + 1 \right)} - 38.066.
    \end{aligned}
\end{equation}
Figure \ref{fig:CacazuFitB} shows that the presented approach is also able to fit this asymmetric yield function with reasonable accuracy.

\subsection{Tresca yield function}
\begin{figure}
\begin{subfigure}[b]{0.45\linewidth}
        \centering
        \includegraphics[scale=0.3]{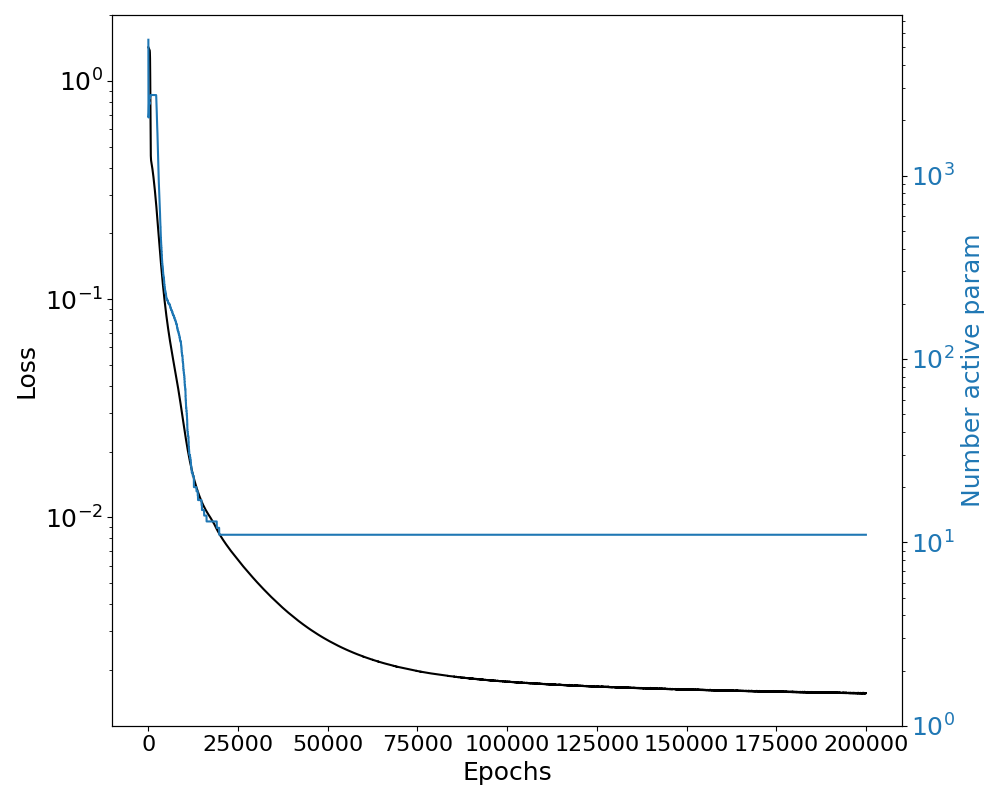}
    \caption{}\label{fig:TrescaFitA}
\end{subfigure}
\begin{subfigure}[b]{0.45\linewidth}
        \centering
    \includegraphics[scale=0.3]{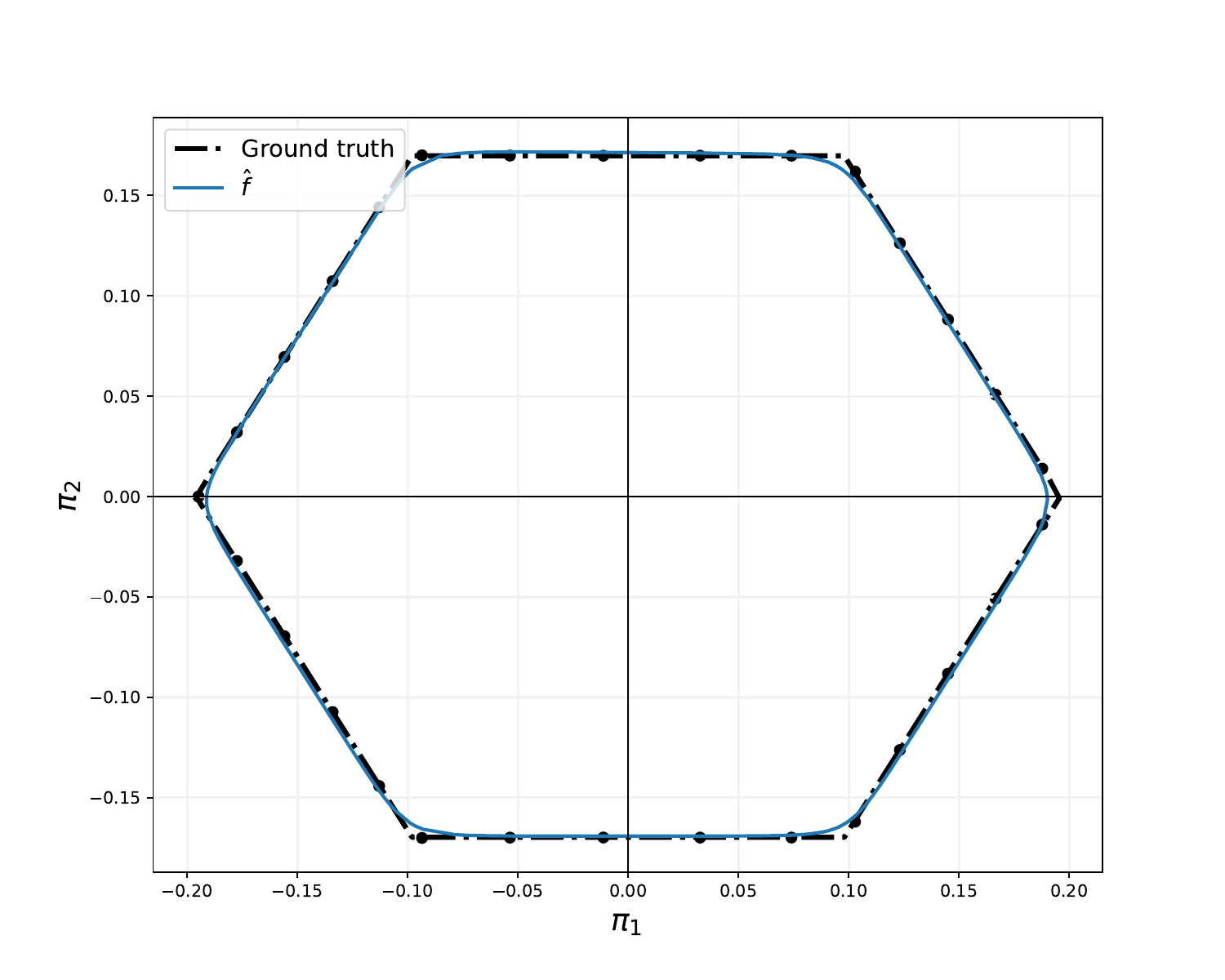}
    \caption{}\label{fig:TrescaFitB}
\end{subfigure}
\caption{Fit of a convex neural network to the Tresca yield function. (a) Training loss behavior and number of active ($\abs{w}>0$) trainable parameters; (b)Data points (black dots), true curve (black dotted), and predicted curve (blue).}\label{fig:TrescaFit}
\end{figure}

In contrast to the other two yield functions, the last example deals with the non-smooth Tresca yield surface as proposed in \cref{tresca1869memoire} which we specify as
\begin{equation}
    f_{T} = \max \left( \abs{\sigma_{1}-\sigma_{2}},
    \abs{\sigma_{1}-\sigma_{3}},
    \abs{\sigma_{3}-\sigma_{2}} \right) - 0.24.
\end{equation}
Figure \ref{fig:TrescaFitA} again shows that both the number of parameters and the loss decrease over the course of the training process.
The final functional representation is given by 
\begin{equation}
    \begin{aligned}
       \hat{f}_{T}&= 0.021 \log{\left(\left(1 + e^{- 116.395 \pi_{2}}\right)^{2.706} + 1 \right)} + 0.019 \log{\left(\left(e^{- 96.41 \pi_{1} + 55.983 \pi_{2}} + 1\right)^{3.117} + 1 \right)} \\ & + 0.023 \log{\left(\left(e^{99.193 \pi_{1}+ 57.03 \pi_{2}} + 1\right)^{2.46} + 1 \right)} - 1.127
    \end{aligned}
\end{equation}
which is reasonably small to allow for some interpretation. Finally, the predicted yield limit $\hat{f}_{T}=0$ is overlaid over the true data in Figure \ref{fig:TrescaFitB}. We can see that the presented approach is also able to find interpretable functional forms when approximating non-smooth yield surfaces. We remark, that due to the smoothness of the activation function, the final yield function is necessarily also smooth. To exactly fit non-smooth yield surfaces other sharper activation functions like the rectified linear unit \cite{schmidt2020nonparametric} could be employed.

\subsection{Isotropic hardening law}
For this application, we only have access to uniaxial monotonic loading tests and we aim to fit the isotropic hardening function $R(r)$ as given in eq. \eqref{eq:yieldFinalIso}. Given this limited data, we assume that the elastic response is isotropic, i.e.
\begin{equation}
    \mathbb{C} = \frac{E \nu}{(1+\nu)(1-2\nu)} \delta_{ij} \delta_{kl} + \frac{E}{2 (1+\nu)} \left( \delta_{ik} \delta_{jl} + \delta_{il} \delta_{jk} \right)
\end{equation}
where the Young's modulus $E$ and the Poisson's ratio $\nu$ are adjustable parameters.
We then aim to find the functional form $R(r)$ depending on the internal hardening variable $r$ which influences the scaling of a von Mises yield function \cite{mises1913mechanik} of the form
\begin{equation}
    f(\frac{1}{R(r)} \, \pi_{1}, \frac{1}{R(r)} \, \pi_{2}) = \frac{3}{2} \sqrt{\frac{1}{R(r)} \, \pi_{1}^{2} + \frac{1}{R(r)} \, \pi_{2}^{2}} - \sigma_{y}
\end{equation}
where the yield stress $\sigma_{y}$ can be directly obtained from the data. If we assume isotropic hardening then the function $R(r)$ is required to be positive and monotonically increasing. We can therefore employ a positive, monotonically increasing neural network as introduced in Section \ref{sec::PosiMonInc} with a sigmoid activation function to approximate the functional form of $R(r)$. We remark that we require $R(r=0.0)=1.0$ when no plastic yielding has occurred. 

We will test the approach out on three different datasets which are summarized in Table \ref{tab:isoHardData}.

\subsubsection{Experimental data - U71Mn rail steel}
\begin{figure}
\begin{subfigure}[b]{0.45\linewidth}
        \centering
    \includegraphics[scale=0.27]{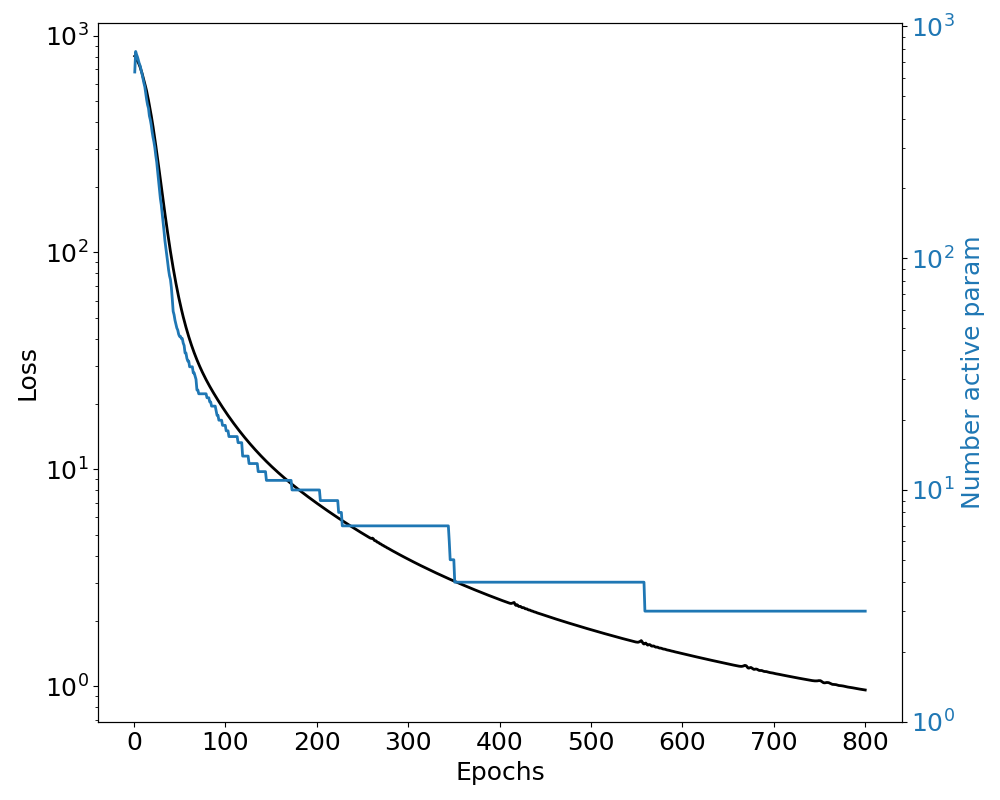}
    \caption{}\label{fig:U71Mna}
\end{subfigure}
\begin{subfigure}[b]{0.45\linewidth}
        \centering
    \includegraphics[scale=0.3]{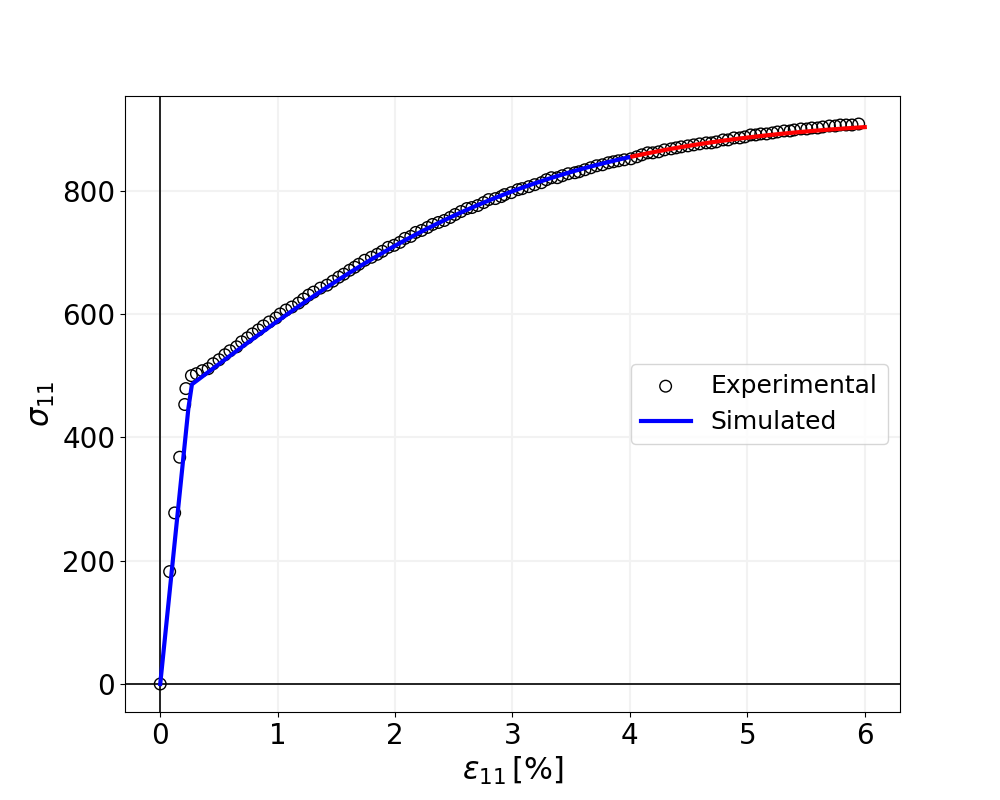}
    \caption{}\label{fig:U71Mnb}
\end{subfigure}
\caption{Fit of experimental monotonic loading curve with the hardening function represented by a sparse neural network  - U71Mn rail steel.  (a) Loss and number of active parameters over epochs (b) Uniaxial tension curve; Black dots represent the experimental data; Blue line indicates prediction in training data; Red line is extrapolation.}\label{fig:U71Mn}
\end{figure}

We start with a monotonic uniaxial loading dataset of U17Mn rail steel discussed in \cref{kang2002uniaxial}. We first fit the elastic parameters and the yield stress to
$E=220\cdot 10^{3} \text{MPa}$, $\nu=0.3$ and  $\sigma_{y}= 484.5 \text{MPa}$ respectively.
Using the network with the median loss after 10 runs, the training evolution and the reduction of the number of active parameters are shown in Figure \ref{fig:U71Mna}. The obtained functional form for the isotropic hardening function reads 
\begin{equation}
    \hat{R}_{U71Mn}(r) = 0.099 + \frac{1.801}{1 + e^{- 194.688 r}}
\end{equation}
which is an easily interpretable function, i.e. we can see that it is always monotonically increasing since $\exp(-r)$ is a monotonically decreasing function in $r$. Figure \ref{fig:U71Mnb} shows the true data, and the predicted response of the monotonic loading process with the fitted $\hat{R}_{U71Mn}(r)$. The blue line indicates the range of the training data, i.e. $R(r)$ was trained until roughly $4\%$ strain. To highlight the extrapolation quality of the model the red line is the model prediction into unseen loading ranges. We can see that both the training and testing data were fitted proficiently well.

\subsubsection{Experimental data - SS316L stainless steel}
\begin{figure}
\begin{subfigure}[b]{0.45\linewidth}
        \centering
    \includegraphics[scale=0.27]{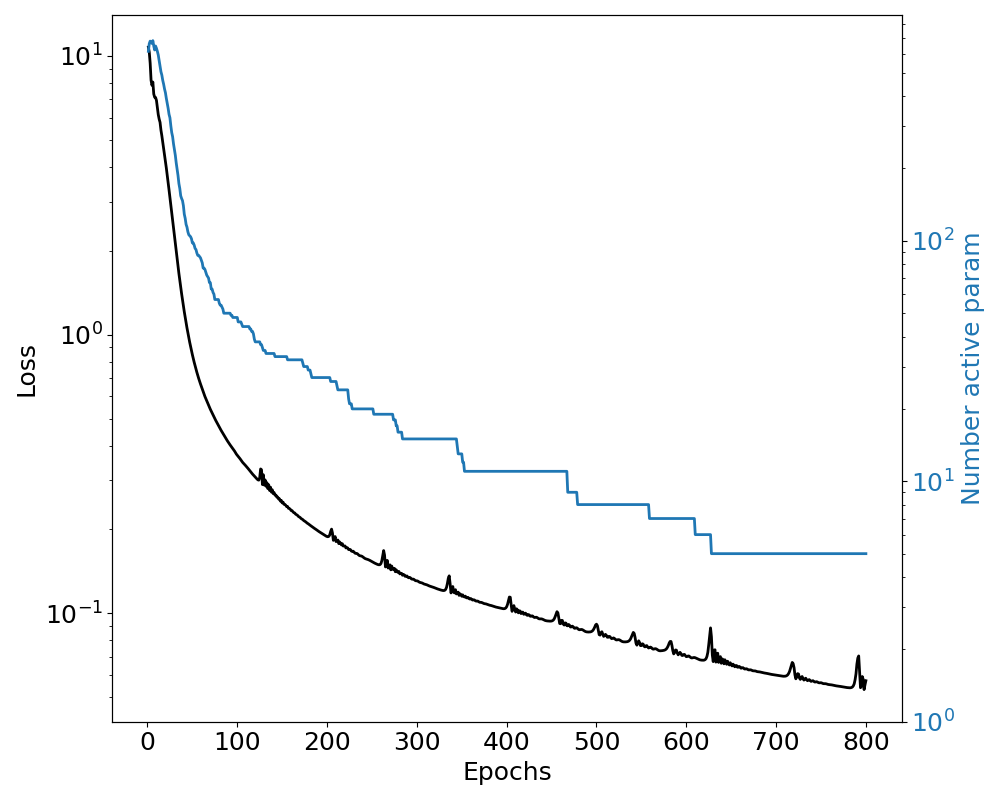}
    \caption{}\label{fig:SS316La}
\end{subfigure}
\begin{subfigure}[b]{0.45\linewidth}
        \centering
        \includegraphics[scale=0.3]{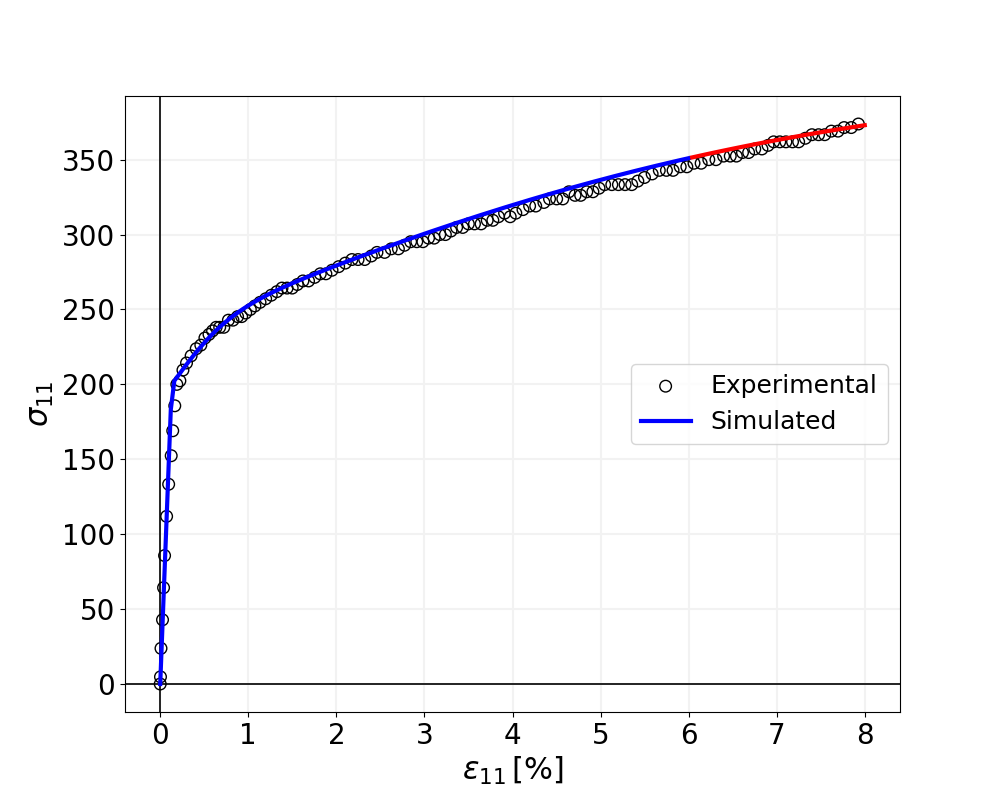}
    \caption{}\label{fig:SS316Lb}
\end{subfigure}
\caption{Fit of experimental monotonic loading curve with the hardening function represented by a sparse neural network   - SS316L stainless steel. (a) Loss and number of active parameters over epochs (b) Uniaxial tension curve; Black dots represent the experimental data; Blue line indicates prediction in training data; Red line is extrapolation.}\label{fig:SS316L}
\end{figure}

Next, we look at data of a uniaxial loading test done on SS316L stainless steel performed by \cref{kang2002uniaxial}. The fitted parameters of the elastic range read
$E=190\cdot 10^{3} \text{MPa}$, $\nu=0.35$ and the yield stress is given by  $\sigma_{y}= 200 \text{MPa}$.
Figure \ref{fig:SS316La} depicts the evolution of the training loss and the parameters while 
Figure \ref{fig:SS316Lb} highlights the ability of the resulting model to fit the training data (up to the end of the blue solid line) and the ability to generalize well beyond the training region (solid red line).
The interpretable functional form of the isotropic hardening function reads
\begin{equation}
\hat{R}_{SS316L}(r) = 0.023 + \frac{1.662}{1 + 1.071 e^{- 190.683 r}} + \frac{0.362}{1 + 1.071 e^{- 2200.640 r}}
\end{equation}
which, for example, allows us to easily see that there is no hardening when the internal variable is zero, i.e. $\hat{R}_{SS316L}(0)\approx0.023+0.802+0.175 \approx 1$.

\subsubsection{Experimental data - 40Cr3MoV bainitic steel}
\begin{figure}
\begin{subfigure}[b]{0.45\linewidth}
        \centering
    \includegraphics[scale=0.27]{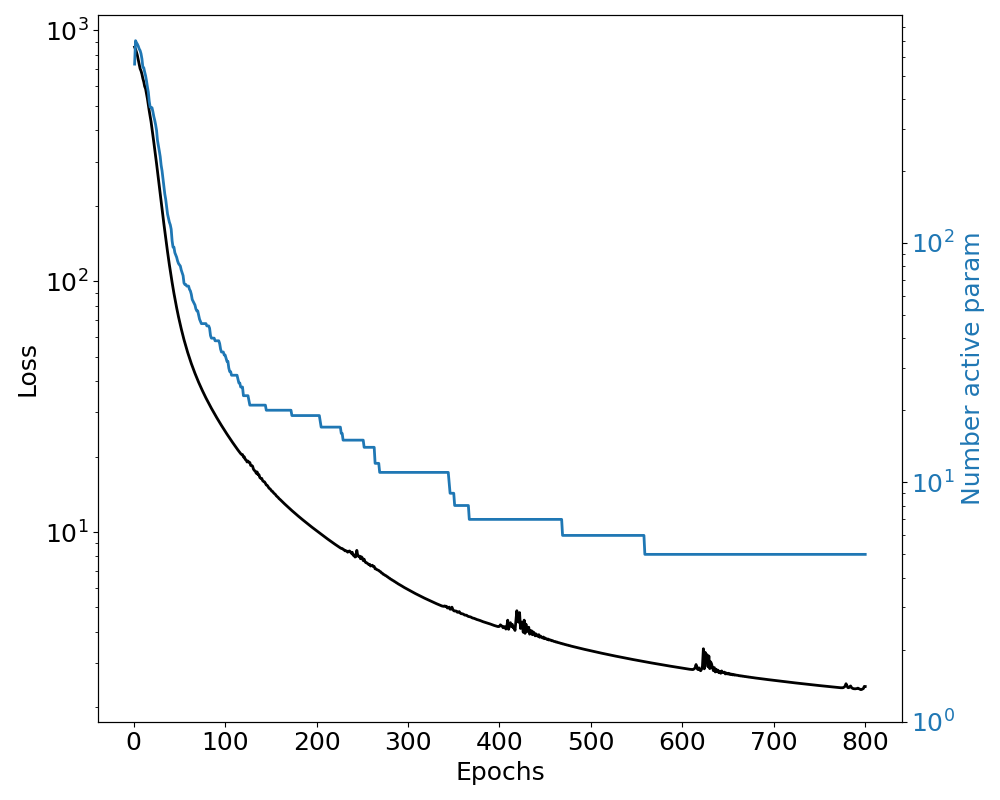}
    \caption{}\label{fig:40Cr3MoVa}
\end{subfigure}
\begin{subfigure}[b]{0.45\linewidth}
        \centering
        \includegraphics[scale=0.3]{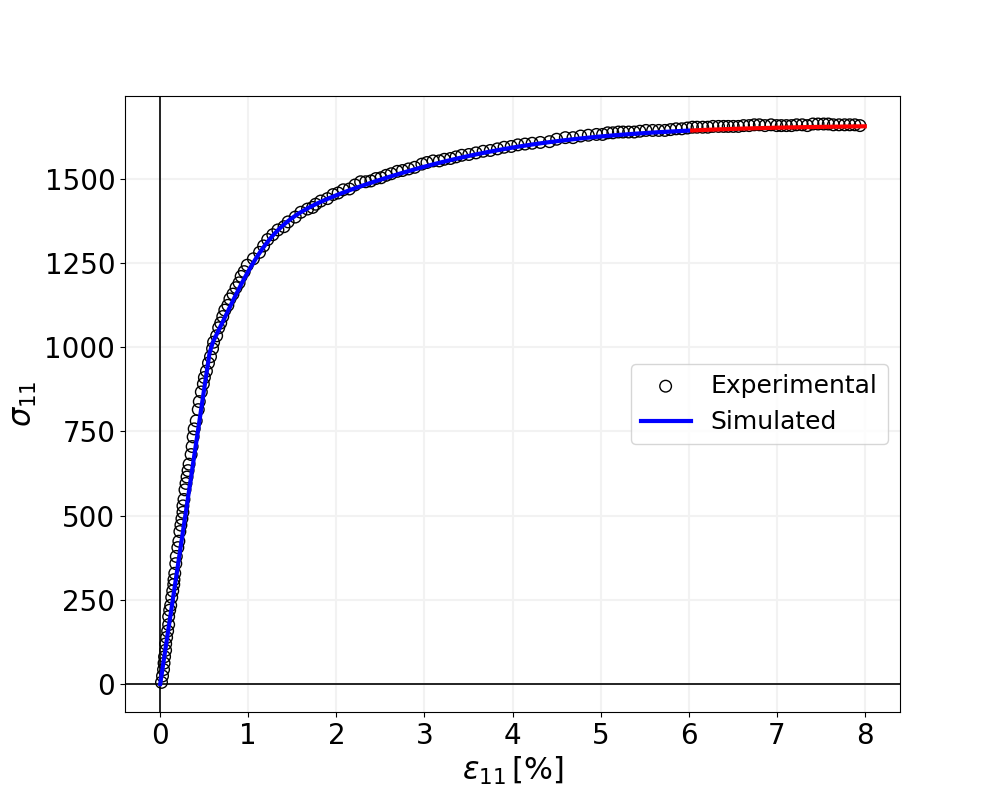}
    \caption{}\label{fig:40Cr3MoVb}
\end{subfigure}
\caption{Fit of experimental monotonic loading curve with the hardening function represented by a sparse neural network   -  40Cr3MoV bainitic steel. (a) Loss and number of active parameters over epochs (b) Uniaxial tension curve; Black dots represent the experimental data; Blue line indicates prediction in training data; Red line is extrapolation. }\label{fig:40Cr3MoV}
\end{figure}

Lastly, we look at monotonic loading data of 40Cr3MoV bainitic steel which was also published \cref{kang2002uniaxial}. We set the material parameters as
$E=207\cdot 10^{3} \text{MPa}$, $\nu=0.3$ and  $\sigma_{y}= 1000 \text{MPa}$.
The loss and parameter evolution are shown in 
Figure \ref{fig:40Cr3MoVa}.
Figure \ref{fig:40Cr3MoVb} depicts the interpolated and extrapolated predictions using the fitted model which is of the form
\begin{equation}
  \hat{R}_{40Cr3MoV}(r)=  -0.501 + \frac{2.904}{1.669 e^{- \frac{0.839}{1 + 1.720 e^{- 102.643 r}} - \frac{0.740}{1 + 1.720 e^{- 667.227 r}}} + 1}.
\end{equation}
We can see that similar to the first two cases the accuracy of the fit is proficiently high.

\begin{equation}
\end{equation}

\section{Discussion and Conclusion}\label{sec::4}
We have proposed to prune physics-augmented neural network-based constitutive models using a smoothed version of $L^{0}$-regularization to enable interpretable and trustworthy model discovery in a wide array of problems in mechanics. The network is trained by simultaneously fitting the training data and penalizing the number of active parameters. On a variety of applications including synthetic and experimental data, we have shown that we are able to obtain accurate, yet interpretable constitutive models for compressible and incompressible hyperelasticity, yield functions, and isotropic hardening functions. The presented approach seems to be highly flexible and has the potential to overcome the restrictions of functional form selection, towards automation of constitutive modeling. Uniquely, i) enforcing physical constraints enables generalization/extrapolation as well as training with limited and low data, ii) utilization of neural networks enables high expressiveness and eliminates the development of specific model form libraries, and iii) pruning, leads to interpretable discovery and also enhances generalization/extrapolation.

In the next steps, we will use this approach to obtain functional forms for other constitutive models such as representations for kinematic hardening or viscoelasticity. We have furthermore (purposefully) only used neural networks with single nonlinear activation functions. Future work could center around finding representations using a different activation function at each neuron as a means to enhance the expressivity of our representations to parsimoniously model even more complex material responses. 
\section*{Acknowledgments}

JF and NB gratefully acknowledge support by the Air Force Office of Scientific Research under award number FA9550-22-1-0075.

Sandia National Laboratories is a multimission laboratory managed and operated by National Technology and Engineering Solutions of Sandia, LLC., a wholly owned subsidiary of Honeywell International, Inc., for the U.S. Department of Energy's National Nuclear Security Administration under contract DE-NA-0003525. This paper describes objective technical results and analysis. Any subjective views or opinions that might be expressed in the paper do not necessarily represent the views of the U.S. Department of Energy or the United States Government. 

\bibliography{bib.bib}

\appendix
\renewcommand{\theequation}{\thesection.\arabic{equation}}
\numberwithin{equation}{section}
\section{Hyperelasticity formulation} \label{app:hyperelasticity} 
\setcounter{equation}{0}
Given the deformation gradient $\bm{F}$ and the right Cauchy-Green tensor $\bm{C} = \bm{F}^{T} \bm{F}$, let the isotropic invariants be defined by 
\begin{equation}
    I_{1} = \text{tr} \, \bm{C}, \quad I_{2} = \text{tr} \, (\text{cof} \bm{C}),  \quad I_{3} = \det \bm{C} 
\end{equation}
and denote $J=\sqrt{I_3}$. Let the strain energy prediction be a function of these invariants $\hat{\Psi}(I_{1},I_{2}, J)$.

\subsection{Compressible hyperelasticity}\label{app:Comhyperelasticity}
Under consideration of the derivatives
\begin{equation}\label{eq:DerivativeInvApp}
\begin{aligned}
         \frac{\partial I_{1}}{\partial \bm{C}} = \bm{I}, \quad
          \frac{\partial I_{2}}{\partial \bm{C}} = I_{1} \bm{I} - \bm{C}, \quad 
           \frac{\partial I_{3}}{\partial \bm{C}}=  J^{2}\bm{C}^{-1}, \quad     \frac{\partial J}{\partial I_{3}}  = \frac{1}{2 J}
    \end{aligned}
\end{equation}
the second Piola-Kirchhoff stress is
\begin{equation}\label{eq:CompStressApp}
\begin{aligned}
       \hat{ \bm{S}} &= 2 \frac{\partial \hat{\Psi}}{\partial \bm{C}} = 2 \left( \sum_{i} \frac{\partial \hat{\Psi}}{\partial I_{i}} \frac{\partial I_{i}}{\partial \bm{C}} \right) = 2 \left(  \frac{\partial \hat{\Psi}}{\partial I_{1}} \frac{\partial I_{1}}{\partial \bm{C}} +  \frac{\partial \Psi}{\partial I_{2}} \frac{\partial I_{2}}{\partial \bm{C}} +  \frac{\partial\hat{\Psi}}{\partial J} \frac{\partial J}{\partial I_{3}} \frac{\partial I_{3}}{\partial \bm{C}} \right)\\
    &= 2 \frac{\partial \hat{\Psi}}{\partial I_{1}} \bm{I} + 2 \frac{\partial \hat{\Psi}}{\partial I_{2}} (I_{1} \bm{I} - \bm{C}) + 2  \frac{\partial \hat{\Psi}}{\partial J} \frac{1}{2 J} J^{2}\bm{C}^{-1}.
\end{aligned}
\end{equation}
Then, at the undeformed configuration $\bm{C}=\bm{I}$, we see that
\begin{equation}
    \begin{aligned}
         \left. \frac{\partial I_{1}}{\partial \bm{C}} \right\rvert_{\bm{C}=\bm{I}}= \bm{I}, \quad 
          \left. \frac{\partial I_{2}}{\partial \bm{C}} \right\rvert_{\bm{C}=\bm{I}} = 3 \bm{I} - \bm{I} = 2 \bm{I}, \quad  
          \left. \frac{\partial I_{3}}{\partial \bm{C}} \right\rvert_{\bm{C}=\bm{I}} = \det (\bm{I}) \bm{\bm{I}}^{-1} = \bm{I}.
    \end{aligned}
\end{equation}
So we define  
\begin{equation}
    \left. \hat{ \bm{S}} (\bm{C})\right\rvert_{\bm{C}=\bm{I}}  =  2 \left(  \frac{\partial \hat{\Psi}}{\partial I_{1}}  \bm{I} +  \frac{\partial \hat{\Psi}}{\partial I_{2}} 2 \bm{I} +  \frac{\partial \hat{\Psi}}{\partial J} \frac{1}{2} \bm{I} \right) = \underbrace{2 \left. \left(  \frac{\partial \hat{\Psi}}{\partial I_{1}}  +  \frac{\partial \hat{\Psi}}{\partial I_{2}} 2  +  \frac{\partial \hat{\Psi}}{\partial J} \frac{1}{2} \right)\right\rvert_{\bm{C}=\bm{I}}}_{=n} \bm{I} = n\bm{I}
\end{equation}
and set the strain energy prediction as 
\begin{equation}
    \hat{\Psi}(I_{1}, I_{2}, J) = \hat{\Psi}^{NN}(I_{1}, I_{2}, J) - \hat{\Psi}^{NN}(3, 3, 1) - \psi^{S}(J),
\end{equation}
where $\psi^{NN}(\bullet)$ is the output of an input convex neural network and the correction is chosen to be $\psi^{S}= n (J-1)$.
We can see that both the strain energy function prediction
\begin{equation}
   \left.  \hat{\psi}(I_{1},I_{2}, J)\right\rvert_{\bm{C}=\bm{I}} = 0
\end{equation}
and the stress
\begin{equation}
    \left. \bm{S}(\bm{C})\right\rvert_{\bm{C}=\bm{I}} =  2 \left(  \frac{\partial \hat{\Psi}^{NN}}{\partial I_{1}}  \bm{I} +  \frac{\partial \hat{\Psi}^{NN}}{\partial I_{2}} 2 \bm{I} +  \frac{\partial \hat{\Psi}^{NN}}{\partial J} \frac{1}{2} \bm{I} \right) - n \bm{I} = \bm{0}
\end{equation}
fulfill the normalization condition.

\subsection{Incompressible hyperelasticity}\label{app:InComhyperelasticity}
For incompressible materials, we have the constraint $J=1$. Using the classical Lagrange multiplier formulation \cite{truesdell2004non,ogden1997non,holzapfel2002nonlinear}, we augment the prediction of the strain energy as
\begin{equation}\label{eq:IncompStrainAppendix}
\begin{aligned}
    \hat{\psi}(I_{1}, I_{2}) &= \hat{\Psi}^{NN}(I_{1}, I_{2}) - p (J-1) - \hat{\Psi}^{NN}(3,3) - \psi^{S}(J) \\
   &= \hat{\Psi}^{NN}(I_{1}, I_{2}) - p (J-1) - \hat{\Psi}^{NN}(3,3) - n (J-1) \\
   &=\hat{\Psi}^{NN}(I_{1}, I_{2}) - (p+n) (J-1) - \hat{\Psi}^{NN}(3,3) ,
\end{aligned}
\end{equation}
where $\hat{\Psi}^{NN}(I_{1},I_{2})$ is the output of an input convex neural network with inputs $I_{1}$ and $I_{2}$, $p$ is the hydrostatic pressure acting as the Lagrange multiplier enforcing $J=1$, and the correction $n$ is
\begin{equation}
    n = 2   \left. \left(\frac{\partial \hat{\Psi}^{NN} }{\partial I_{1}} + 2 \frac{\partial \hat{\Psi}^{NN} }{\partial I_{2}} \right)\right\rvert_{\bm{C}=\bm{I}}.
\end{equation}
We can see from eq. \eqref{eq:IncompStrainAppendix} that the normalization condition of the strain energy is fulfilled, i.e.
\begin{equation}
      \left. \hat{\Psi}\right\rvert_{\bm{C}=\bm{I}} = 0.
\end{equation}
Furthermore, by using the derivative of eq. \eqref{eq:DerivativeInvApp} and the definition of the stress of eq. \eqref{eq:CompStressApp},  the stress prediction yields
\begin{equation}
\begin{aligned}
       \hat{ \bm{S}}&=  2 \left( \frac{\partial \hat{\Psi}^{NN} }{\partial I_{1}} \frac{\partial I_{1}}{\partial \bm{C}} + \frac{\partial \hat{\Psi}^{NN} }{\partial I_{2}} \frac{\partial I_{2}}{\partial \bm{C}} \right) - 2 \frac{\partial ((p+n) (J-1))}{\partial \bm{C}} \\
       &= 2 \left( \frac{\partial \hat{\Psi}^{NN} }{\partial I_{1}}  \bm{I} + \frac{\partial \hat{\Psi}^{NN} }{\partial I_{2}}  (I_{1} \bm{I}- \bm{C}) \right) - 2(p+n) \frac{\partial J}{\partial I_{3}} \frac{\partial I_{3}}{\partial \bm{C}} \\
&=  2 \left( \left[ \frac{\partial \hat{\Psi}^{NN} }{\partial I_{1}} + I_{1} \frac{\partial \hat{\Psi}^{NN} }{\partial I_{2}} \right]  \bm{I} - \frac{\partial \hat{\Psi}^{NN} }{\partial I_{2}}   \bm{C} \right) - (p+n) J \bm{C}^{-1} .
\end{aligned}
\end{equation}
At the undeformed configuration, we then get
\begin{equation}
\begin{aligned}
       \left. \bm{S}(\bm{C})\right\rvert_{\bm{C}=\bm{I}}&=  2  \left. \left( \left[ \frac{\partial \hat{\Psi}^{NN} }{\partial I_{1}} + 3 \frac{\partial \hat{\Psi}^{NN} }{\partial I_{2}} \right]  \bm{I} - \frac{\partial \hat{\Psi}^{NN} }{\partial I_{2}}   \bm{I} \right)\right\rvert_{\bm{C}=\bm{I}} - (p+n) \bm{I} \\
        &= 2  \left. \left(  \left[ \frac{\partial \hat{\Psi}^{NN} }{\partial I_{1}}+ 2\frac{\partial \hat{\Psi}^{NN} }{\partial I_{2}} \right]\right)\right\rvert_{\bm{C}=\bm{I}} - (p+n) \bm{I} \\
       &= 2  \left. \left(  \left[ \frac{\partial \hat{\Psi}^{NN} }{\partial I_{1}}+ 2\frac{\partial \hat{\Psi}^{NN} }{\partial I_{2}} \right]\right)\right\rvert_{\bm{C}=\bm{I}} - p \bm{I} -  2   \left. \left(\frac{\partial \hat{\Psi}^{NN} }{\partial I_{1}} + 2 \frac{\partial \hat{\Psi}^{NN} }{\partial I_{2}}  \right)\right\rvert_{\bm{C}=\bm{I}} \bm{I}  \\
       &= -p \bm{I}
\end{aligned}
\end{equation}
For an incompressible material, the pressure is not determined by the deformation, hence $\bm{S}=\bm{0}$ at $\bm{C}=\bm{I}$ requires that the applied tractions are consistent with $p=0$.
The first Piola-Kirchoff stress can be derived as
\begin{equation}\label{eq:PForIncApp}
\begin{aligned}
        \bm{P} = 2 \bm{F} \frac{\partial \hat{\Psi}}{\partial \bm{C}} 
        = 2 \left( \left[ \frac{\partial \hat{\Psi}^{NN} }{\partial I_{1}} + I_{1} \frac{\partial \hat{\Psi}^{NN} }{\partial I_{2}} \right]  \bm{F} - \frac{\partial \hat{\Psi}^{NN} }{\partial I_{2}}   \bm{F} \bm{C} \right) - (p+n) J \bm{F}^{-T}.
\end{aligned}
\end{equation}
\subsubsection{Deformation modes}
It is easier to study different modes of deformation, in principal space. Hence define the three principal strains $\lambda_{1} \leq \lambda_{2} \leq \lambda_{3}$. In terms of these quantities, the invariants read
\begin{equation}
    I_{1} = \lambda_{1}^{2}+\lambda_{2}^{2}+\lambda_{3}^{2}, \quad I_{2}= \lambda_{1}^{2}\lambda_{2}^{2}+\lambda_{1}^{2}\lambda_{3}^{2}+\lambda_{2}^{2}\lambda_{3}^{2}, \quad I_{3}=\lambda_{1}^{2}\lambda_{2}^{2}\lambda_{3}^{2} \equiv 1.
\end{equation}
From eq. \eqref{eq:PForIncApp} we see that
\begin{equation}
\begin{aligned}
        P_{i} &= 2 \left(\left[ \frac{\partial \hat{\Psi}^{NN} }{\partial I_{1}} + (\lambda_{1}^{2}+\lambda_{2}^{2}+\lambda_{3}^{2}))\frac{\partial \hat{\Psi}^{NN} }{\partial I_{2}} \right] \lambda_{i} - \frac{\partial \hat{\Psi}^{NN} }{\partial I_{2}} \lambda_{i}^{3} \right) - (p+n) \frac{1}{\lambda_{i}},
\end{aligned}
\end{equation}
and hence,
\begin{equation}\label{eq:DefoModes1PKApp}
    \begin{aligned}
        P_{i} \lambda_{i}
     &=   2 \left( \lambda_{i}^{2} \frac{\partial \hat{\Psi}^{NN} }{\partial I_{1}} +  \frac{\partial \hat{\Psi}^{NN} }{\partial I_{2}}  \left[ (\lambda_{1}^{2}+\lambda_{2}^{2}+\lambda_{3}^{2})\lambda_{i}^{2}-  \lambda_{i}^{4}\right] \right) - p- n.
    \end{aligned}
\end{equation}

Next, we look at the resulting responses for different forms of deformation.

\subsubsection{Uniaxial tension}
Given that $J=\lambda_{1}\lambda_{2}\lambda_{3}=1$, uniaxial tension is defined by 
\begin{equation}
    \bm{F} = \text{diag} \lbrace \lambda_{1}, \frac{1}{\sqrt{\lambda_{1}}}, \frac{1}{\sqrt{\lambda_{1}}} \rbrace, \quad \text{and} \quad \bm{P} = \text{diag} \lbrace P_{1}, 0,0\rbrace
\end{equation}
which means
\begin{equation}
    I_{1} = \lambda^{2} + \frac{2}{\lambda_{1}}, \quad I_{2} = 2 \lambda_{1} + \frac{1}{\lambda_{1}^{2}} .
\end{equation}
We can obtain an expression for $p$ from the assumed stress state
\begin{equation}
    \begin{aligned}
      P_{3} \lambda_{3} = 0 &=  2 \left( \frac{1}{\lambda_{1}} \frac{\partial \hat{\Psi}^{NN} }{\partial I_{1}} +  \frac{\partial \hat{\Psi}^{NN} }{\partial I_{2}}  \left[ (\lambda_{1}^{2}+\frac{1}{\lambda_{1}}+\frac{1}{\lambda_{1}})\frac{1}{\lambda_{1}}-  \frac{1}{\lambda_{1}^{2}} \right] \right) - p- n \\
 \leftrightarrow     p&=  2 \left( \frac{1}{\lambda_{1}} \frac{\partial \hat{\Psi}^{NN} }{\partial I_{1}} +  \frac{\partial \hat{\Psi}^{NN} }{\partial I_{2}}  \left[ \lambda_{1}+  \frac{1}{\lambda_{1}^{2}} \right] \right) - n
    \end{aligned}
\end{equation}
which is consistent with lateral traction-free surfaces.
Hence, this determines the tension component
\begin{equation}
    \begin{aligned}
        P_{1} &= \frac{2}{\lambda_{1}} \left( \lambda_{1}^{2} \frac{\partial \hat{\Psi}^{NN} }{\partial I_{1}} +  \frac{\partial \hat{\Psi}^{NN} }{\partial I_{2}}  \left[ (\lambda_{1}^{2}+\frac{1}{\lambda_{1}}+\frac{1}{\lambda_{1}}) \lambda_{1}^{2}-  \lambda_{1}^{4}\right] \right) - p \frac{1}{\lambda_{1}}- n \frac{1}{\lambda_{1}}\\
&= 2 \left( \frac{\partial \hat{\Psi}^{NN} }{\partial I_{1}} + \frac{1}{\lambda_{1}} \frac{\partial \hat{\Psi}^{NN} }{\partial I_{2}} \right) \left[ \lambda_{1} - \frac{1}{\lambda_{1}^{2}}\right]
    \end{aligned}
\end{equation}

\subsubsection{Equibiaxial tension}
For equibiaxial tension, the deformation gradient and the first Piola-Kirchhoff stress read
\begin{equation}
    \bm{F} = \text{diag} \lbrace \lambda_{1}, \lambda_{1}, \frac{1}{\lambda_{1}^{2}} \rbrace, \quad \text{and} \quad \bm{P} = \text{diag} \lbrace P_{1}, P_{2},0\rbrace.
\end{equation}
Again we get an expression for the pressure from the fact that $P_{3}=0$, which is consistent with traction-free lateral surfaces, i.e.
\begin{equation}
\begin{aligned}
        P_{3} = 0 &=   2 \left(  \frac{1}{\lambda_{1}^{4}} \frac{\partial \hat{\Psi}^{NN} }{\partial I_{1}} +  \frac{\partial \hat{\Psi}^{NN} }{\partial I_{2}}  \left[ (\lambda_{1}^{2}+\lambda_{1}^{2}+\frac{1}{\lambda_{1}^{4}})\frac{1}{\lambda_{1}^{4}}-  \frac{1}{\lambda_{1}^{8}}\right] \right) - p- n \\
 \leftrightarrow       p &= 2 \left(  \frac{1}{\lambda_{1}^{4}} \frac{\partial \hat{\Psi}^{NN} }{\partial I_{1}} +  \frac{\partial \hat{\Psi}^{NN} }{\partial I_{2}}  \frac{2}{\lambda_{1}^{2}}\right) - n.
\end{aligned}
\end{equation}
Hence, we see that
\begin{equation}
    \begin{aligned}
        P_{1} = P_{2}&= \frac{2}{\lambda_{1}} \left( \lambda_{1}^{2} \frac{\partial \hat{\Psi}^{NN} }{\partial I_{1}} +  \frac{\partial \hat{\Psi}^{NN} }{\partial I_{2}}  \left[ (\lambda_{1}^{2}+\lambda_{1}^{2}+\frac{1}{\lambda_{1}^{4}}) \lambda_{1}^{2}-  \lambda_{1}^{4}\right] \right) - p \frac{1}{\lambda_{1}}- n \frac{1}{\lambda_{1}}\\
    &= 2 \left( \frac{\partial \hat{\Psi}^{NN} }{\partial I_{1}} + \lambda_{1}^{2}  \frac{\partial \hat{\Psi}^{NN} }{\partial I_{2}} \right) \left[ \lambda_{1} - \frac{1}{\lambda_{1}^{5}}\right]
\end{aligned}
\end{equation}

\subsubsection{Pure shear stress}
Pure shear stress is defined by
\begin{equation}
    \bm{F} = \text{diag} \lbrace \lambda_{1}, 1, \frac{1}{\lambda_{1}} \rbrace, \quad \text{and} \quad \bm{P} = \text{diag} \lbrace P_{1}, P_{2},0\rbrace.
\end{equation}
The invariants yield
\begin{equation}
    I_{1} = I_{2} = \lambda_{1}^{2} + 1 + \frac{1}{\lambda_{1}^{2}}.
\end{equation}
Similarly, to the last two cases, we can obtain an expression for the pressure from $P_{3}=0$ which leads to
\begin{equation}
    p = 2 \left( \frac{2}{\lambda^{2}} \frac{\partial \hat{\Psi}^{NN} }{\partial I_{1}} + \frac{\partial \hat{\Psi}^{NN} }{\partial I_{2}} \left[ 1 + \frac{1}{\lambda^{2}} \right] \right) - n.
\end{equation}
Then, we see that
\begin{equation}
    P_{1} = 2 \left(\frac{\partial \hat{\Psi}^{NN} }{\partial I_{1}} + \frac{\partial \hat{\Psi}^{NN} }{\partial I_{2}} \right) \left[ \lambda - \frac{1}{\lambda^{3}} \right]
\end{equation}
and
\begin{equation}
    P_{2} = 2 \left( \frac{\partial \hat{\Psi}^{NN} }{\partial I_{1}} + \lambda^{2} \frac{\partial \hat{\Psi}^{NN} }{\partial I_{2}} \right) \left[ 1 - \frac{1}{\lambda^{2}} \right].
\end{equation}

\subsubsection{Simple shear deformation}
Simple shear deformation is defined by the deformation gradient
\begin{equation}
    \bm{F} =\begin{bmatrix}
        1 & \gamma  & 0 \\
        0 & 1 & 0 \\
        0 & 0 & 1
    \end{bmatrix}
  \end{equation}  
which yields a right Cauchy-Green tensor of the form
\begin{equation}
    \bm{C} = \begin{bmatrix}
        1 & \gamma  & 0 \\
        \gamma & 1+\gamma^{2} & 0 \\
        0 & 0 & 1
    \end{bmatrix}
    .
\end{equation}
The resulting invariants read
\begin{equation}
    I_{1} = 3+ \gamma^{2}, \quad I_{2}  = 3+ \gamma^{2}.
\end{equation}
The shear stress which is independent of the pressure can then be derived as
  \begin{equation}
      P_{12} = 2 \gamma \left(  \frac{\partial \hat{\Psi}^{NN} }{\partial I_{1}} +  \frac{\partial \hat{\Psi}^{NN} }{\partial I_{2}} \right) .
  \end{equation}

\subsubsection{Simple torsion}
The deformation gradient for simple torsion can be defined as 
\begin{equation}
    \bm{F} = \begin{bmatrix}
        1& 0 & 0 \\
        0 & 1& \rho \, \phi \\
        0 &0 & 1
    \end{bmatrix}
\end{equation}
where $\rho$ is a normalized radial coordinate and $\phi$ describes the normalized twist.
The invariants are defined as
\begin{equation}
    I_{1} = I_{2} = 3 + (\rho \phi)^{2}.
\end{equation}
We follow \cref{hartmann2001parameter} and \cref{flaschel2023automatedBrain} to obtain the normalized torque defined by
\begin{equation}
\begin{aligned}
        \tau &= \int_{0}^{1} 2 \pi \rho^{2} \frac{\partial \hat{\Psi}}{\partial (\rho \phi)} d \rho \\
        &= \int_{0}^{1} 2 \pi \rho^{2} \left( \frac{\partial \hat{\Psi}}{\partial I_{1}} \frac{\partial I_{1}}{\partial (\rho \phi)} + \frac{\partial \hat{\Psi}}{\partial I_{2}} \frac{\partial I_{2}}{\partial (\rho \phi)} \right) d \rho \\
        &= \int_{0}^{1} 4 \pi \rho^{3} \phi \left( \frac{\partial \hat{\Psi}}{\partial I_{1}} + \frac{\partial \hat{\Psi}}{\partial I_{2}} \right) d \rho
\end{aligned}
\end{equation}
where we can use a trapezoidal rule to solve the integral numerically \cite{suli2003introduction}.

\clearpage
\section{Datasets}
\setcounter{table}{0}
\renewcommand{\thetable}{A\arabic{table}}

The six datasets used for training are given here.

\begin{table}[b]
\begin{center}
\begin{tabular}{||c| c | c||}
    \hline
$I_{1}$
            & $I_{2}$
                    & $J$\\
    \hline
        \hline
  \makecell{ $[-]$ }  &       \makecell{ $[-]$ }     &       \makecell{ $[-]$ }   \\
      \hline
    \hline
  3.0 & 3.0 & 1.0 \\ 3.095 & 3.186 & 1.045 \\ 2.721 & 2.372 & 0.817 \\ 3.209 & 3.326 & 1.048 \\ 2.494 & 2.01 & 0.723 \\ 3.089 & 3.069 & 0.983 \\ 2.968 & 2.828 & 0.923 \\ 3.016 & 2.964 & 0.97 \\ 3.399 & 3.839 & 1.201 \\ 3.289 & 3.603 & 1.147 \\ 3.458 & 3.92 & 1.204 \\ 2.866 & 2.714 & 0.921 \\ 3.366 & 3.696 & 1.154 \\ 3.432 & 3.855 & 1.193 \\ 2.759 & 2.52 & 0.873 \\ 2.649 & 2.253 & 0.781 \\ 3.5 & 4.081 & 1.259 \\ 2.876 & 2.649 & 0.883 \\ 3.432 & 3.924 & 1.223 \\ 3.555 & 4.152 & 1.261 \\ 3.194 & 3.322 & 1.064 \\ 3.391 & 3.759 & 1.163 \\ 2.568 & 2.126 & 0.754 \\ 3.466 & 4.002 & 1.241 \\ 3.248 & 3.516 & 1.126 \\ 2.961 & 2.882 & 0.962 \\ 3.297 & 3.54 & 1.116 \\ 3.072 & 3.069 & 1.002 \\ 2.921 & 2.737 & 0.902 \\ 3.152 & 3.199 & 1.016 \\ 2.634 & 2.313 & 0.822 \\ 3.136 & 3.201 & 1.034 \\ 3.25 & 3.421 & 1.073 \\ 3.14 & 3.286 & 1.071 \\ 3.047 & 3.095 & 1.024 \\ 3.604 & 4.323 & 1.314 \\ 2.796 & 2.494 & 0.846 \\ 3.225 & 3.419 & 1.093 \\ 3.329 & 3.67 & 1.158 \\ 2.917 & 2.835 & 0.958 \\ 3.49 & 4.005 & 1.227 \\ 2.817 & 2.644 & 0.909 \\ 3.547 & 4.193 & 1.285 \\ 3.18 & 3.37 & 1.091 \\ 3.321 & 3.59 & 1.119 \\ 3.361 & 3.764 & 1.185 \\ 2.703 & 2.434 & 0.855 \\ 2.443 & 1.981 & 0.73 \\ 3.528 & 4.086 & 1.248 \\ 3.583 & 4.24 & 1.287 \\ 
    \hline
\end{tabular}
    \end{center}
    \caption{$50$ space-filling samples in $20\%$ strain range in terms of the invariants; obtained with the sampling algorithm proposed in \cref{fuhg2021physics}.}
    \label{tab:Compressible50}
\end{table}

\begin{table}[b]
\begin{center}
\begin{tabular}{|c| c||c|c||c|c||c|c||c|c||c|c| }
    \hline
\multicolumn{2}{|c||}{\makecell{ \textbf{UT}\\\textbf{rubber}\\Treloar $20^{\circ}$ C}}
            & \multicolumn{2}{c||}{\makecell{ \textbf{ET}\\\textbf{rubber}\\Treloar $20^{\circ}$ C}}
                    & \multicolumn{2}{c|}{\makecell{ \textbf{PS}\\\textbf{rubber}\\Treloar $20^{\circ}$ C}}
                            & \multicolumn{2}{|c||}{\makecell{ \textbf{UT}\\\textbf{rubber}\\Treloar $50^{\circ}$ C}}
            & \multicolumn{2}{c||}{\makecell{ \textbf{ET}\\\textbf{rubber}\\Treloar $50^{\circ}$ C}}
                    & \multicolumn{2}{c|}{\makecell{ \textbf{PS}\\\textbf{rubber}\\Treloar $50^{\circ}$ C}}       \\
    \hline
        \hline
  \makecell{ $\lambda$\\$[-]$ }  &     \makecell{ $P$\\$[\text{MPa}]$ } &     \makecell{ $\lambda$\\$[-]$ }  &     \makecell{ $P$\\$[\text{MPa}]$ }  &     \makecell{ $\lambda$\\$[-]$ }  &     \makecell{ $P$\\$[\text{MPa}]$ }  &     \makecell{ $\lambda$\\$[-]$ }  &     \makecell{ $P$\\$[\text{MPa}]$ }&    \makecell{ $\lambda$\\$[-]$ }  &     \makecell{ $P$\\$[\text{MPa}]$ }&    \makecell{ $\lambda$\\$[-]$ }  &     \makecell{ $P$\\$[\text{MPa}]$ }  \\
      \hline
    \hline
  1	&0& 1	& 0& 1 & 0 &1 & 0 & 1 & 0 & 1 & 0 \\
1.01 &	0 &1.04	& 0.09 & 1.05 & 0.06 &    1.11 & 0.17 & 1.02 & 0.15 & 1.04 & 0.17 \\
1.13 & 0.14 &1.08	& 0.16& 1.13 & 0.16 &    1.23 & 0.29 & 1.08 & 0.3 &1.23 & 0.4 \\ 
1.23 & 0.24 &1.12	& 0.24& 1.2 & 0.24 &     1.57 & 0.54 & 1.16 & 0.48& 1.48 & 0.63 \\
1.41 & 0.33&1.15	& 0.26& 1.33 & 0.33 & 2.12 & 0.8 & 1.37 & 0.74 &2.52 & 1.03 \\
1.61 & 0.43&1.21	& 0.33& 1.45 & 0.42 &   2.73 & 1.03 & 1.57 & 0.92& 3.51 & 1.49 \\
1.89 & 0.52&1.32	& 0.44& 1.86 & 0.59 &    3.36 & 1.3 & 1.96 & 1.17& 4.33 & 1.9 \\
2.17 & 0.59&1.43	& 0.51& 2.4 & 0.77 & 3.95 & 1.57 & 2.46 & 1.49 &5.07 & 2.36 \\
2.45 & 0.68&1.7	& 0.66& 2.99 & 0.95 &  4.39 & 1.79 & 2.79 & 1.78& 5.74 & 2.74 \\
3.06 & 0.87&1.95	& 0.77& 3.5 & 1.13 &   5.29 & 2.29 & 3.14 & 2.04& 6.24 & 3.22 \\
3.62 & 1.06&2.5	& 0.97& 3.98 & 1.29 &  6.11 & 2.8 & 3.45 & 2.33 &6.36 & 3.63 \\
4.06 & 1.24&3.04	& 1.26& 4.39 &1.48 &  6.54 & 3.75 & 3.6 & 2.53& 6.65 & 4.49 \\ 
4.82 & 1.6&3.44	& 1.47& 4.72 & 1.65 &  6.95 & 5.27 & 3.86 & 2.96 &6.91 & 5.34 \\
5.41 & 1.95&3.76	& 1.73& 4.99 & 1.82 &   7.43 & 7.73 & 4.11 & 3.24&  7.06 & 6.23 \\
5.79 & 2.3&4.03	& 1.97& & &  7.76 & 10.21 & 4.6 & 4.24 & 7.26 & 7 \\
6.23 & 2.68&4.26	& 2.23& & &   & & 5.06 & 6.15 & 7.42 & 7.89 \\
6.46 & 3.03&4.45	& 2.45& & &   & & 5.28 & 6.99 & 7.56 & 9.18 \\
6.67& 3.4& & && &   & &               5.42 & 8.18 & 7.83 & 10.9 \\
6.96& 3.78&& & &  & & & 5.59&9.87 & & \\
7.14& 4.16&& && & & &5.67&11.59&    &  \\     
7.25& 4.49&& && && && &    &  \\     
7.36 &4.86&& && && && &    &  \\     
7.49&	5.24&& && && && &    &  \\     
7.6&	5.6&& && && && &    &  \\     
7.69	&6.33&& && && && &    &  \\     
    \hline
\end{tabular}
    \end{center}
    \caption{Treloar dataset describing the response of vulcanized rubber under different loading conditions and temperatures. Data adapted from \cite{linka2023new}.}
    \label{tab:treloarRubber}
\end{table}

\begin{table}
\begin{center}
\begin{tabular}{|c| c||c|c||c|c||c|c||c|c||c|c| }
    \hline
\multicolumn{2}{|c|}{\makecell{ \textbf{UT}\\\textbf{Cortex}\\Budday \cite{budday2017mechanical}}}        & \multicolumn{2}{c|}{\makecell{ \textbf{UC}\\\textbf{Cortex}\\Budday \cite{budday2017mechanical}}}   & \multicolumn{2}{c|}{\makecell{ \textbf{SS}\\\textbf{Cortex}\\Budday}}      
                            & \multicolumn{2}{c|}{\makecell{ \textbf{UT}\\\textbf{Corona Radiata}\\Budday \cite{budday2017mechanical}}}        & \multicolumn{2}{c|}{\makecell{ \textbf{UC}\\\textbf{Corona Radiata}\\Budday \cite{budday2017mechanical}}}   & \multicolumn{2}{c|}{\makecell{ \textbf{SS}\\\textbf{Corona Radiata}\\Budday \cite{budday2017mechanical}}}          \\
    \hline
        \hline
  \makecell{ $\lambda$\\$[-]$ }  &     \makecell{ $P$\\$[\text{MPa}]$ } &     \makecell{ $\lambda$\\$[-]$ }  &     \makecell{ $P$\\$[\text{MPa}]$ }  &     \makecell{ $\lambda$\\$[-]$ }  &     \makecell{ $P$\\$[\text{MPa}]$ }  &     \makecell{ $\lambda$\\$[-]$ }  &     \makecell{ $P$\\$[\text{MPa}]$ }&    \makecell{ $\lambda$\\$[-]$ }  &     \makecell{ $P$\\$[\text{MPa}]$ }&    \makecell{ $\lambda$\\$[-]$ }  &     \makecell{ $P$\\$[\text{MPa}]$ }  \\
      \hline
    \hline
  1 & 0 & 1 & 0 & 0 & 0& 1 & 0 &1 & 0 &  0  & 0 \\
1.0063 & 0.0251 & 0.9938 & -0.0308 & 0.0125 & 0.0147 &1.0063 & 0.0157 & 0.9938 & -0.0193 & 0.0125 & 0.0079\\  
1.0125 & 0.0462 & 0.9875 & -0.0659 & 0.025 & 0.0294 & 1.0125 & 0.0235 & 0.9875 & -0.0387 & 0.025 & 0.0159\\   
1.0188 & 0.0666 & 0.9812 & -0.104 & 0.0375 & 0.0486 & 1.0188 & 0.0345 & 0.9812 & -0.0543 & 0.0375 & 0.0238\\    
 1.025 & 0.0838 & 0.975 & -0.1479 &0.05 & 0.0633 & 1.025 & 0.0423 & 0.975 & -0.08 & 0.05 & 0.0318\\     
1.0312 & 0.101 & 0.9688 & -0.1908 & 0.0625 & 0.0814 &1.0312 & 0.0509 & 0.9688 & -0.104 & 0.0625 & 0.0409\\
1.0375 & 0.1175 & 0.9625 & -0.2375 & 0.075 & 0.0983 & 1.0375 & 0.0572 & 0.9625 & -0.1305 & 0.075 & 0.0488 \\     
1.0437 & 0.1324 & 0.9563 & -0.292 & 0.0875 & 0.1186 &     1.0437 & 0.0642 & 0.9563 & -0.1674 & 0.0875 & 0.0601\\
1.05 & 0.1488 & 0.95 & -0.3504 & 0.1 & 0.1412 &1.05 & 0.0721 & 0.95 & -0.2024  & 0.1 & 0.0681\\
1.0562 & 0.1661 & 0.9437 & -0.4127 & 0.1125 & 0.1649 &    1.0562 & 0.0791 & 0.9437 & -0.2453 & 0.1125 & 0.0817\\ 
1.0625 & 0.1856 & 0.9375 & -0.4866 & 0.125 & 0.1942 & 1.0625 & 0.0869 & 0.9375 & -0.2959 & 0.125 & 0.0964\\
1.0688 & 0.2091 & 0.9313 & -0.5684 & 0.1375 & 0.2292 &     1.0688 & 0.094&0.9313 & -0.3543 & 0.1375 & 0.1133\\
1.075 & 0.2366 & 0.925 & -0.6579 & 0.15 & 0.2698 &    1.075 & 0.105 & 0.925 & -0.4127 & 0.15 & 0.1347\\
1.0813 & 0.271 & 0.9187 & -0.763 & 0.1625 & 0.3227 &   1.0813 & 0.1151 & 0.9187 & -0.4827 & 0.1625 & 0.1596\\  
1.0875 & 0.3125 & 0.9125 & -0.8837 & 0.175 & 0.3791 &     1.0875 & 0.1292 & 0.9125 & -0.5723 & 0.175 & 0.1878\\
1.0938 & 0.365 & 0.9062 & -1.0005 & 0.1875 & 0.4557 &     1.0938 & 0.1418 & 0.9062 & -0.6657  & 0.1875 & 0.2227\\
1.1 & 0.4151 & 0.9 & -1.1484 & 0.2 & 0.5435 &   1.1 & 0.1582 & 0.9 & -0.7591 & 0.2 & 0.2611\\    
    \hline
\end{tabular}
    \end{center}
    \caption{Data from mechanical tests on two parts of the human brain. Data adapted from \cref{budday2017mechanical} and \cref{pierre2023principal}.}
    \label{tab:CortexRadiataData}
\end{table}

\begin{table}
\begin{center}
\scalebox{0.9}{
\begin{tabular}{|c| c||c|c||c|c||c|c||}
    \hline
\multicolumn{2}{|c||}{\makecell{ \textbf{UT}\\Midbrain\\ Section 1}}
            & \multicolumn{2}{c||}{\makecell{ \textbf{ST}\\Midbrain\\ Section 1}}
               & \multicolumn{2}{c||}{\makecell{\textbf{UT}\\Midbrain\\ Section 2}}         & \multicolumn{2}{c||}{\makecell{\textbf{ST}\\Midbrain\\ Section 2}}       \\
    \hline
        \hline
 \makecell{ $\epsilon_{11}$\\$[-]$ }  &     \makecell{ $\sigma_{11}$\\$[\text{MPa}]$ } &    \makecell{ Angle\\$[-]$ }  &     \makecell{ $\tau$\\$[\text{MPa}]$ } &
 \makecell{ $\epsilon_{11}$\\$[-]$ }  &     \makecell{ $\sigma_{11}$\\$[\text{MPa}]$ } &
   \makecell{ Angle\\$[-]$ }  &     \makecell{ $\tau$\\$[\text{MPa}]$ }\\
      \hline
    \hline
0.85 & -1002.346 & -0.325 & -182.192 & 0.849 & -1071.268 & -0.354 & -89.688 \\ 0.855 & -958.358 & -0.315 & -169.521 & 0.854 & -1035.775 & -0.344 & -82.344 \\ 0.86 & -859.824 & -0.305 & -155.479 & 0.859 & -944.507 & -0.334 & -76.558 \\ 0.865 & -778.886 & -0.295 & -142.808 & 0.864 & -857.465 & -0.323 & -71.662 \\ 0.87 & -708.504 & -0.284 & -131.164 & 0.869 & -782.254 & -0.314 & -67.433 \\ 0.875 & -637.243 & -0.274 & -119.178 & 0.874 & -714.648 & -0.303 & -63.427 \\ 0.88 & -569.501 & -0.264 & -109.247 & 0.879 & -647.042 & -0.293 & -59.421 \\ 0.885 & -501.76 & -0.254 & -100.0 & 0.884 & -584.507 & -0.283 & -55.638 \\ 0.89 & -443.695 & -0.244 & -91.096 & 0.889 & -525.352 & -0.273 & -52.522 \\ 0.895 & -419.062 & -0.233 & -82.877 & 0.895 & -478.028 & -0.262 & -49.184 \\ 0.9 & -387.39 & -0.223 & -75.342 & 0.9 & -429.014 & -0.252 & -46.068 \\ 0.905 & -343.402 & -0.213 & -69.178 & 0.905 & -384.225 & -0.242 & -43.175 \\ 0.91 & -293.255 & -0.203 & -63.014 & 0.91 & -341.127 & -0.232 & -40.504 \\ 0.916 & -252.786 & -0.193 & -57.192 & 0.915 & -305.634 & -0.222 & -37.611 \\ 0.921 & -224.633 & -0.183 & -52.397 & 0.92 & -272.676 & -0.212 & -35.163 \\ 0.926 & -209.677 & -0.173 & -47.945 & 0.925 & -237.183 & -0.201 & -32.715 \\ 0.931 & -192.082 & -0.162 & -42.123 & 0.93 & -205.07 & -0.191 & -30.267 \\ 0.936 & -170.088 & -0.153 & -38.356 & 0.936 & -174.648 & -0.181 & -27.819 \\ 0.941 & -139.296 & -0.142 & -34.589 & 0.941 & -155.211 & -0.171 & -25.593 \\ 0.946 & -113.783 & -0.132 & -30.822 & 0.946 & -138.31 & -0.161 & -23.591 \\ 0.951 & -91.789 & -0.122 & -27.397 & 0.951 & -123.944 & -0.15 & -21.588 \\ 0.956 & -90.909 & -0.112 & -24.315 & 0.956 & -108.732 & -0.141 & -20.03 \\ 0.961 & -80.352 & -0.101 & -21.575 & 0.961 & -91.831 & -0.13 & -18.027 \\ 0.966 & -58.358 & -0.092 & -18.493 & 0.966 & -78.31 & -0.12 & -16.469 \\ 0.971 & -44.282 & -0.081 & -16.096 & 0.971 & -67.324 & -0.11 & -14.911 \\ 0.976 & -25.806 & -0.071 & -14.041 & 0.976 & -60.563 & -0.1 & -13.798 \\ 0.981 & -24.927 & -0.061 & -11.986 & 0.981 & -50.423 & -0.09 & -12.24 \\ 0.986 & -21.408 & -0.051 & -9.932 & 0.986 & -36.901 & -0.08 & -11.35 \\ 0.992 & -9.091 & -0.041 & -7.534 & 0.992 & -18.31 & -0.07 & -9.57 \\ 0.997 & -1.173 & -0.03 & -5.822 & 0.997 & -5.634 & -0.059 & -8.902 \\ 1.001 & 0.587 & -0.02 & -3.082 & 1.002 & -3.944 & -0.049 & -6.899 \\ 1.007 & 1.466 & -0.01 & -1.027 & 1.007 & -3.099 & -0.039 & -5.341 \\ 1.012 & 2.346 & 0.0 & 0.0 & 1.012 & 3.662 & -0.028 & -4.006 \\ 1.017 & 3.226 & 0.01 & 3.425 & 1.017 & 8.732 & -0.018 & -2.893 \\ 1.022 & 4.106 & 0.02 & 5.479 & 1.022 & 11.268 & -0.008 & -1.558 \\ 1.027 & 4.106 & 0.03 & 7.534 & 1.027 & 12.958 & 0.002 & -0.223 \\ 1.032 & 15.543 & 0.041 & 10.274 & 1.033 & 18.028 & 0.012 & 1.113 \\ 1.037 & 19.941 & 0.05 & 13.356 & 1.038 & 23.099 & 0.022 & 2.448 \\ 1.042 & 20.821 & 0.061 & 16.438 & 1.043 & 24.789 & 0.032 & 3.783 \\ 1.047 & 23.46 & 0.071 & 19.178 & 1.048 & 28.169 & 0.043 & 5.564 \\ 1.052 & 32.258 & 0.081 & 22.603 & 1.053 & 29.014 & 0.052 & 6.454 \\ 1.058 & 32.258 & 0.091 & 26.37 & 1.058 & 30.704 & 0.063 & 8.457 \\ 1.063 & 32.258 & 0.102 & 30.137 & 1.063 & 34.93 & 0.073 & 10.682 \\ 1.068 & 33.138 & 0.112 & 34.589 & 1.068 & 37.465 & 0.083 & 12.24 \\ 1.073 & 37.537 & 0.121 & 38.699 & 1.073 & 37.465 & 0.093 & 14.466 \\ 1.078 & 48.094 & 0.132 & 43.493 & 1.078 & 40.0 & 0.103 & 16.914 \\ 1.083 & 52.493 & 0.142 & 48.63 & 1.084 & 40.845 & 0.113 & 19.585 \\ 1.088 & 53.372 & 0.152 & 54.11 & 1.089 & 41.69 & 0.124 & 22.033 \\ 1.093 & 55.132 & 0.162 & 60.274 & 1.094 & 46.761 & 0.134 & 24.481 \\ 1.098 & 56.012 & 0.173 & 66.096 & 1.099 & 47.606 & 0.144 & 27.596 \\ 1.103 & 56.012 & 0.183 & 73.288 & 1.104 & 49.296 & 0.154 & 31.157 \\ 1.108 & 56.891 & 0.193 & 80.479 & 1.109 & 49.296 & 0.164 & 34.941 \\ 1.113 & 57.771 & 0.203 & 89.041 & 1.114 & 50.141 & 0.175 & 38.279 \\ 1.118 & 57.771 & 0.213 & 97.603 & 1.119 & 51.831 & 0.184 & 42.062 \\ 1.123 & 62.17 & 0.223 & 107.877 & 1.125 & 56.056 & 0.195 & 46.291 \\ 1.128 & 64.809 & 0.234 & 117.808 & 1.129 & 56.901 & 0.205 & 50.964 \\ 1.133 & 64.809 & 0.244 & 129.11 & 1.135 & 56.901 & 0.215 & 56.083 \\ 1.138 & 65.689 & 0.254 & 141.438 & 1.14 & 56.901 & 0.225 & 61.647 \\ 1.143 & 65.689 & 0.265 & 154.795 & 1.145 & 60.282 & 0.236 & 66.988 \\ 1.149 & 69.208 & 0.274 & 167.466 & 1.15 & 62.817 & 0.245 & 72.774 \\ 
    \hline
\end{tabular}
}
    \end{center}
    \caption{Data from mechanical tests on two sections of the midbrain. Data adapted from \cref{flaschel2023automatedBrain}.}
    \label{tab:Midbrain}
\end{table}

\begin{table}
\begin{center}
\begin{tabular}{||c| c||c| c||c|c||}
    \hline
    \multicolumn{2}{||c||}{\makecell{ \textbf{Drucker}\\yield function}} &
\multicolumn{2}{|c||}{\makecell{ \textbf{Cacazu}\\yield function}}
            & \multicolumn{2}{c||}{\makecell{ \textbf{Tresca}\\yield function}}
                           \\
    \hline
        \hline
  \makecell{ $\pi_{1}$\\$[\text{MPa}]$ }  &     \makecell{ $\pi_{2}$\\$[\text{MPa}]$ } &  
  \makecell{ $\pi_{1}$\\$[\text{MPa}]$ }  &     \makecell{ $\pi_{2}$\\$[\text{MPa}]$ } &     \makecell{ $\pi_{1}$\\$[\text{MPa}]$ }  &     \makecell{ $\pi_{2}$\\$[\text{MPa}]$ }   \\
      \hline
    \hline
-1.0783 & 0.0 & -0.5117 & 0.0 & -0.1948 & 0.0 \\ -1.0652 & -0.2352 & -0.4844 & -0.1052 & -0.1774 & -0.032 \\ -1.0092 & -0.4671 & -0.4155 & -0.1906 & -0.1559 & -0.0695 \\ -0.8835 & -0.6711 & -0.3325 & -0.2514 & -0.134 & -0.1073 \\ -0.7041 & -0.8285 & -0.2557 & -0.3008 & -0.1129 & -0.1441 \\ -0.5052 & -0.9531 & -0.1823 & -0.3443 & -0.0935 & -0.17 \\ -0.2925 & -1.0548 & -0.1063 & -0.3858 & -0.0536 & -0.1698 \\ -0.0603 & -1.1123 & -0.0231 & -0.4273 & -0.0113 & -0.1698 \\ 0.1789 & -1.0916 & 0.0783 & -0.465 & 0.0326 & -0.1698 \\ 0.4009 & -1.0074 & 0.1875 & -0.4682 & 0.074 & -0.1698 \\ 0.6061 & -0.8937 & 0.2855 & -0.4224 & 0.1028 & -0.1619 \\ 0.7982 & -0.755 & 0.3512 & -0.333 & 0.1232 & -0.1262 \\ 0.9553 & -0.5747 & 0.3785 & -0.2288 & 0.145 & -0.0882 \\ 1.0447 & -0.3524 & 0.3858 & -0.1313 & 0.1666 & -0.0508 \\ 1.0753 & -0.117 & 0.389 & -0.0427 & 0.1878 & -0.0139 \\ 1.0753 & 0.117 & 0.389 & 0.0427 & 0.1878 & 0.0139 \\ 1.0447 & 0.3524 & 0.3858 & 0.1313 & 0.1666 & 0.0508 \\ 0.9553 & 0.5747 & 0.3785 & 0.2288 & 0.145 & 0.0882 \\ 0.7982 & 0.755 & 0.3512 & 0.333 & 0.1232 & 0.1262 \\ 0.6061 & 0.8937 & 0.2855 & 0.4224 & 0.1028 & 0.1619 \\ 0.4009 & 1.0074 & 0.1875 & 0.4682 & 0.074 & 0.1698 \\ 0.1789 & 1.0916 & 0.0783 & 0.465 & 0.0326 & 0.1698 \\ -0.0603 & 1.1123 & -0.0231 & 0.4273 & -0.0113 & 0.1698 \\ -0.2925 & 1.0548 & -0.1063 & 0.3858 & -0.0536 & 0.1698 \\ -0.5052 & 0.9531 & -0.1823 & 0.3443 & -0.0935 & 0.17 \\ -0.7041 & 0.8285 & -0.2557 & 0.3008 & -0.1129 & 0.1441 \\ -0.8835 & 0.6711 & -0.3325 & 0.2514 & -0.134 & 0.1073 \\ -1.0092 & 0.4671 & -0.4155 & 0.1906 & -0.1559 & 0.0695 \\ -1.0652 & 0.2352 & -0.4844 & 0.1052 & -0.1774 & 0.032 \\ -1.0783 & 0.0 & -0.5117 & 0.0 & -0.1948 & -0.0 \\ 
    \hline
\end{tabular}
    \end{center}
    \caption{Yield function data in $\pi$-plane}
    \label{tab:yieldFunData}
\end{table}

\begin{table}
\begin{center}
\begin{tabular}{|c| c||c|c||c|c||}
    \hline
\multicolumn{2}{|c||}{\makecell{ \textbf{40Cr3MoV} \cite{kang2006uniaxial}\\bainitic steel}}
            & \multicolumn{2}{c||}{\makecell{ \textbf{SS316L} \cite{kang2006uniaxial}\\stainless steel}}
               & \multicolumn{2}{c||}{\makecell{ \textbf{U71Mn} \cite{kang2002uniaxial}\\rail steel}}             \\
    \hline
        \hline
 \makecell{ $\epsilon_{11}$\\$[-]$ }  &     \makecell{ $\sigma_{11}$\\$[\text{MPa}]$ } &    \makecell{ $\epsilon_{11}$\\$[-]$ }  &     \makecell{ $\sigma_{11}$\\$[\text{MPa}]$ } &
   \makecell{ $\epsilon_{11}$\\$[-]$ }  &     \makecell{ $\sigma_{11}$\\$[\text{MPa}]$ }\\
      \hline
    \hline
0.01 & 4.76 & 0.0 & 0.0 & 0.0 & 0.0 \\ 0.05 & 80.95 & 0.04 & 64.29 & 0.21 & 453.23 \\ 0.08 & 157.14 & 0.12 & 152.38 & 0.36 & 508.06 \\ 0.12 & 233.33 & 0.22 & 202.38 & 0.55 & 533.87 \\ 0.16 & 309.52 & 0.41 & 223.81 & 0.74 & 561.29 \\ 0.2 & 404.76 & 0.59 & 235.71 & 0.93 & 587.1 \\ 0.25 & 490.48 & 0.77 & 242.86 & 1.12 & 611.29 \\ 0.28 & 576.19 & 0.97 & 247.62 & 1.31 & 635.48 \\ 0.33 & 652.38 & 1.2 & 257.14 & 1.52 & 659.68 \\ 0.39 & 757.14 & 1.44 & 264.29 & 1.69 & 680.65 \\ 0.47 & 866.67 & 1.68 & 269.05 & 1.89 & 701.61 \\ 0.55 & 952.38 & 1.95 & 276.19 & 2.08 & 722.58 \\ 0.64 & 1033.33 & 2.25 & 283.33 & 2.28 & 740.32 \\ 0.73 & 1109.52 & 2.55 & 288.1 & 2.47 & 756.45 \\ 0.86 & 1176.19 & 2.84 & 295.24 & 2.65 & 772.58 \\ 0.99 & 1242.86 & 3.11 & 297.62 & 2.85 & 787.1 \\ 1.22 & 1319.05 & 3.37 & 304.76 & 3.05 & 801.61 \\ 1.45 & 1371.43 & 3.64 & 307.14 & 3.25 & 812.9 \\ 1.73 & 1414.29 & 3.91 & 311.9 & 3.42 & 824.19 \\ 1.96 & 1452.38 & 4.19 & 319.05 & 3.62 & 833.87 \\ 2.21 & 1480.95 & 4.5 & 323.81 & 3.82 & 845.16 \\ 2.45 & 1500.0 & 4.77 & 328.57 & 4.01 & 851.61 \\ 2.69 & 1521.43 & 5.05 & 333.33 & 4.19 & 861.29 \\ 2.97 & 1542.86 & 5.35 & 333.33 & 4.39 & 869.35 \\ 3.23 & 1557.14 & 5.67 & 342.86 & 4.59 & 875.81 \\ 3.5 & 1571.43 & 5.98 & 345.24 & 4.79 & 882.26 \\ 3.83 & 1588.1 & 6.31 & 350.0 & 4.98 & 887.1 \\ 4.14 & 1602.38 & 6.61 & 354.76 & 5.16 & 891.94 \\ 4.5 & 1616.67 & 6.9 & 359.52 & 5.36 & 896.77 \\ 4.86 & 1628.57 & 7.18 & 361.9 & 5.55 & 901.61 \\ 5.14 & 1635.71 & 7.48 & 366.67 & 5.75 & 904.84 \\ 5.38 & 1638.1 & 7.77 & 371.43 & 5.95 & 908.06 \\ 5.66 & 1642.86 &   &   &   &   \\ 5.92 & 1647.62 &   &   &   &   \\ 6.16 & 1652.38 &   &   &   &   \\ 6.4 & 1654.76 &   &   &   &   \\ 6.62 & 1657.14 &   &   &   &   \\ 6.87 & 1657.14 &   &   &   &   \\ 7.11 & 1657.14 &   &   &   &   \\ 7.35 & 1659.52 &   &   &   &   \\ 7.59 & 1659.52 &   &   &   &   \\ 7.83 & 1661.9 &   &   &   &   \\ 
    \hline
\end{tabular}
    \end{center}
    \caption{Uniaxial tension data for the fit of isotropic hardening function}
    \label{tab:isoHardData}
\end{table}

\end{document}